\documentclass[]{raa}            

\usepackage{graphicx,times}             
\usepackage{natbib}
\usepackage{amssymb,amsmath}
\usepackage{multirow}
\usepackage{tabularx}
\usepackage{enumitem}
\bibpunct{(}{)}{;}{a}{}{,}

\usepackage[pagebackref=true]{hyperref}

\begin{document}

  \title{Full-frame data reduction method: a data mining tool to detect the potential variations in optical photometry
}

   \volnopage{Vol.0 (20xx) No.0, 000--000}      
   \setcounter{page}{1}          

   \author{Zhi-Bin Dai 
      \inst{1,2,3,4}
   \and Hao Zhou
      \inst{5}
   \and Jin Cao
      \inst{5}
   }

   \institute{Yunnan Observatories, Chinese Academy of Sciences, 396 Yangfangwang, Guandu District, Kunming, 650216, P. R. China; {\it zhibin\_dai@ynao.ac.cn}\\
        \and
             Key Laboratory for the Structure and Evolution of Celestial Objects, Chinese Academy of Sciences, P. R. China.\\
        \and
             Center for Astronomical Mega-Science, Chinese Academy of Sciences, 20A Datun Road, Chaoyang District, Beijing, 100012, P. R. China.\\
        \and
             University of Chinese Academy of Sciences, No.19(A) Yuquan Road, Shijingshan District, Beijing, 100049, P.R. China.\\
        \and
             School of Information Science and Engineering of Yunnan University, South Section, East Outer Ring Road, Chenggong District, Kunming, 650500, P. R. China.\\
\vs\no
   {\small Received 20xx month day; accepted 20xx month day}}

\abstract{A Synchronous Photometry Data Extraction (SPDE) program, performing indiscriminate monitors of all stars appearing at the same field of view of astronomical image, is developed by integrating several Astropy affiliated packages to make full use of time series observed by the traditional small/medium aperture ground-based telescope. The complete full-frame stellar photometry data reductions implemented for the two time series of cataclysmic variables: RX\,J2102.0+3359 and Paloma\,J0524+4244 produce 363 and 641 optimal light curves, respectively. A cross-identification with the SIMBAD finds 23 known stars, of which 16 red giant-/horizontal-branch stars, 2 W\,UMa-type eclipsing variables, 2 program stars, a X-ray source and 2 Asteroid Terrestrial-impact Last Alert System variables. Based on the data productions of the SPDE program, a followup Light Curve Analysis (LCA) program identifies 32 potential variable light curves, of which 18 are from the time series of RX\,J2102.0+3359, and 14 are from that of Paloma\,J0524+4244. They are preliminarily separated into periodical, transient, and peculiar types. By querying for the 58 VizieR online data catalogs, their physical parameters and multi-band brightness spanning from X-ray to radio are compiled for future analysis.
}

   \authorrunning{Z.-B. Dai, H. Zhou \& J. Cao}            
   \titlerunning{Full-frame data reduction method}  

   \maketitle

%
%
\section{Introduction}           
\label{sec:sec1}

Logically, all celestial objects are in principle variables within the certain timescales and amplitudes of variations. On the benefits of the current space/ground-based all-sky surveys in optical band (e.g., Kepler \citep{bor10}, GAIA \citep{gai16}, CRTS \citep{dra09}, ASAS \citep{poj05}, and Pan-STARRS \citep{cham16}), the number of General Catalogue of Variable Stars (GCVS, \citep{sam17}) continues increasing. From the version of the GCVS4.2 published on November 2016 to the GCVS5.1, the increment of variable stars is almost 5500 \citep{sam17}. Although these surveys enormously enriched the observation data, their exposure stratagems (i.e., $\sim$\,1-2\,obs/24\,hr at best) usually provide little information on brightness variations for most of variables \citep{dai16}. This means that many evidences on the system light variability of the variable candidates are obtained from their customized followup ground-based observations. Based on them, some of newfound variables may be not first detected, but re-clarified from the known ``constant" stars. For example, the spectrophotometric standard star G24-9 as a DQ7 white dwarf, originally classified by \citet{fil84} was reported to be a potential white-dwarf eclipsing system \citep{lan85}. The followup observations \citep[cf.][]{car88,zuc88} confirmed this eclipse feature leading to G24-9's designation as V1412\,Aql \citep{kho89}. \citet{lan07} further pointed out several misclassified standard stars due to the apparent variability. For most of variable stars, the classification is mainly based on the light curve features extracted from the time-series photometry data. Thus, the combination of a great variety of optical survey data and the archival data is helpful to update the guidelines of classification. To suppress this subjective classification, \citet{pan22} and \citet{abd22} proposed many high-efficient variable star hierarchical classification methods to automatically classify several million celestial sources produced from the immense amount of time-series photometry. Cataclysmic variables (CVs), as the champion of variability, exhibiting the miscellaneous photometric variations blurring the traditional classification scheme, more direct observational properties explored from the time-series photometry are desired to characterize their variations on all time-scales \citep{mun95,mas04,bru20}. Since the quasi-periodical large-amplitude luminosity variation (i.e., outburst $\sim$\,4-9\,mag in optical band) is the remarkable variable phenomenon in CVs, many new CV candidates are discovered from multiple transient surveys \citep{pat19}. Hence, the supplementary observation data taken by the typical meter-class ground-based telescopes are equally important for these optical surveys.

To derive the brightness variations from the typical photometric data (i.e., time series\footnote{Time series is a sequence of continuous two-dimensional digital images (i.e., astronomical images) in a standard Flexible Image Transport System (FITS; \url{https://fits.gsfc.nasa.gov/fits\_home.html}) format with a single exposure recorded by charge-coupled devices (CCD; \citep[e.g.][]{boy70,kri82}) generally containing an astronomical object at least.}), the differential photometry as an useful measurement method is commonly applied to the time series observations. As often as not, the traditional optical photometry only focus on several program stars\footnote{In general, three stars composed of an object, a comparison and check star are enough to carry out the differential photometry.}, while the leftover stars contained within the same field of view (FOV, from several to dozens arc minutes of night sky taken by the typical meter-class ground-based telescope) are mostly abandoned. Obviously, this is a huge waste of the observation data. To reduce this waste as much as possible, it is substantial to seek an available method simultaneously measuring as many stars as possible appearing in the same FOV.

Considering that it is impossible to carry out a specialized observation program detecting the unknown brightness variations in ``constant" stars, a full-frame stellar photometry data reduction method, synchronously extracting the light curves of all stars more than the program star of time series appearing in the same FOV, provides a low-cost and available chance to indiscriminately monitor any potential variability of the claimed ``constant" stars during each time series photometry. This is also a good solution to reduce the waste of observation data caused by the traditional data reduction pipeline. However, the problem of the following acquisition and analysis resulting from the vast quantities of data produced by the CCD \citep{ste87} cannot be neglected. Accompanying with the developments of the computer technique spanning over the last thirty years, this problem is hopefully overcome using the high-powered hardware and sophisticated software. A more efficient method with the ability of synchronously implementing synthetic aperture photometry on a sequence of stars to takes full advantage of the observed time series deserves some detailed discussions.

In this paper, we focus on the performance of a newly developed Synchronous Photometry Data Extraction (hereafter: SPDE) program on ground-based time series photometry. Due to the seeing variations and telescope tracking errors, a sequence of images continuously taken during a time series is hard to stabilize the same patch of night sky. This means that the image excursions are inevitable in the time series photometry. Therefore, a batch reduction for an astronomical object in time series cannot be simply achieved using the synthetic aperture photometry algorithm on a fixed position of the CCD image (FITS), since the same object may locate at the different position on the different FITS due to the image excursion. Although almost every robotic survey (e.g., Dark Energy Survey (DES; \citep{mor18}), Zwicky Transient Facility (ZTF; \citep{mas19}), Pan-STARRS \citep{mag20}) over the last decade has automated data processing pipelines implementing the data acquisition, calibration, flat fielding, astrometry, photometry, image subtraction, and additional analyses like deblending, the SPDE program is not designed for any specified telescope/survey, but for the general astronomical time series lasting only a few hours at night (i.e., the customized photometry using the small/medium aperture ground-based telescopes). Compared with the powerful capabilities of the data-processing pipelines and enormous data volumes produced in the large-scale surveys, the SPDE program does not pursue any novel algorithm used for any specific robotic survey, but attempt to develop a core tool integrating several mature Astropy affiliated packages to carry out the key functionalities of a full-frame stellar photometry data reduction method. 

Compared with the other traditional photometry data reduction pipelines, the SPDE program expects to produce a large volume of light curves hiding many potential variable light curves. A surge in the quantity of light curve indicates that an automated and fast identification of these authentic variations in light curves is essential. Furthermore, an on-line cross-identification to preliminarily figure out the detected variations and corresponding celestial objects is desired. Therefore, a followup light curve analysis (hereafter: LCA) program is also important and necessary to be combined with the SPDE program for implementing a complete full-frame stellar photometry data reduction.

A general overview of the SPDE and LCA programs, composed of the pivotal functionalities of the main tasks (5 and 4 modules in the SPDE and LCA program, respectively), is described in Section~\ref{sec:sec2}. Section~\ref{sec:sec3} shows the results of full-frame stellar photometry data reductions on the two time series observed by two different ground-based telescopes. The preliminary followup light curve analysis results of all the suspected brightness variations discovered from the data productions of the SPDE program are detailed in Section~\ref{sec:sec4}.

\section{Overview of the SPDE and LCA Programs}
\label{sec:sec2}

\subsection{History of the Batch Process}

\begin{figure}
	\centering
	\includegraphics[width=10cm,angle=0]{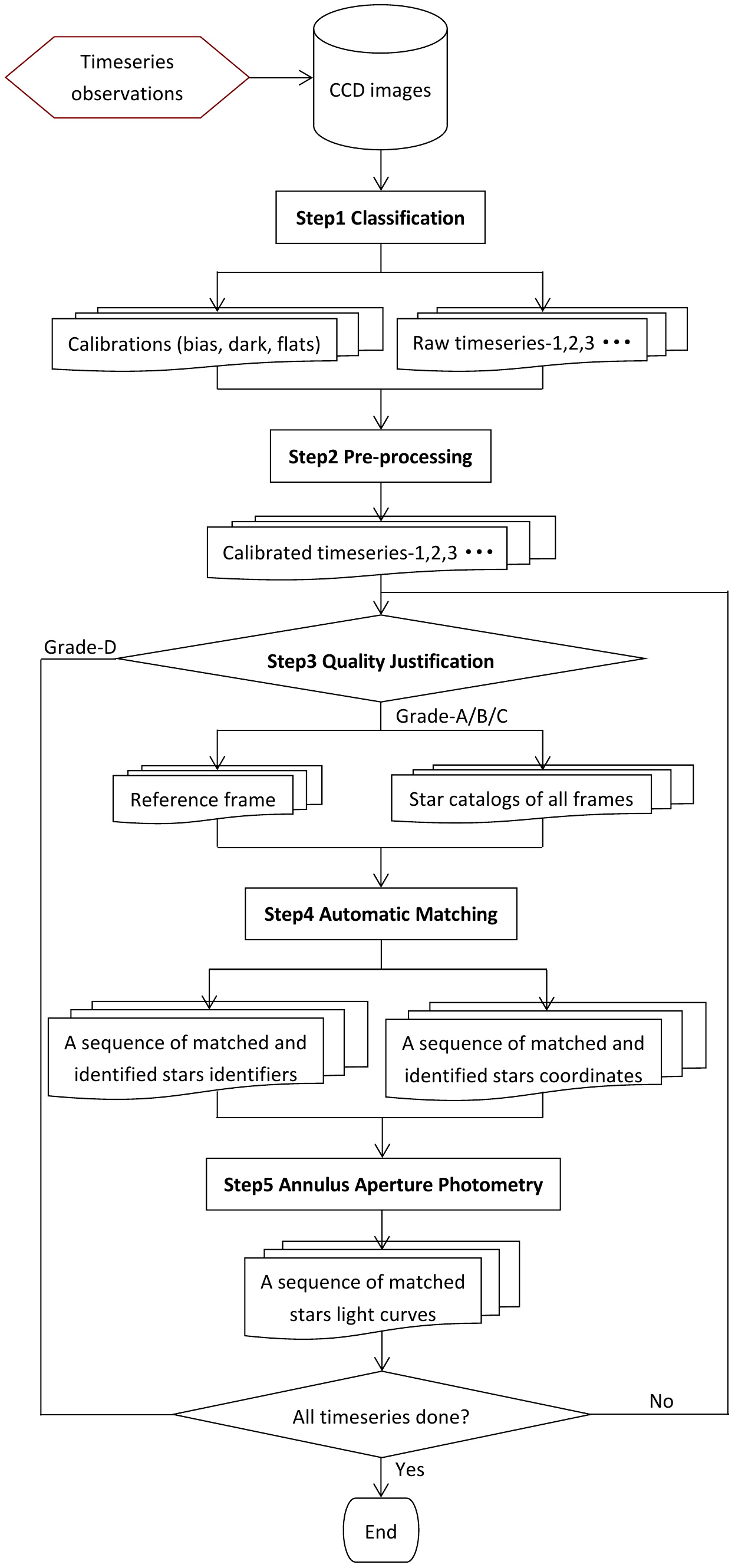}
	\caption{The SPDE program flowchart presents a 5-step pipeline performing a full-frame stellar photometry data reduction on time series.}
	\label{fig1}
\end{figure}

Many computer programs for the general customized-photometric data-processing \citep[cf.][]{tod81,pen86,lup86} were proposed since over 30 years ago. However, the details of their batch processes developed to carry out the automated data reduction and analysis indicate that few of them can authentically carry out a batch process. A general purpose software system, Image Reduction and Analysis Facility (IRAF\footnote{IRAF is distributed by the National Optical Astronomy Observatory (NOAO), which is operated by the Association of Universities for Research in Astronomy under cooperative agreement with the National Science Foundation.}) implements a primary batch jobs for time series using the package IMMATCH, which is able to directly counteract the image excursions by determining the shifts of all images in units of pixel with respect to a reference image preset from the time series by the user, and then overlapping the same patch of image (i.e., a pattern consists of several reference stars) as a reference image. Since the matching ability of the package IMMATCH seriously depends on a set of experiential iteration parameters and an appropriate reference image with a batch of the selected reference stars, the interactive model is commonly required to prevent a program crash in most cases. To reduce the interactivity during the data reductions, the other two packages of IRAF, TAPEREDGE and CROSSCOR provides an indirect method using a two-dimensional cross-correlations algorithm to search for a mutual correspondence in two list of stars coordinates on two images. This matching process outputs a catalog of the reference stars coordinates on each of image. Like IMMATCH, the inputted reference stars coordinates are required to manually pre-mark on a reference image arbitrarily picked out from the time series. IRAF provides a simple programming command language (CL) to make script allowing the user to perform a sequence of tasks in an enclosed and special runtime environment. A customized IRAF script (a functional prototype) wrote by the CL may partially execute a batch processing routine on the basis of some relevant packages mentioned above. Note that NOAO at present is transitioning IRAF to an end-of-support state, and took NOAO's IRAF distribution offline\footnote{The details can be found on the webpage: \url{http://ast.noao.edu/data/software}}.

Aside from IRAF, there were many other general astronomical image processing software, for instance, European Southern Observatory - Munich Image Data Analysis System (ESO-MIDAS\footnote{\url{http://eso.org/sci/software/esomidas/}}, \citet{war92}), Source-Extractor(SExtractor\footnote{\url{https://www.astromatic.net/software/sextractor/}}, \citet{ber96}), and C-Munipack\footnote{\url{http://c-munipack.sourceforge.net/}} (the follow-up Munipack\footnote{\url{http://munipack.physics.muni.cz/munipack.html}}, \citet{hro14}). The former two software cannot perform the batch process, while C-Munipack provides a matching function for time series similar to that of the package IMMATCH of IRAF. However, the main goal of C-Munipack is the portability and comfortability, rather than automaticity reducing the human intervention as much as possible. Thus, C-Munipack with a simple and intuitive graphical user interface is designed to be a semi-automatic procedure for the data reduction of time series.

\subsection{Pipeline of the SPDE Program}

\begin{figure}
	\centering
	\begin{minipage}{13cm}
		\includegraphics[width=12cm,angle=0]{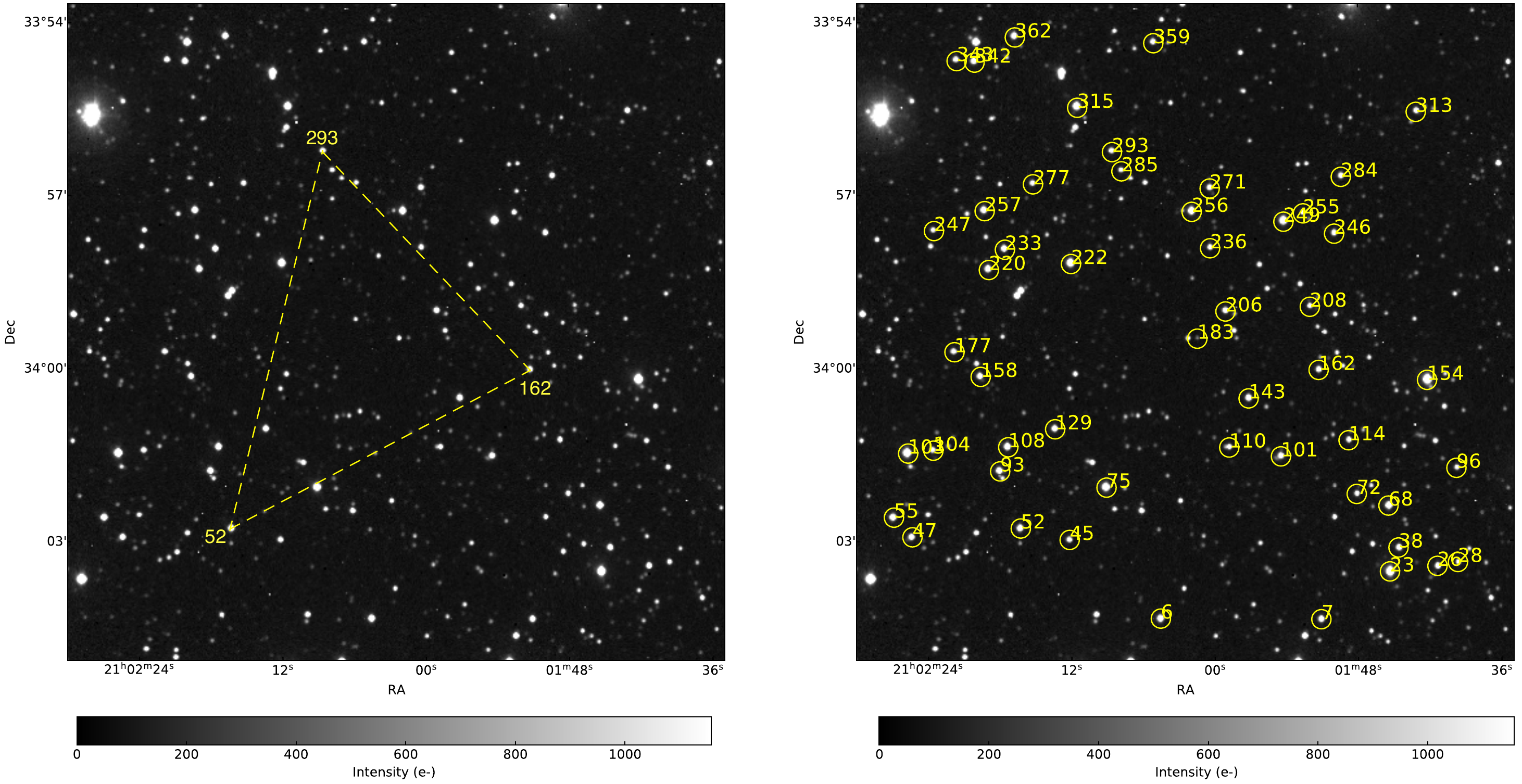}
	\end{minipage}	
	\begin{minipage}{13cm}
		\includegraphics[width=12cm,angle=0]{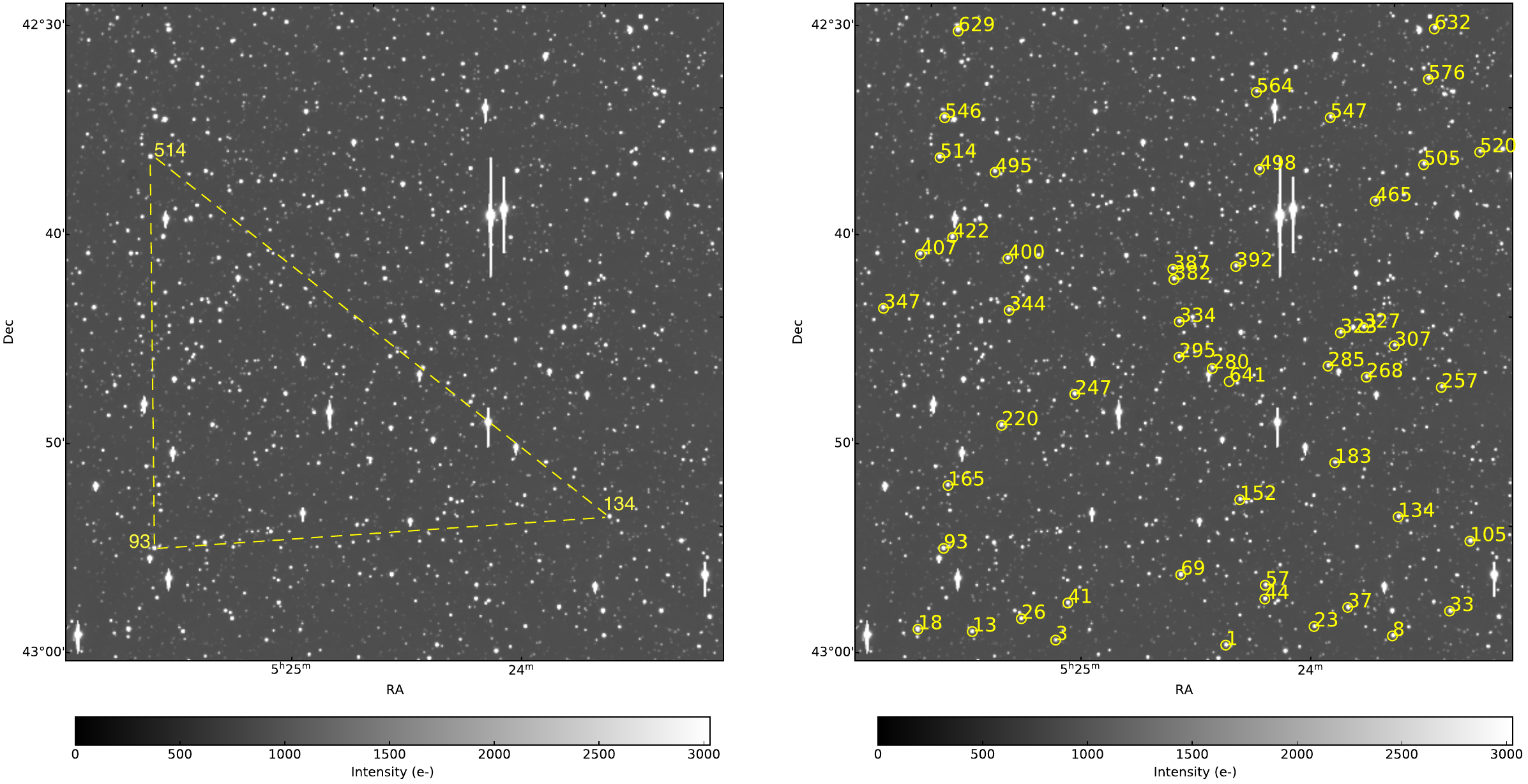}
	\end{minipage}
	\caption{From top to bottom, the mapped reference images in the time series of RX\,J2102.0+3359 and Paloma\,J0524+4244 with the transformed celestial coordinates (RA and Dec) are shown. The north and east of sky is towards the bottom and left, respectively. The triangle consist of three landmark stars with the indexed Nos. shown in the left-hand panels refers to the standard triangle used in Step4 Automatic matching. The brightest 50 stars with the indexed Nos. are marked on the right-hand panels.}
	\label{fig2}
\end{figure}

Based on a group of popular and professional astronomy Python packages of the Astropy Project\footnote{\url{http://www.astropy.org}} \citep{ast13,ast18}, the SPDE program is developed by using the Python programming language. The two Astropy affiliated packages: ccdproc\footnote{\url{https://ccdproc.readthedocs.io/en/latest/index.html}} \citep{cra21} and photutils\footnote{\url{https://photutils.readthedocs.io/en/stable/index.html}} \citep{bra19}, are combined to be a core package accomplishing the whole pipeline of the SPDE program. Moreover, the package astroquery \citep{gin19} is used for querying astronomical web forms and databases, e.g., SIMBAD\footnote{\url{http://simbad.cds.unistra.fr/simbad/}} and VizieR\footnote{It provides the most complete on-line library of published astronomical catalogues at the homepage: \url{https://vizier.cds.unistra.fr/viz-bin/VizieR}.} \citep{och00}. The ccdproc package performs the classifications and the basic calibrated processes, while the photutils package implements DAOPHOT (i.e., the DAOFIND algorithm) using the API DAOStarFinder and annulus aperture photometry. Considering that the popular CCD instrument attached on a common ground-based telescope has the square FOV, the SPDE program is designed for the square CCD images. Compared with the big survey data, the small data volumes of a customized time series can be easily reduced by a common desktop/laptop, rather than a specialized computer. Hence, the hardware used to implement the SPDE program is not considered any more.

Although the common pipeline of photometry data reduction conducted by a general astronomical image processing software only has two steps: pre-processing and aperture photometry, there are massive and trivial manual operations for the preparations of the input data and adjustments of the parameters in the initial each step. To achieve a highly-automated data reduction of time series, we designed a 5-step pipeline shown in Figure~\ref{fig1}, performing a complete processing from a sequence of raw CCD images to a bundle of optimal light curves. The detailed functions and tasks of this pipeline are presented in Appendix~\ref{app1}.

\subsection{Functionalities of the LCA Program}

Due to the complexity of variations in light curves, it is impossible to make any thorough or complete analyses for these variations based on once/twice time series photometry. Thus, the main purposes of the LCA program include the demonstrations and separations of variations in light curves detected from a time series, markings of corresponding stars on the sky (i.e., the reference image of time series as the finding chart), and compilations of the known stellar information for the references of further analysis in the future. Four basic modules: (1)\,separations, (2)\,demonstrations, (3)\,markings, and (4)\,cross-identifications constitute the LCA program. The 1st and 4th modules are the two key functionalities of the LCA program (the details in Appendix~\ref{app2}), while the 2nd and 3rd modules are designed for reducing the burdensome manual operations as much as possible.

\section{Full-frame Data Reductions for Two Time Series}
\label{sec:sec3}

\begin{table}[htpb]
	\begin{center}
		\caption{Log of time series photometry for two CVs.}
		\label{tab1}
		\begin{tabular}{cccccc}
			\hline\hline
			Program star&UT Date&Telescopes&Filters&Exposures&FOV\\
			\hline
			RX\,J2102.0+3359&2016 Sep 24&ARCSAT 0.5m&sdss g$^{'}$&157$\times$60\,s + 1$\times$90\,s images&11$^{'}\times$11$^{'}$\\
			Paloma\,J0524+4244&2020 Dec 01&XLO 0.85m&no-filter&182$\times$60\,s images&32$^{'}\times$32$^{'}$\\
			\hline\hline
		\end{tabular}
	\end{center}
\end{table}

\begin{figure}
	\centering
	\begin{minipage}{7.2cm}
		\includegraphics[width=7.1cm,angle=0]{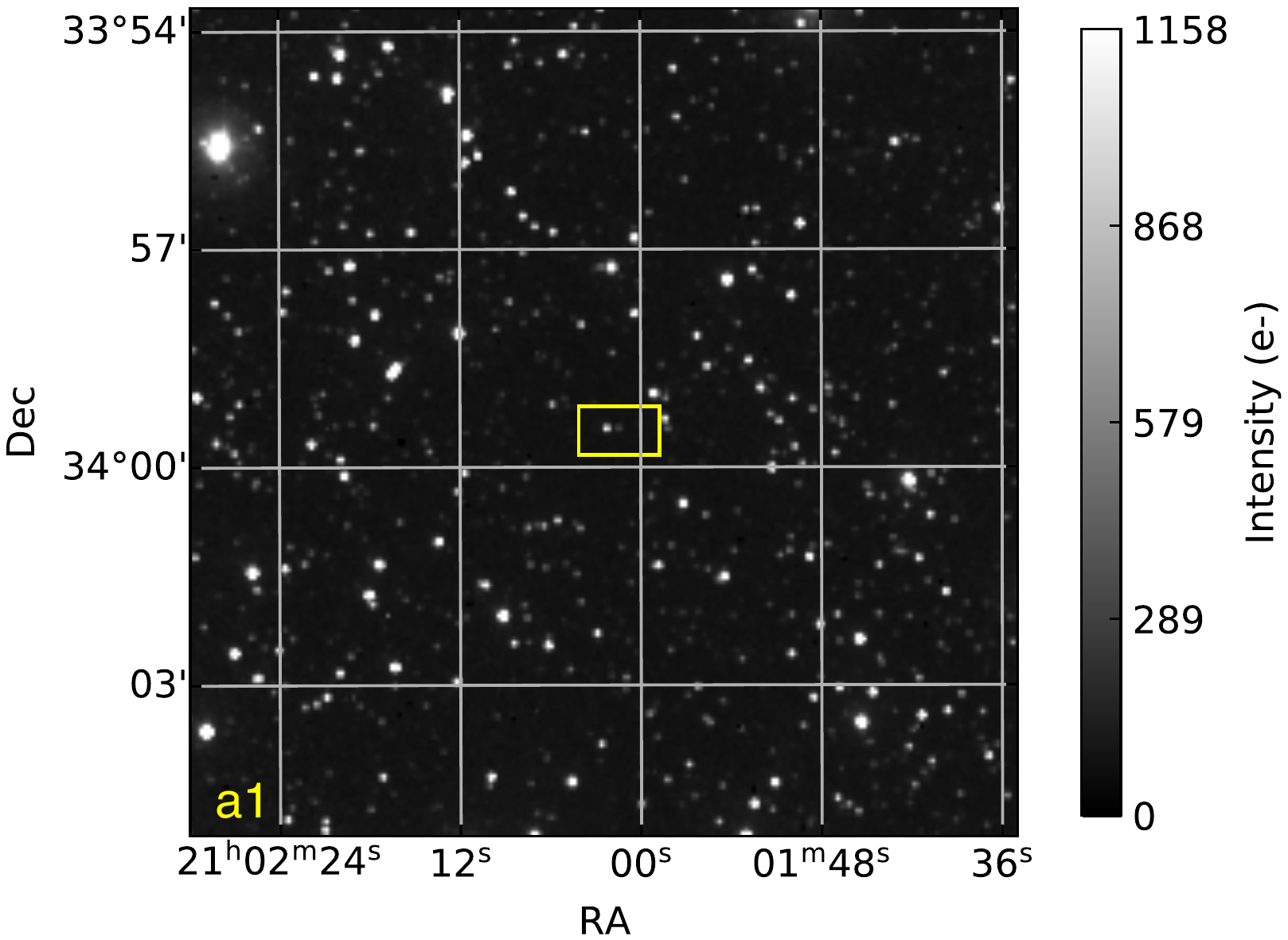}
	\end{minipage}
	\begin{minipage}{7.2cm}
		\includegraphics[width=7.1cm,angle=0]{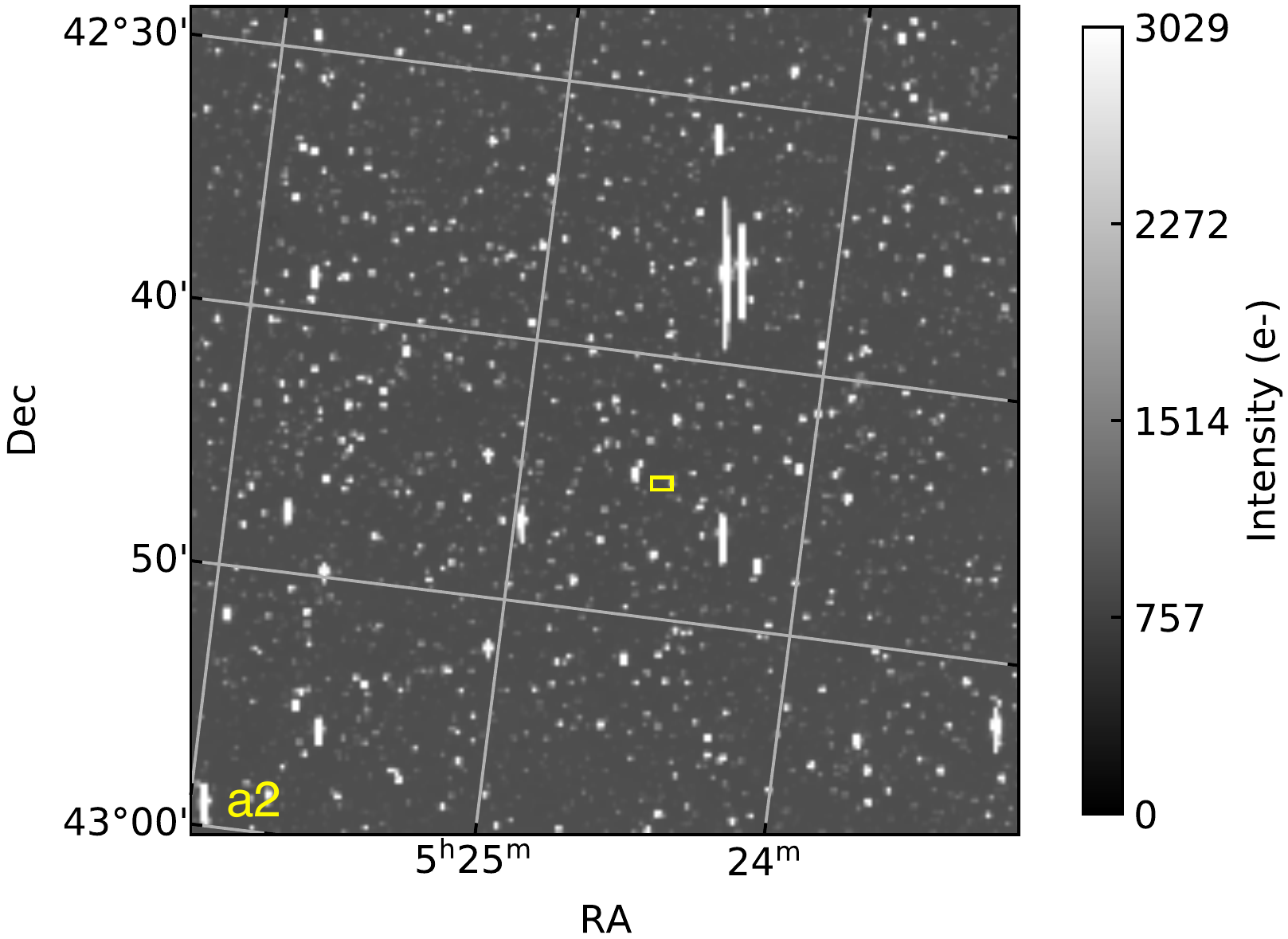}
	\end{minipage}	
	\begin{minipage}{7.2cm}
		\includegraphics[width=7.1cm,angle=0]{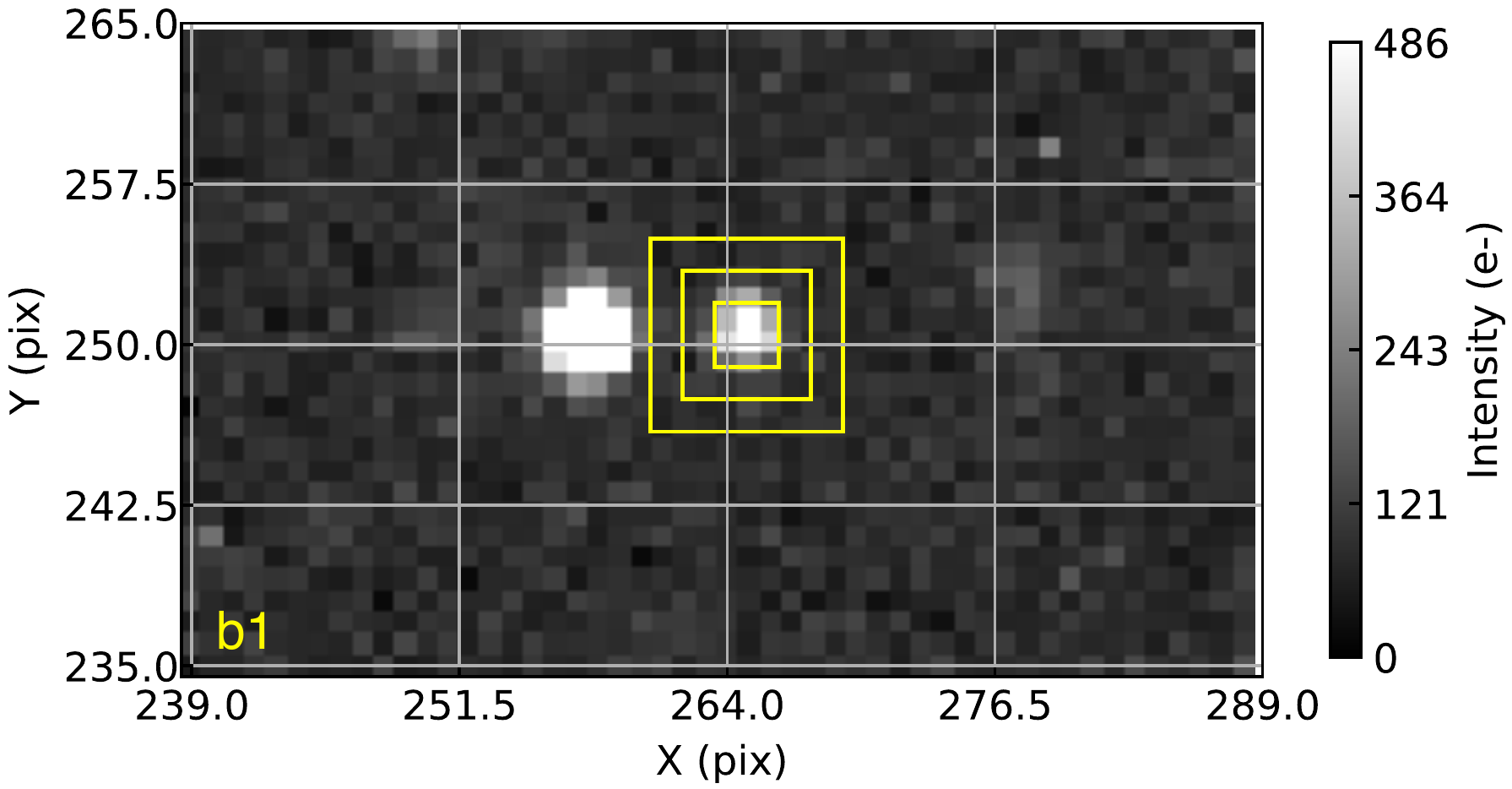}
	\end{minipage}
	\begin{minipage}{7.2cm}
		\includegraphics[width=7.1cm,angle=0]{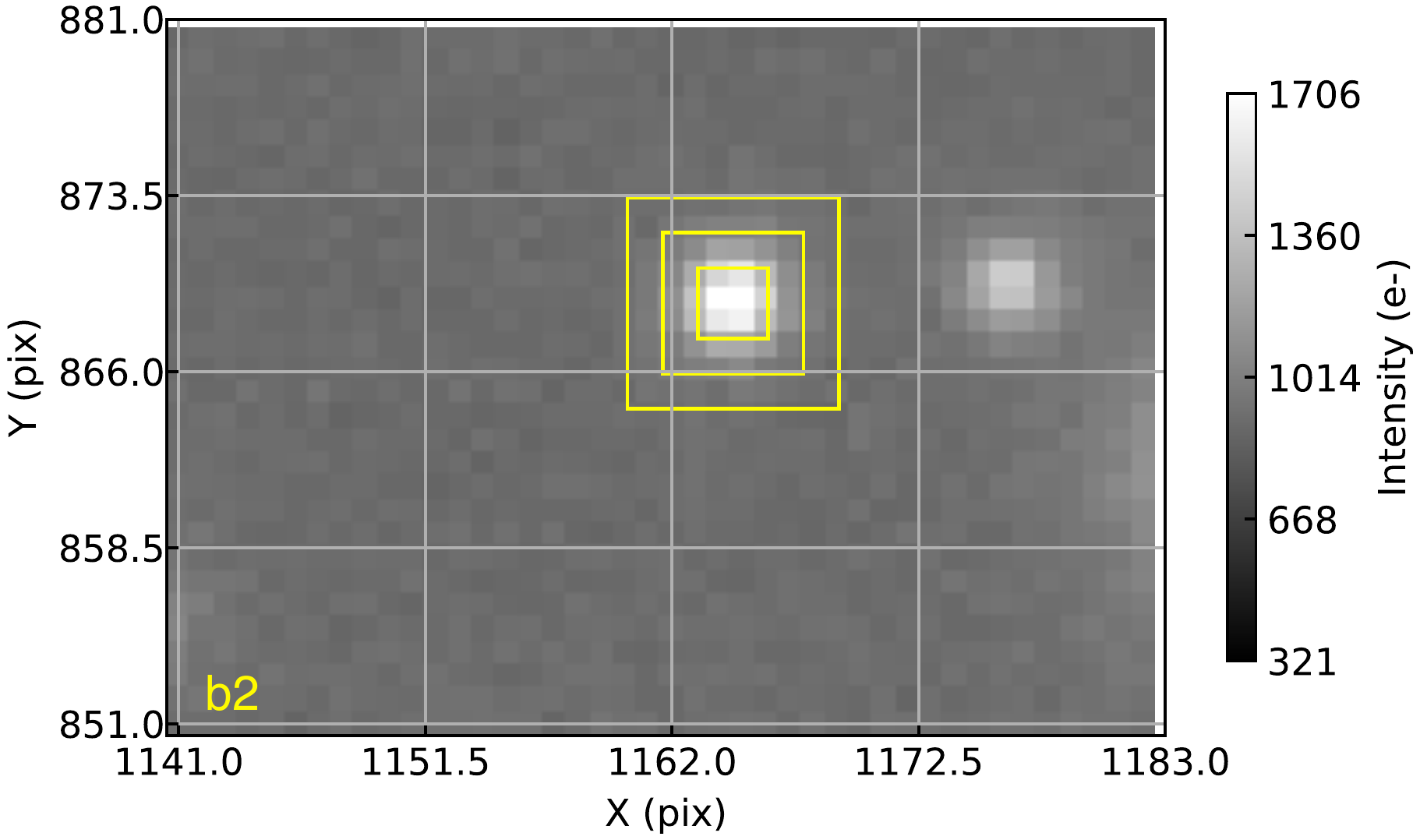}
	\end{minipage}	
	\begin{minipage}{7.2cm}
		\includegraphics[width=7.1cm,angle=0]{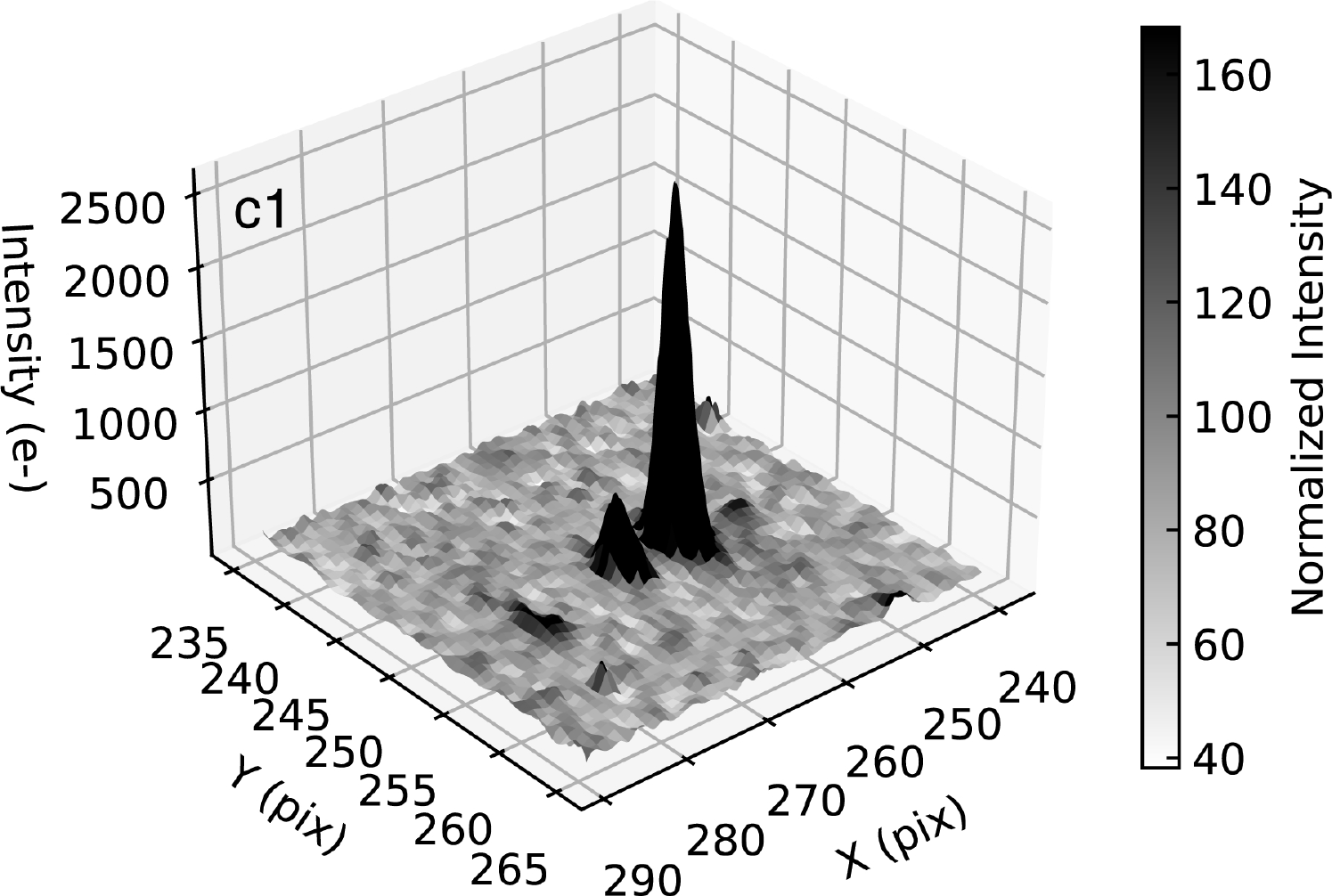}
	\end{minipage}
	\begin{minipage}{7.2cm}
		\includegraphics[width=7.1cm,angle=0]{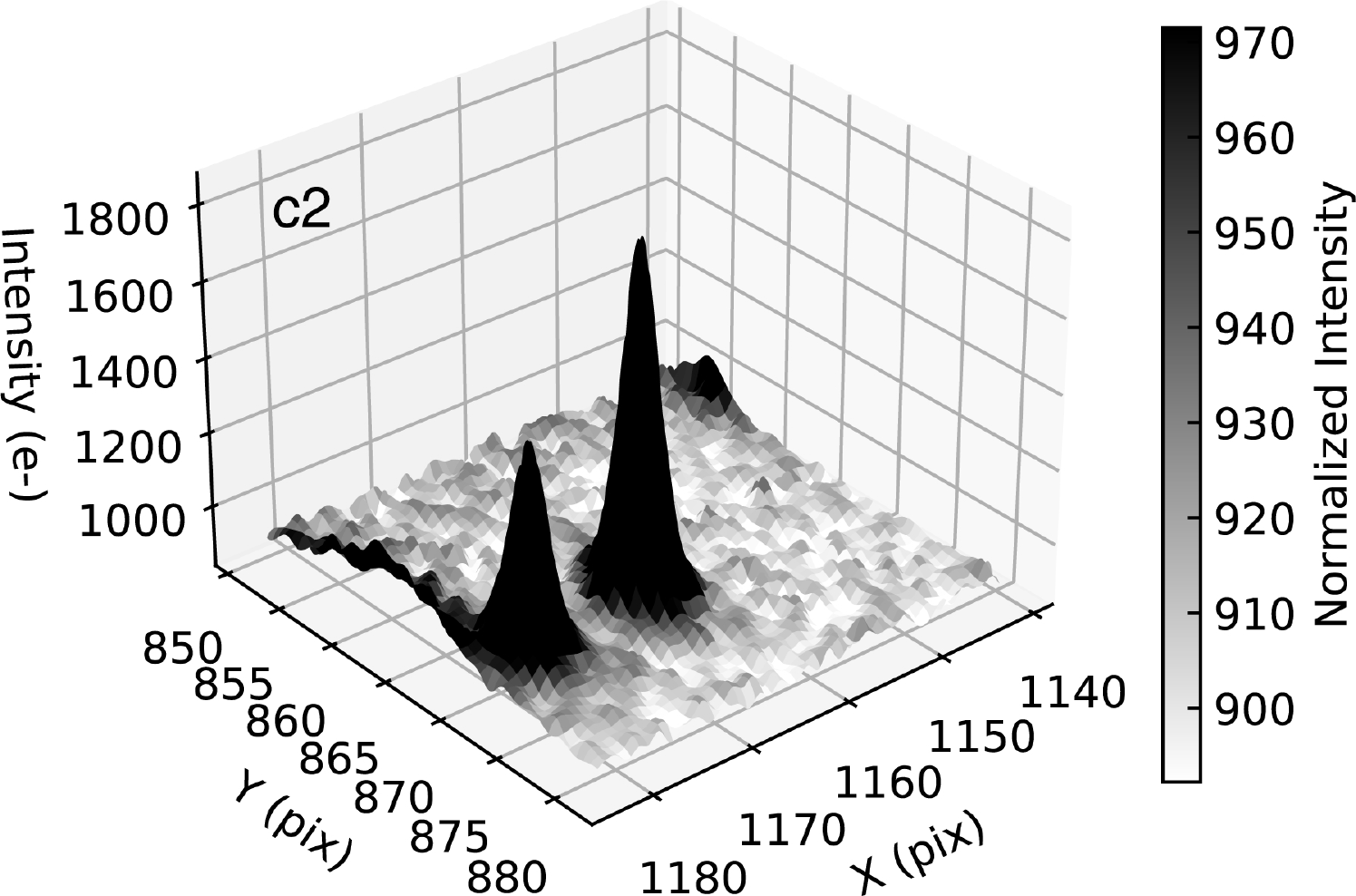}
	\end{minipage}	
	\begin{minipage}{7.2cm}
		\includegraphics[width=7.1cm,angle=0]{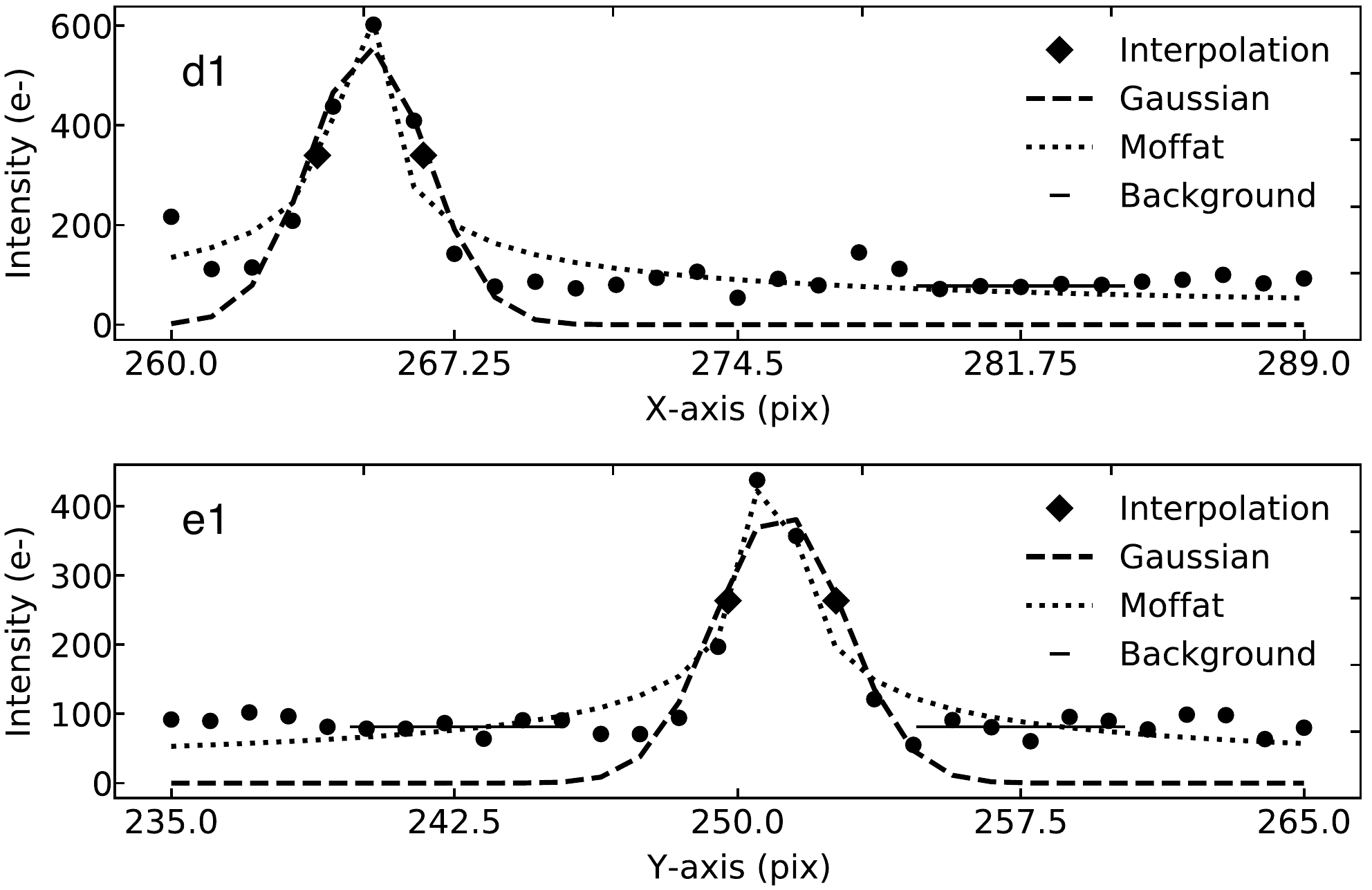}
	\end{minipage}
	\begin{minipage}{7.2cm}
		\includegraphics[width=7.1cm,angle=0]{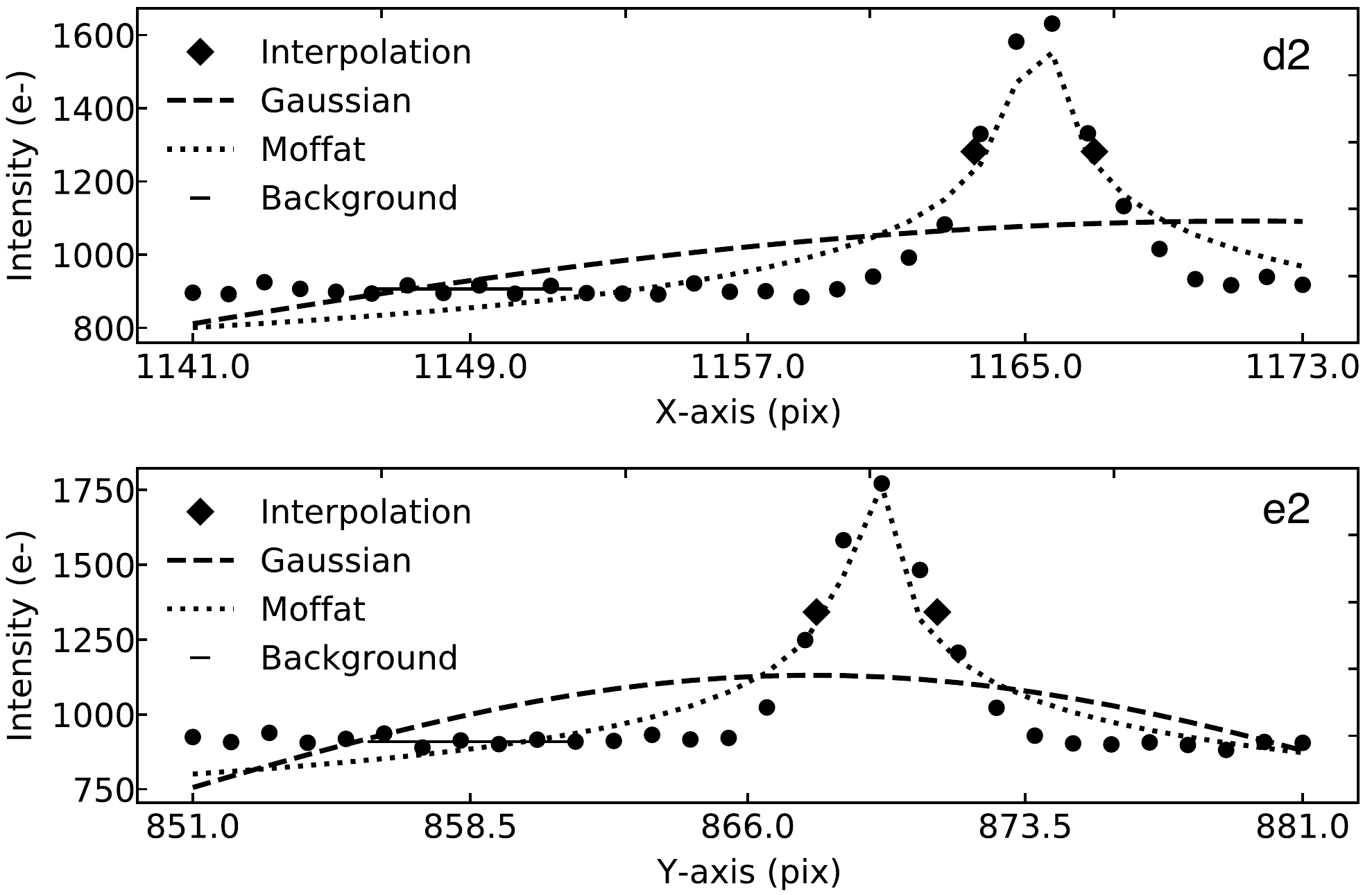}
	\end{minipage}
	\caption{The left- and right-hand plots are corresponding to the time series of RX\,J2102.0+3359 and Paloma\,J0524+4244, respectively. The rectangles shown in panels a1 and a2 denotes an sky area with the same directions as that of Figure~\ref{fig2} in the world coordinates nearby the program star of time series. Panels b1 and b2 zoom in the rectangle subimages of panels a1 and a2. The normalized intensity plots of the subimages b1 and b2 are demonstrated in panels c1 and c2. The three methods estimating the FWHMs of the radial intensity profiles of the program stars along X- and Y-axis are plotted in panels d1 and d2, and panels e1 and e2, respectively.}
	\label{fig3}
\end{figure}

Table~\ref{tab1} gives a log of two time series photometry for two CVs, randomly picked out from a historical photometry database demonstrating this full-frame data reduction method.

\subsection{Time Series of RX\,J2102.0+3359}

The differential time series photometry for CV RX\,J2102.0+3359 is from the 1\,K FlareCam with 1$^{''}$.3 pixel$^{-1}$ when binned 2$\times$2 mounted on the Apache Point Observatory (APO) 0.5\,m Astrophysical Research Consortium Small Aperture Telescope (ARCSAT\footnote{https://www.apo.nmsu.edu/Telescopes/ARCSAT/index.html}). A time series lasting 3\,hrs is composed of N$_{\rm f}$=158 continuous CCD images. After the first two steps of the pipeline shown in Figure~\ref{fig1}, an iteration of the DAOPHOT for this calibrated time series figures out the three optimal DAOPHOT parameters: Sigma, FWHM, and Threshold listed in Table~\ref{tab2}, and No.\,63 image is set to be the reference image with N$_{\rm max}$=484. Due to 30 ``median-quality" images ($\sim$18\% of this time series length) with the averaged number of the DAOPHOT stars $\sim$\,76\%\,N$_{\rm max}$, this time series is classified as Grade-B. In order to save the runtime, the brightest 50 stars picked out from all 484 identified DAOPHOT stars marked on the top right-hand panel of Figure~\ref{fig2}, are used in the next-step matching calculations. Using the three landmark stars shown in the top left-hand panel of Figure~\ref{fig2}, all of 158 FITS files are successfully matched with the reference image. N$_{\rm cm}$=363 indexed candidate stars (75\%\,N$_{\rm max}$) including their corresponding coordinates in pixels are prepared for the Step5 annulus aperture photometry. Since the WCS are recorded in the header of ARCSAT FITS files, the celestial coordinates of all 363 candidate stars including the program star are assigned.

Based on the celestial coordinates of the program star CV RX\,J2102.0+3359, the SPDE program downloads its finding chart from the SIMBAD, with an image size similar to FOV of ARCSAT FITS, and localize the exact positions of program star in the reference image on a zoom-in rectangle subimage with a size of 50$\times$30\,pixels shown in the panels a1 and b1 of Figure~\ref{fig3}. The normalized intensity plot of this subimage shown in the panel c1 of Figure~\ref{fig3} demonstrates that our program star is a fainter star, but clearly separated from its neighbor. Although the panels d1 and e1 of Figure~\ref{fig3} show that the Gaussian and Moffat functions roughly fit the radial intensity profile of the program star, the large fitting uncertainties imply that both fitting methods may be failed to derive an appropriate FWHM for this faint star. Hence, an averaged FWHM $\sim$\,1.39\,pixel for the program star is calculated by the interpolation method. Moreover, a circular annulus aperture shape is preferred due to the approximately symmetric brightness distribution of the program star indicated by R$_{\rm fwhm}\sim$\,0.95. From a total of 3.3$\times$10$^{5}$ differential light curves produced in the Step5 Annulus aperture photometry, the optimal 363 differential light curves and their corresponding reference light curves are searched out.

\begin{table}[htpb]
	\begin{center}
		\caption{Data reduction parameters of two time series photometry.}\label{tab2}
		\begin{tabular}{cccc}
			\hline\hline
			Parameters&RX\,J2102.0+3359&Paloma\,J0524+4244&Statements\\
			\hline
			Sigma&3.0&10.0&Background subtraction of the DAOPHOT\\
			FWHM&3.0&6.0&Full width at half maximum\\
			Threshold&5.0&30.0&Lower limit of detection\\
			Reference image No.&63&169&Image with the maximal DAOPHOT stars\\
			N$_{\rm max}$&484&1051&Number of DAOPHOT stars on the reference image\\
			N$_{\rm cm}$&363&641&Number of stars ready for aperture\\
			Grade$^{a}$&B&C&Quality grade of time series\\
			TLN&52-162-293&93-134-514&Three landmark star Nos.\\
			Aperture range (pix)&1.39-6.97&1.68-8.41&Used to search the optimal aperture size\\
			N$_{\rm a}$&12&14&Number of grids in Aperture range\\
			R$_{\rm fwhm}$&0.95&0.94&Ratio of FWHM along X- and Y-axis)\\
			Aperture shape$^{b}$&C&C&(C)ircular/(E)lliptical/(R)ectangular\\
			\hline\hline
		\end{tabular}
	\end{center}
	$^{a}$ Four quality grades listed in Table~\ref{taba1}.\\
	$^{b}$ Provided by a Astropy affiliated package: Photutils.\\
\end{table}

\subsection{Time Series of Paloma\,J0524+4244}

With the 2\,K Andor CCD cameras\footnote{More details are listed in \citet{zho09} and \citet{bai18}} mounted on the Xinglong Observatory (XLO) 0.85\,m telescope, we obtained a time series of CV Paloma\,J0524+4244 lasting 3.3\,hrs. Using the best three DAOPHOT parameters listed in Table~\ref{tab2}, No.\;169 image with N$_{\rm max}$=1051 is identified as the reference image of the time series composed of N$_{\rm f}$=182 wider-field CCD FITS. Due to $\sim$71\%\,N$_{\rm f}$ images with the averaged $\sim$64\%\,N$_{\rm max}$, this is a Grade-C time series triggering one-by-one manual check for the ``suspected" CCD images. Then, the SPDE program selects the brightest 50 stars marking on the reference image (the bottom right-hand panel of Figure~\ref{fig2}) to successfully perform the automatic matching calculations for all of 182 FITS using the landmark triangle (the bottom left-hand panel of Figure~\ref{fig2}), and identify 640 stars, $\sim$61\%\,N$_{\rm max}$ for the Step5 Annulus aperture photometry. However, the faint program star is never included in the 640 candidate stars\footnote{Although the three smaller parameters reset in the DAOPHOT is able to detect more faint stars including our program star, the sharp increase in N$_{\rm max}$ and runtime seriously lowers the possibility of matching success due to the large degeneracy of landmark triangle, and even crash the SPDE program.}. Thus, we only used the three default DAOPHOT parameters, and directly appended this faint program star into the candidate star list (i.e., N$_{\rm cm}$=641). 

In spite of the lack of the WCS in the header of XLO FITS file, the SPDE program successfully maps the reference image to the world coordinates shown in the panel a2 of Figure~\ref{fig3}. Although our program star is too faint to be detected by the DAOPHOT, it still can be find out from the mapped reference image according to the transformed celestial coordinates, and marked in a zoom-in rectangle subimage with a size of 42$\times$30\,pixels shown in the panel b2 of Figure~\ref{fig3}. Inspections of the panel c2 of Figure~\ref{fig3} indicate that our program star is not affected by the neighbor star, despite of the closest, right-side fainter star slightly affected by a much brighter star outside of this subimage. For the radial intensity profile of the program star plotting in the panels d2 and e2 of Figure~\ref{fig3}, the Gaussian-fit is completely failed, while the Moffat-fit shows the large deviation. Like CV RX\,J2102.0+3359, an averaged FWHM$\sim$\,1.68\,pixel is measured by the interpolation method, and the similar R$_{\rm fwhm}$ close to 1 means that the circular annulus aperture shape is appropriate. All 5.2$\times$10$^{7}$ differential light curves are extracted in the Step5 annulus aperture photometry. On the basis of this, all the 641 optimal differential light curves and their corresponding reference light curves are derived.

\begin{table}[htpb]
	\begin{center}
		\caption{23 known stellar objects in the same FOVs of two time series.}\label{tab3}
		\begin{tabular}{cccc}
			\hline\hline
			time series$^{a}$&No.&SIMBAD Name&Type$^{b}$\\
			\hline
			1&102&V2746\,Cyg&EW\\
			1&143&V2743\,Cyg&EW\\
			1&183&RX\,J2102.0+3359&CV\\
			1&203&1RXS\,J210209.4+335906&Xs\\
			2&2&2MASS\,J05234123+4256139&RGBs\\
			2&36&2MASS\,J05241536+4255275&RGBs\\
			2&85&2MASS\,J05234299+4252135&HBs\\
			2&130&ATO\,J081.1876+42.8659&MPULSE\\
			2&139&2MASS\,J05241773+4250546&HBs\\
			2&200&2MASS\,J05251178+4248567&RGBs\\
			2&299&2MASS\,J05244268+4243429&RGBs\\
			2&306&ATO\,J081.3906+42.7440&MSINE\\
			2&315&2MASS\,J05254448+4244300&RGBs\\
			2&316&2MASS\,J05242815+4242397&RGBs\\
			2&337&2MASS\,J05245476+4242196&RGBs\\
			2&463&2MASS\,J05234305+4235061&RGBs\\
			2&477&2MASS\,J05245674+4236075&HBs\\
			2&533&2MASS\,J05250747+4233386&HBs\\
			2&536&2MASS\,J05251299+4233389&HBs\\
			2&571&2MASS\,J05254319+4232064&RGBs\\
			2&600&2MASS\,J05235861+4228345&HBs\\
			2&632&TYC\,2917-1121-1&N\\
			2&641&Paloma\,J0524+4244&CV\\
			\hline\hline
		\end{tabular}
	\end{center}
	$^{a}$ The number of 1 and 2 denotes the time series of RX\,J2102.0+3359 and Paloma\,J0524+4244,respectively.\\
	$^{b}$ EW: W\,UMa-type eclipsing variables; CV: Cataclysmic variables; Xs: X-ray source; RGBs: Red Giant Branch star; HBs: Horizontal Branch Star; MPULSE: Modulated pulsating star with multiple pulses in the ATLAS catalog; MSINE: Sinusoidal variables (low-amplitude $\delta$\;Scuti stars or ellipsoidal variables) in the ATLAS catalog; N: Unknown.\\
\end{table}

\begin{figure}
	\centering
	\begin{minipage}{13cm}
		\includegraphics[width=13cm,angle=0]{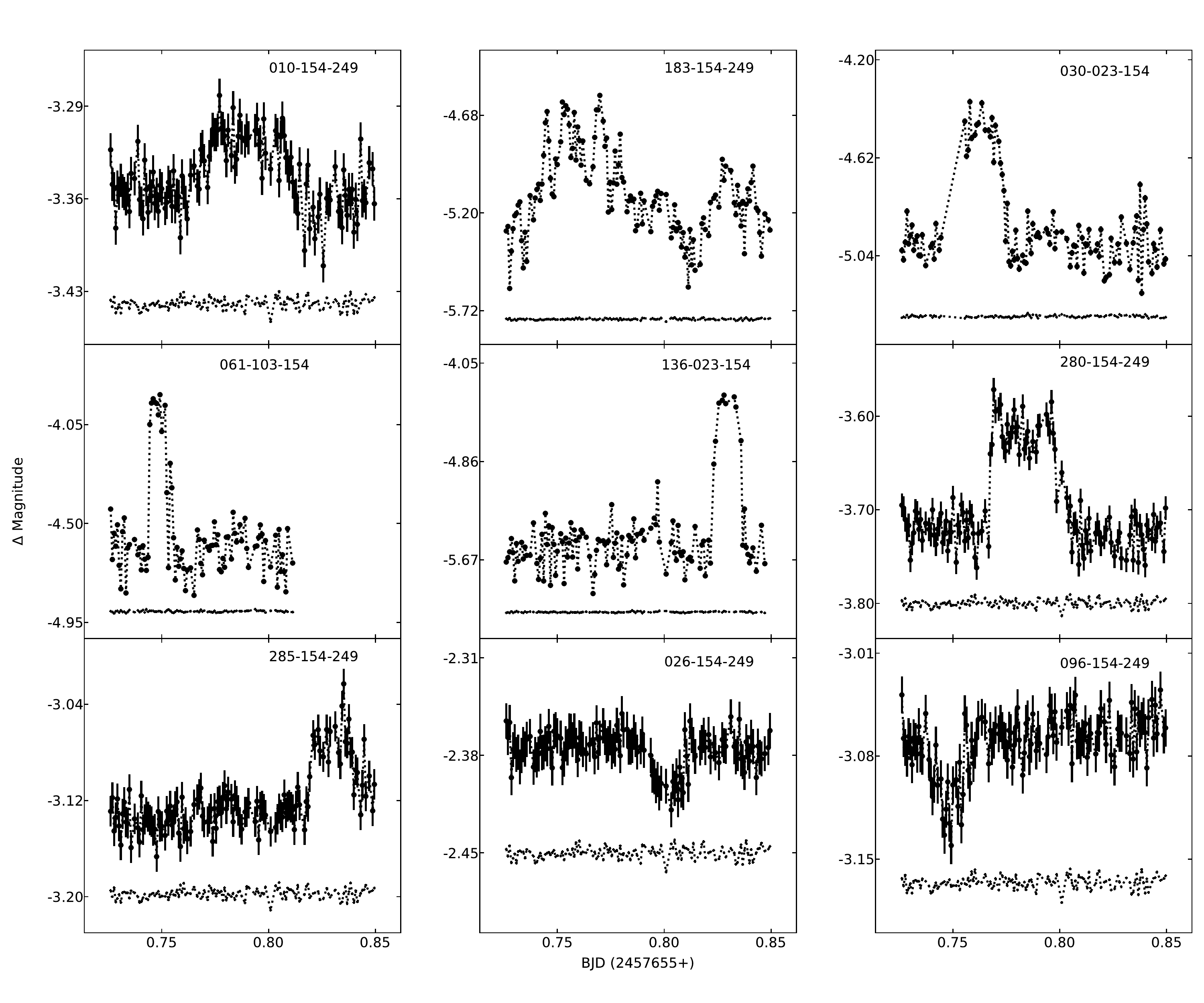}
	\end{minipage}	
	\begin{minipage}{13cm}
		\includegraphics[width=13cm,angle=0]{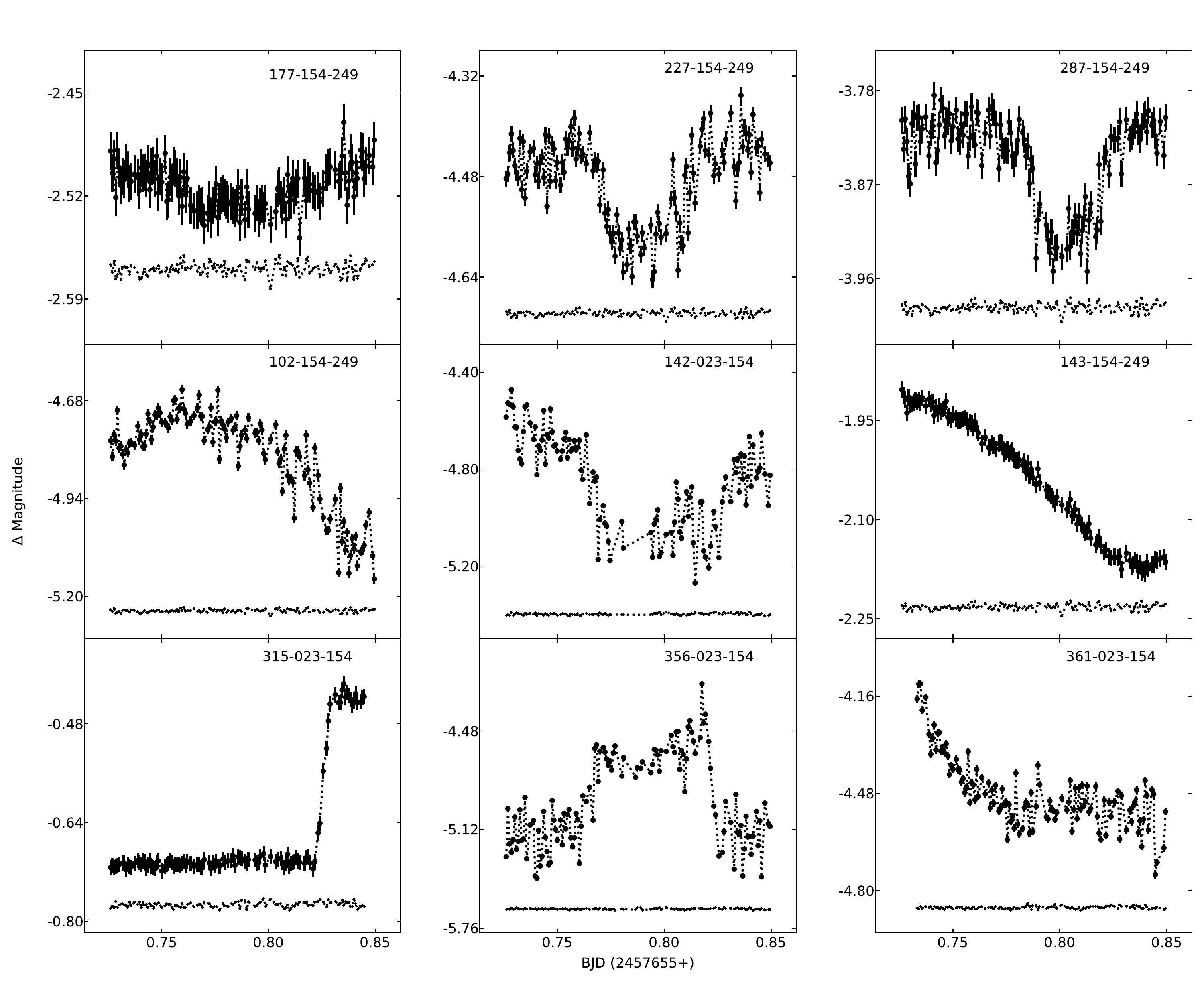}
	\end{minipage}
	\caption{18 suspected variable light curves with the errorbars and corresponding optimal reference light curves picked out by the LCA program from the same FOV of time series of RX\,J2102.0+3359. The three digits in the top right corner of panel are corresponding to the three indexed Nos. of variable, comparison, and check stars from left to right, respectively.}
	\label{fig4}
\end{figure}

\begin{figure}
	\centering
	\includegraphics[width=17cm,angle=0]{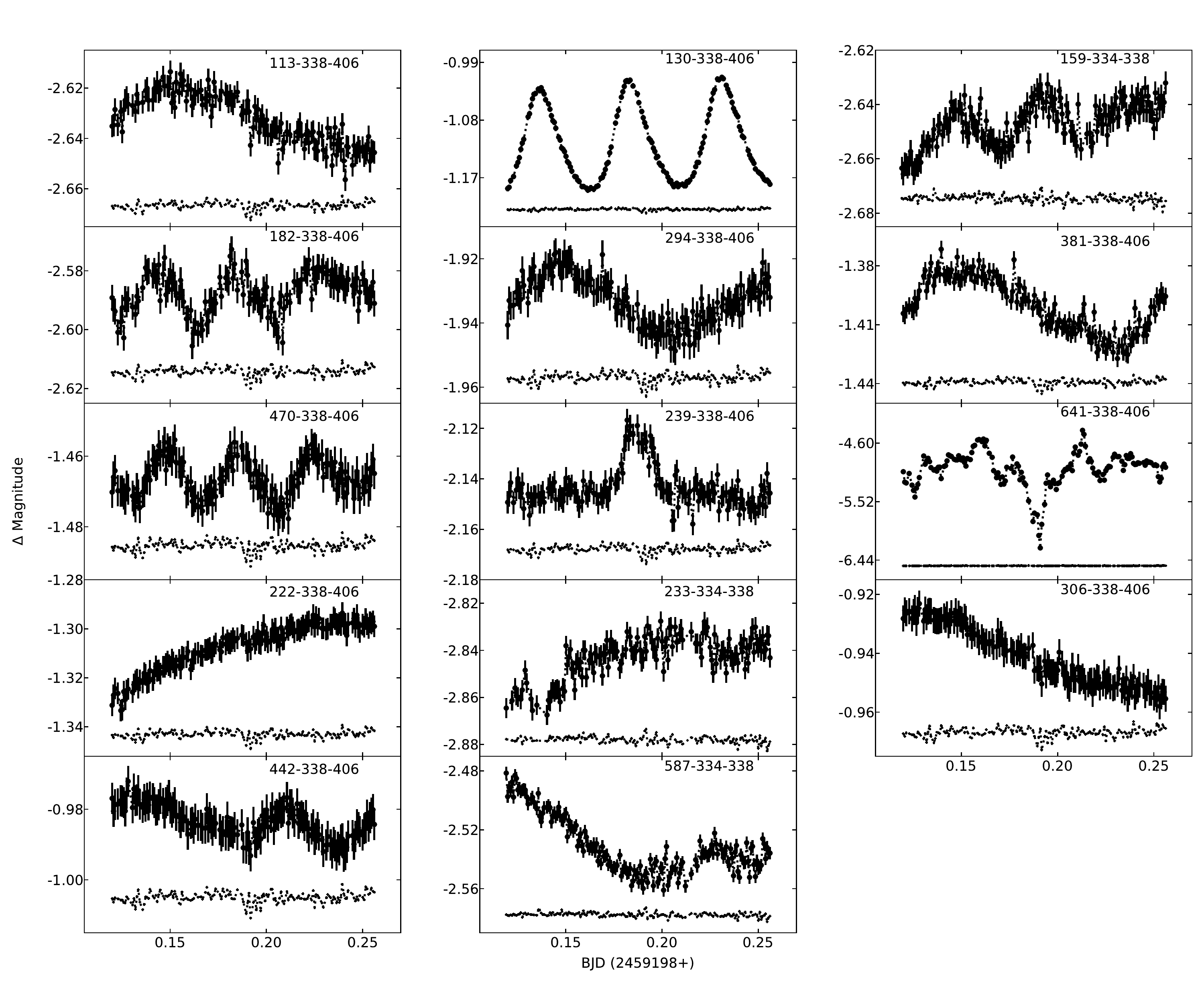}
	\caption{14 suspected variable light curves from the same FOV of time series of Paloma\,J0524+4244.}
	\label{fig5}
\end{figure}

\begin{figure}
	\centering
	\begin{minipage}{14cm}
		\includegraphics[width=14cm,angle=0]{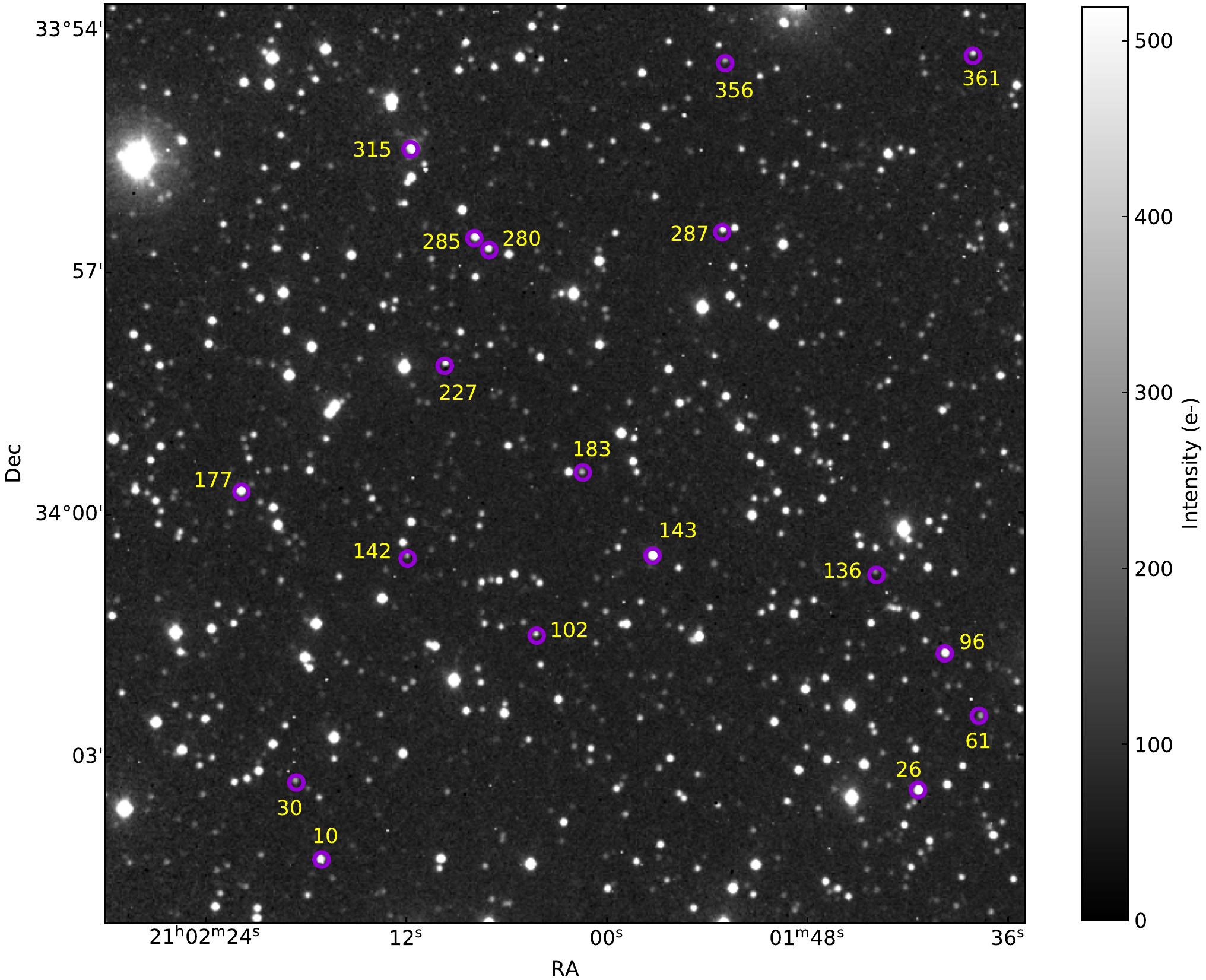}
	\end{minipage}
	\begin{minipage}{14cm}
		\includegraphics[width=14cm,angle=0]{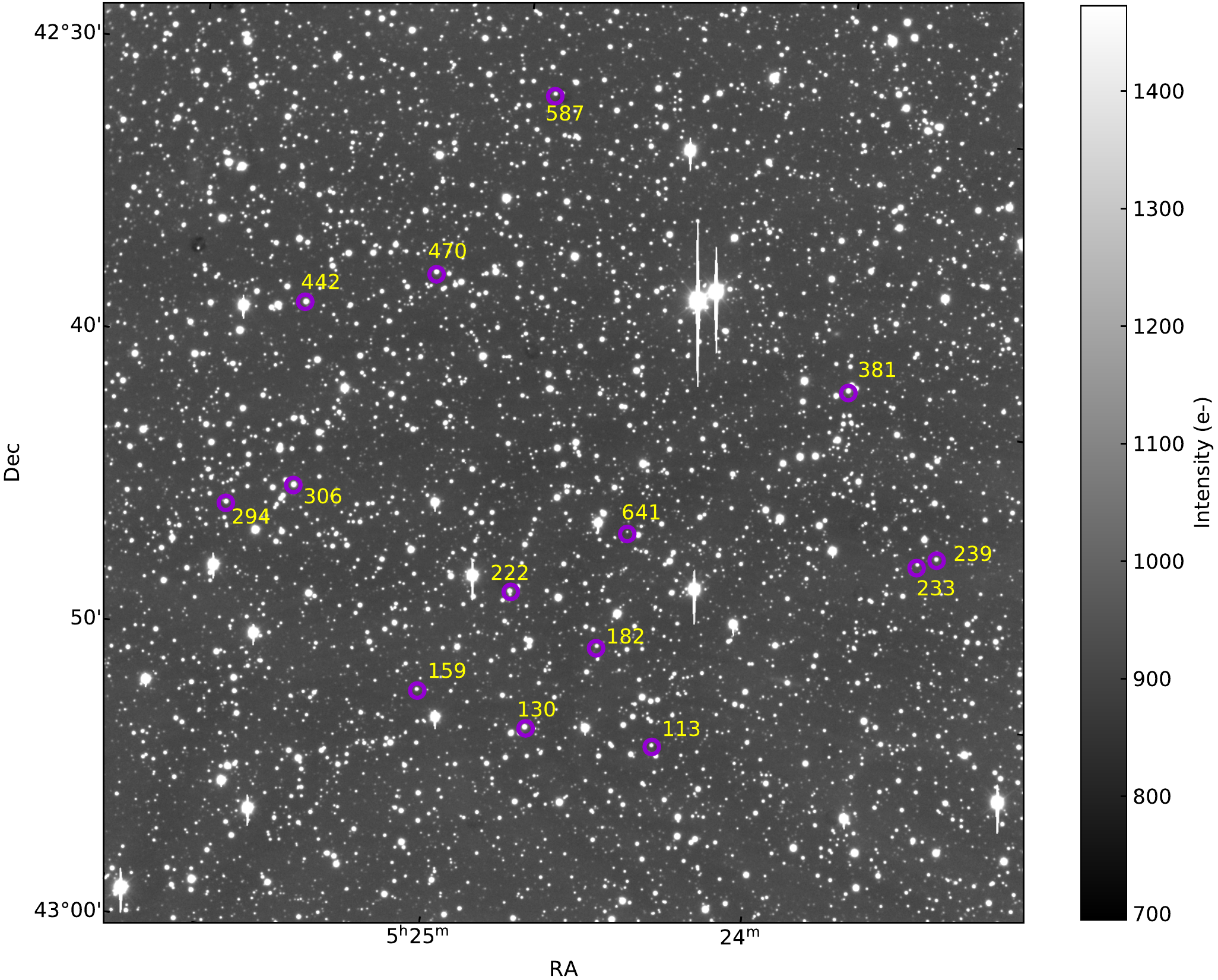}
	\end{minipage}
	\caption{The finding charts of the 18 and 14 stars marked with the blue open circles and labeled with the indexes corresponding to the light curves shown in Figures~\ref{fig4} and \ref{fig5} are shown in the upper and lower panel, respectively. Both of plots have the same directions as that in Figure~\ref{fig2}.}
	\label{fig6}
\end{figure}

\section{Discussions of Potential Variability}
\label{sec:sec4}

The cross-identification with the SIMBAD made by the SPDE program indicates 23 known stars listed in Table~\ref{tab3}, of which 4 found from the 11$^{'}\times$11$^{'}$ FOV of time series of RX\,J2102.0+3359, 19 detected from the larger FOV (i.e., 32$^{'}\times$32$^{'}$) of time series of Paloma\,J0524+4244. Except for 2 EWs, 2 programe stars (CVs), and 2 variables listed in the Asteroid Terrestrial-impact Last Alert System (ATLAS) catalog, all of the other 17 light curves of the known stellar objects are automatically neglected by the separation routine of the LCA program, since they are almost flat with the variation amplitude smaller than the default S$_{\rm t}$ listed in Table~\ref{taba2}. This implies that 9 RGBs may not be pulsating RGBs or just Long Secondary Periods \citep[cf.][]{sos21}, and the 6 HBs may not be RR\,Lyr variables.

The separation routine implemented by the LCA program automatically singled out 46 and 142 light curves with the suspected ``variations" from all of 363 and 641 light curves extracted from the time series of RX\,J2102.0+3359 and Paloma\,J0524+4244, respectively. Although all the known variables are successfully identified, a manual confirmation for this preliminary result is required to further filtrate the false-alarm variable light curves. Finally, 18 and 14 light curves with the potential variability shown in Figures~\ref{fig4} and \ref{fig5} are picked out from the FOVs of RX\,J2102.0+3359 and Paloma\,J0524+4244, respectively. Their positions locating on the sky are marked in Figure~\ref{fig6} (i.e., finding charts). Tables~\ref{tab4} and \ref{tab5} collect their physical parameters, while Tables~\ref{tab6} and \ref{tab7} list their multi-band brightness spanning from X-ray to radio by compiling from the 58 queried VizieR online data catalogs (the details of the catalogs are listed in Table~\ref{tab8}). Due to the sporadic observations in X-ray and radio band, Tables~\ref{tab6} and \ref{tab7} mainly list the observations from far-ultraviolent (FUV) to mid-infrared (MIR) observations. But the X-ray and radio observations for three stars are exclusively recorded in the footnotes of both Tables. Note that the celestial coordinates listed in Tables~\ref{tab5} and \ref{tab6} are not compiled from the data catalogs, but transformed from the CCD pixels.

\begin{figure}
	\centering
	\includegraphics[width=9cm,angle=0]{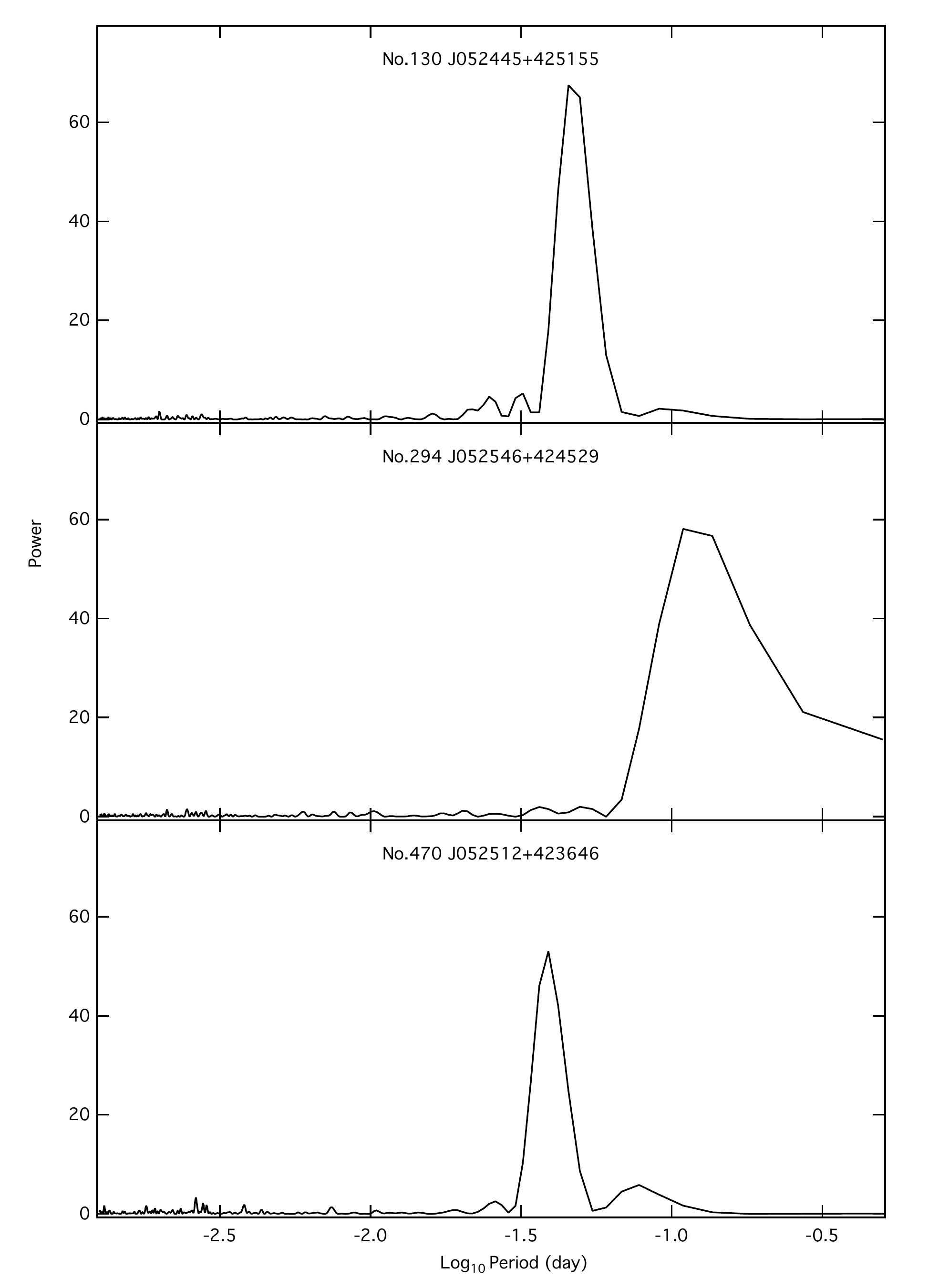}
	\caption{The periodograms of 3 stars with the significant periodical variable light curves detected from the time series of Paloma\,J0524+4244 using the LSP method.}
	\label{fig7}
\end{figure}

\subsection{Periodical Variations}

Since many autoregressive moving average (ARMA) models may produce temporary quasi-periodicities resulting in spurious Lomb-Scargle periodogram peaks (LSP; \citet{lom76}; and \citet{sca82}), which never represent any physical periodic behavior \citep{bal08}, the calculations of the significance levels of periodogram peaks (i.e., the quantitative measures of false-alarm probabilities) are difficult \citep[e.g.][]{koe90,suv15,vio19,del20,koe21}. Therefore, Figure~\ref{fig7} only demonstrates 3 LSPs with the higher significance levels of periodogram peaks than the others\footnote{In next version of the LCA program, an improved LSP with the Thomson multitaper proposed by \cite{spr20} may be used to estimate the periodicities of light curves.}. Among these 3 stars, No.130\;ATO\,J081.1876+42.8659 is the only known periodical variable with a strict periodicity, while the other two stars: No.294\;J052546+424529 and No.470\;J052512+423646 need more available data to support their identifications of periodical variables. The features of their variations are detailed as follows.
\begin{itemize}
	\item[$\bullet$]{\textbf{No.130\;ATO\,J081.1876+42.8659:}\\
		It is already identified as a modulated pulsating star with multiple pulses (MPULSE), and also identified to be a $\delta$\,Scuti variable (DSCT, listed in the GAIAv3, VSX, and ZTFVS catalogs) with the typical amplitude and period, similar to that in the ASASSN and VSX catalogs. Note that the period listed in the ATLAS catalog is close to the second harmonic. Its physical parameters and observations mainly in optical- and infrared-band are listed in Tables~\ref{tab6} and \ref{tab8}, respectively. Inspections of Figure~\ref{fig5} indicate that the shapes of two continuous minimum light are not repeatable. The latter seems to be slightly flatter than the former. Moreover, the system light shows a slow increase with a rate of $\sim$\,0.17\,mag\,day$^{-1}$. This significant periodical light curve seems to testify the availability of data reduction using the SPDE program.}
	\item[$\bullet$]{\textbf{No.294\;J052546+424529:}\\
		It is first time to detect this sinusoidal-like variation in the optical light curve. A typical sinusoidal formula,
		\begin{equation}
			\begin{split}
				\Delta\rm Magnitude=-1.9336(3)+0.0098(4)\,\sin[53.1(9)T_{\rm BJD}-0.05(18)],
			\end{split}
		\end{equation}
		can describe the modulation shown in Figure~\ref{fig5} with a period of 0$^{d}$.118(2) and an amplitude of 0$^{m}$.0196(4). This best-fit period, 0$^{d}$.118, with a precise of 3\,min, is $\sim$8\,min shorter than that listed in Table~\ref{tab9}. Note that this star may not be a star pointed out by the UKIDSS6 and GSC242 catalog, but classified to be a dwarf by the TIC82 catalog.}
	\item[$\bullet$]{\textbf{No.470\;J052512+423646:}\\
		As a Radio+Opt source listed in the S-F catalog, it is interesting that the low integrated 1.4\,GHz flux density observed by the NVSS\footnote{\url{http://www.cv.nrao.edu/nvss/}} is accompanied with a low-amplitude and rapid sinusoidal-like optical oscillation. In spite of the high similarity between the shape of this periodical modulation and a strict sinusoidal curve, the third dip around BJD\,2459198.2 seems to be the deepest, while the fourth dip at the end of this light curve seems to be the shallowest. Thus, this optical light curve may imply the complexity of this star worthy further investigations in the future.}
\end{itemize}
Moreover, there are 6 more stars showing the visible modulations with the larger scatters and weaker LSPs. A linear-plus-sinusoidal fitting method is further used to verify their modulation periods derived using the LSP method. They may be not the authentic periodical variables, but the stars with the short-lived quasi-periodicities. In spite of this, Table~\ref{tab9} lists all 9 modulations amplitudes and periods.

Based on the stellar parameters of No.10\;J210217+340417 listed in Table~\ref{tab4}, it may also be a dwarf star with a smaller modulation period. Due to the short time-base line of light curve, this suspected modulation may be just a hump. The LCA program separates the non-monotonic light curve of No.113\;J052421+425160 with a duration of $\sim$\,3.3\,hr covering 70-80\,\% of modulation period listed in Table~\ref{tab9} into the periodical type. In Table~\ref{tab5}, the stellar parameters derived from the GAIA data seem to support that it is a normal dwarf star identified in the two catalogs: TIC82 and GSC242. Although No.183\;RX\,J2102.0+3359 as a program star never demonstrates any significant evidence of orbital modulation, the distinct double-hump light curve with some large-amplitude and irregular variations appearing at two peaks seems to be a typical ellipsoidal modulation, similar to the other subtype CVs, such as Intermediate Polar (IP) RZ\,Leo \citep{szk17,dai18}, Dwarf Nova (DN) KZ\,Gem \citep{dai17,dai20}, and TW\,Vir \citep{dai21}. Assuming that the periods listed in the VSX and RKCat catalogs represent its orbital motion, this observed similar modulation period may further support that it is a period-gap CV with an amplitude almost twice that in the ASASSN catalog. Table~\ref{tab6} lists all observations of this period-gap CV from X-ray to Near Infrared (NIR). The SWIFT's UV observations (FUV-, NUV- and U-band) denote that No.159\;J052506+425105 is a faint UV source. Its optical light curve similar to that of No.182\;J052434+424851 with a different constant system light shows a low-amplitude modulation superimposing on a slow increase of system light with a rate of $\sim$\,0.12\,mag\,day$^{-1}$. The small Welch-Stetson variability index \citep{wel93} listed in the TASS catalog indicates that No.381\;J052353+423903 may be a constant rather than a variable, the saw-like modulation light curve with a larger amplitude than that of No.294\;J052546+424529 shown in Figure~\ref{fig5} implies its variability. Compared with No.294\;J052546+424529, the best-fit sinusoidal curve shows the larger deviations. In the GAIAsv3 catalog, the modulation period of No.381\;J052353+423903 is smaller, but the amplitude is almost the same.

\subsection{Transient Variations}

Table~\ref{tab10} lists the amplitudes and durations of all the 11 transient events estimated by the LCA program including 6 brightening (hump) and 5 darkening (dip) events shown in Figures~\ref{fig4} and \ref{fig5}. Considering that most of transients are serendipitous and unrepeatable, the recorded and identified events are the important references and data accumulations for the future researches. Although there are plenty of complex and specialized explanations proposed to interpret various different transients, the subtile models seem to make no sense for these 11 poorly understood stars. Therefore, it may be reasonable that the simple and common mechanisms (e.g., stellar flares and occasional occultation by uncertain invisible clumps corresponding to the brightening and darkening events, respectively.) are used to preliminarily explain these detected transient events.

\subsubsection{Brightening events}

All the 6 optical brightening events with the shapes significantly deviating from exponential forms indicate that they may be not accompanied by the high-energy events (e.g., the X-ray flares commonly appearing in polar \citet{ter10}, blazars \citep{sta21} and Supergiant high-mass X-ray transient \citep{sgu06,sid22}). Except for the basically symmetric hump detected in No.136\;J210144+340044 with the largest amplitude in Table~\ref{tab10}, the other 5 brightening events are asymmetric (i.e., the steep rise and moderate fall branches), like the V-band OPTIC flare found during the low state of the polar prototype AM\,Her \citep{kaf05}. Although most of the R-band flares in cool stars have low amplitudes of less than 0.15\,mag, accompanied by occasional large amplitudes over 0.2\,mag \citep[cf.][]{vid09,zha10,dai12}, the three low-amplitude humps detected in three stars: No.280\;J210207+335644 and No.285\;J210208+335636 in time series of RX J2102.0+3359, and No.239\;J052332+424424 in time series of Paloma J0524+4244 may not be the flares in cool stars due to their high T$_{\rm eff}$ listed in Tables~\ref{tab4} and \ref{tab5}. The event appearing in No.280\;J210207+335644 holds a plateaus lasting 39\,min, then suddenly decays in an exponential-like form, while the other two events found in No.30\;J210219+340318 and No.61\;J210138+340231 seem to be the larger optical flares than that detected in the M-type eclipsing binary CU\,Cnc \citep{qia12}.

\subsubsection{Darkening events}

The sharp V-shaped dip with the deepest depth listed in Table~\ref{tab10} may indicate that the program star CV No.641\;Paloma\,J0524+4244 with a period of $\sim$2.6\,hr is first identified as an eclipsing CV inside period gap. Like the oscillations appearing at the two peaks of modulation light curve of the other program star period-gap CV No.183\;RX\,J2102.0+3359, the outside-of-eclipse light curve of No.641\;Paloma\,J0524+4244 shown in Figure~\ref{fig5} demonstrates the similar large-amplitude and irregular variations (i.e., flickering) possibly caused by the same mechanism: the unstable accretion process around the primary white dwarf \citep{bru92}. Since this large-amplitude flickering never appears in the outside-of-eclipse light curves of the other two eclipsing polars: UZ\,For and V348\,Pup inside period gap \citep{dai10}, the V-shaped eclipse of No.641\;Paloma\,J0524+4244 totally different from the flat-bottomed minima in UZ\,For, may manifest the existence of an excited accretion disk (the active source of flickering) surrounding the primary white dwarf. Note that the XMM11 catalogs present the positive variability in X-ray conflicting with the negative result listed in the R-HR catalog. In spite of this, the comprehensive X-ray observations for No.641\;Paloma\,J0524+4244 indicate that it is suspected to be an IP inside period gap. Among the other four dips with the depth less than 0.2\,mag, the two deeper and longer dips in No.227\;J210210+335809 and No.287\;J210153+335630 may be caused by the unknown grazing obscurations. Assuming that the shallow dip in No.26\;J210141+340327 is an intrinsic rather than artificial phenomenon, based on the similarity in morphology of light curve with the exoplanet transits \citep[cf.][]{tra22,bis22}, we boldly suspected that this may be a transit-like dip despite the insufficient evidence provided by this observation data alone.

\subsection{Peculiar Variations}

The light curves of 12 stars asterisked in Tables~\ref{tab4} and \ref{tab5} are temporarily classified as the peculiar type by the LCA program.

A sudden luminosity transition with a duration of $\sim$14\,min and a brightness difference of $\sim$\,0.27\,mag switching from a low to high state appears in the light curve of No.315\;J210211+335529. Both of luminosity states are considerably stable (i.e., the strikingly flat light curves before and after this transition event). A nearly linear switching progress with a rate of $\sim$\,0.02\,mag\,min$^{-1}$ is clearly shown in Figure~\ref{fig5}. Note that this seems to be a Non-star listed in the GSC242 catalog, although a giant star identified in the TIC82 catalog. Therefore, No.315\;J210211+335529 would be worthy of further investigation.

Since the short duration of time series of RX\,J2102.0+3359 ($\sim$3\,hr) cannot cover the complete orbital modulations of two EWs: V2746\,Cyg (No.102\;J210204+340130) and V2743\,Cyg (No.143\;J210157+340032), their light curves shown in Figure~\ref{fig4} are only the declination branches. This is why both light curves are automatically separated into the peculiar type by the LCA program, despite of their known EW identifications via the on-line cross-identifications with the SIMBAD. Moreover, the seven light curves of No.177, No.361 in time series of RX\,J2102.0+3359, No.222, No.233, No.306, No.442, and No.587 in time series of Paloma\,J0524+4244 demonstrate the curvilinear variations similar to the monotonic light curves of two EWs. Thereinto, No.222\;J052451+424720 with an average period of 0$^{d}$.5(5) listed in the ZTFPV catalog and No.306\;ATO\,J081.3906+42.7440 with a period of 1$^{d}$.07 listed in the ATLAS catalog are already identified as a periodical variables. The monotonic light curves shown in both stars are only the branches of their complete modulation variations like the two EWs, due to the time series length of Paloma\,J0524+4244 far short of their own modulation periods.

In the leftover three peculiar type light curves, the large-amplitude and irregular light curves of No.142\;J210212+340032 and No.356\;J210153+335425 seem to be the parts of transient events. while an ambiguous low-amplitude modulation appearing after BJD\,2459198.19 shown in Figure~\ref{fig5} seems to support a weak evidence of a short-lived quasi-periodicity in No.442\;J052536+423820.

\section{Conclusions}
\label{sec:sec5}

Accompanying with the greatly improvement of the computer technique nowadays, the SPDE program, an authentic full-frame data reduction method for the traditional optical photometry data usually taken by the small/medium aperture ground-based telescopes, is developed by using several astronomy Python packages of the Astropy Project to solve the problems of CCD photometry data quantity early claimed by \citet{ste87}. Combining with the LCA program, a preliminary followup light curve analysis tool for the data productions of the SPDE program, a complete automated data reduction pipeline can be achievable. Compared with the final results of the other photometry data reduction pipelines (i.e., several light curves of program stars at most for a single research topic), this complete full-frame data reduction method is able to capture a wide range of science objectives, e.g., data accumulations of known variables, serendipitous optical transient source survey, and detection of new potential variables etc.

Both programs are successfully capable of performing the photometric monitoring campaigns for two time series of faint CVs RX\,J2102.0+3359 and Paloma\,J0524+4244 with 158 and 182 CCD images lasting 3\,hr and 3.3\,hr, observed by the ARCSAT 0.5\,m and XLO 0.85\,m telescopes, respectively. The time series of RX\,J2102.0+3359 is identified as Grade-B, while the time series of Paloma\,J0524+4244 is regarded as Grade-C. In spite of the different grade for two time series, both of them smoothly pass through the matching processes. In total, 363 and 641 indexed stars with the transformed celestial coordinates on the mapped reference images are detected from the time series of RX\,J2102.0+3359 and Paloma\,J0524+4244, respectively. By automatically setting the applicable aperture shape and size, their optimal differential light curves and the corresponding reference light curves are extracted.

Based on the data productions of the SPDE program, the LCA program finally picks out 18 and 14 light curves with the potential variability from the time series of RX\,J2102.0+3359 and Paloma\,J0524+4244, respectively. All of 32 selected light curves are preliminarily separated into three types, in which 9 are periodical type, 11 are transient type, and 12 are peculiar type. The on-line cross-identifications with the SIMBAD made by the SPDE program indicate 23 known stars, of which 16 red giant-/horizontal-branch stars and a X-ray source only show the almost flat light curves. Due to incomplete coverage for the modulations, two known EWs, a ZTFPV variable No.222\;J052451+424720, and an ATLAS variable No.306\;ATO\,J081.3906+42.7440 showing the monotonic curvilinear-like variations, are temporarily identified to be the peculiar type by the LCA program. The other ATLAS variable No.130\;ATO\,J081.1876+42.8659 with a significant periodical variation demonstrates the reliability of this full-frame data reduction method. A collection from the 58 queried VizieR online data catalogs produces the complete catalogs of all 32 stars including their physical parameters and multi-band brightness spanning from X-ray to radio for the future researches.

\begin{acknowledgements}
This work was partly supported by CAS Light of West China Program, the Yunnan Youth Talent Project, the Yunnan Fundamental Research Projects (grant No.2016FB007, No.202201AT070180), the Chinese Natural Science Foundation (No.11933008). We acknowledge the support of the staff of the Xinglong 85cm telescope. This work was partially supported by the Open Project Program of the CAS Key Laboratory of Optical Astronomy, National Astronomical Observatories, Chinese Academy of Sciences. Based on observations obtained with Apache Point Observatory 0.5\,m Astrophysical Research Consortium Small Aperture Telescope (ARCSAT). HZ acknowledges support from the Yunnan Fundamental Research Key Projects (grant No.202001BB050032). We thank Mr. F.-X. Shen for the valuable discussions and testings. This research made use of ccdproc, an Astropy package for image reduction \citep{cra21}, and Photutils, an Astropy package for detection and photometry of astronomical sources \citep{bra19}.
\end{acknowledgements}

\begin{table}[htpb]
	\renewcommand\arraystretch{1.3} 
	\begin{center}
		\caption{18 stars in time series of RX\,J2102.0+3359$^{a}$.}
		\label{tab4}\fontsize{6pt}{7pt}\selectfont
		\begin{tabularx}{18.4cm}{>{\setlength{\hsize}{0.3cm}}X>{\setlength{\hsize}{2.5cm}}X>{\setlength{\hsize}{1.8cm}}X>{\setlength{\hsize}{1.5cm}}X>{\setlength{\hsize}{1.5cm}}X>{\setlength{\hsize}{2.45cm}}X>{\setlength{\hsize}{1.6cm}}X>{\setlength{\hsize}{1.3cm}}X>{\setlength{\hsize}{2.35cm}}X}
			\hline\hline
			\multirow{2}{*}{No.}\centering&RAJ2000\hspace*{3pt}(degree)\centering&\multirow{2}{*}{Variability$^{b}$}\centering&\multirow{2}{*}{Period\hspace*{3pt}(day)}\centering&\multirow{2}{*}{T$_{\rm eff}$\hspace*{3pt}(K)}\centering&\multirow{2}{*}{Distance$^{c}$\hspace*{3pt}(pc)}\centering&\multirow{2}{*}{Radius\hspace*{3pt}(R$_{\odot}$)}\centering&\multirow{2}{*}{Mass\hspace*{3pt}(M$_{\odot}$)}\centering&\multirow{2}{*}{Classification}\\
			&DEJ2000\hspace*{3pt}(degree)\centering&&&&&&&\\\hline
			10&\;\;21h02m17.1s\;(315$^{\circ}$.571) +34d04m16.7s\;(34$^{\circ}$.071)&GAIA1:N\newline WISE:N&&BAI:6493(332)\newline GAIA2:5744\newline S-H21:6359&S-H21:2184\newline GAIA3D:2234(115/108)&GAIA2:1.86\newline TIC82:1.75&S-H21:1.17\newline TIC82:1.19&GSC242:Star\newline TIC82:DWARF\\
			26&\;\;21h01m41.4s\;(315$^{\circ}$.422) +34d03m26.5s\;(34$^{\circ}$.057)&GAIA1:N\newline WISE:N&&BAI:8123(174)\newline GAIA2:8035\newline S-H21:7376&S-H21:3229\newline GAIA3D:3179(179/169)&TIC82:1.93&S-H21:1.45\newline TIC82:1.93&GSC242:Star\newline TIC82:DWARF\\
			30&\;\;21h02m18.6s\;(315$^{\circ}$.577) +34d03m18.2s\;(34$^{\circ}$.055)&GAIA1:N\newline WISE:N&&S-H:6187&S-H:4189\newline GAIA3D:4605(1391/1250)&&S-H:1.14&GSC242:Star\\
			61&\;\;21h01m37.7s\;(315$^{\circ}$.407) +34d02m31.3s\;(34$^{\circ}$.042)&GAIA1:N&&S-H21:6022&S-H21:3188\newline GAIA3D:3260(528/421)&TIC82:1.02&S-H21:0.93\newline TIC82:1.06&GSC242:Star\newline TIC82:DWARF\\
			96&\;\;21h01m39.8s\;(315$^{\circ}$.416) +34d01m44.6s\;(34$^{\circ}$.029)&GAIA1:N\newline WISE:N&&BAI:6177(299)\newline GAIA2:5424\newline S-H21:5781&S-H21:1339\newline GAIA3D:1384(37/32)&GAIA2:1.36\newline TIC82:1.28&S-H21:1.04\newline TIC82:1.03&GSC242:Star\newline TIC82:DWARF\\
			$^{*}$102$^{d}$&\;\;21h02m04.2s\;(315$^{\circ}$.518) +34d01m29.9s\;(34$^{\circ}$.025)&GAIA1:N\newline WISE:N\newline IBVS81:V0.34\newline ZTFVS:V0.35(2)$^{m}$&GCVS:0.328\newline VSX:0.328\newline ATLAS:0.328\newline ZTFVS:0.328&BAI:5580(278)\newline GAIA2:4900\newline S-H21:5702&S-H21:2126\newline GAIA3D:2200(222/183)&TIC82:1.53&S-H21:1.05\newline TIC82:0.91&ATLAS:SINE\newline VSX:EW\newline ZTFVS:EW\newline IBVS81:EW\newline GCVS:EW\newline GSC242:Star\newline TIC82:DWARF\\
			136&\;\;21h01m44.0s\;(315$^{\circ}$.433) +34d00m44.4s\;(34$^{\circ}$.012)&GAIA1:N\newline WISE:N&&S-H21:5964&S-H21:3412\newline GAIA3D:3478(699/568)&TIC82:1.3&S-H21:1.14\newline TIC82:0.95&GSC242:Star\newline TIC82:DWARF\\
			$^{*}$142&\;\;21h02m11.9s\;(315$^{\circ}$.550) +34d00m31.5s\;(34$^{\circ}$.009)&GAIA1:N\newline WISE:N&&BAI:5734(257)$^{m}$\newline GAIA2:5064$^{m}$\newline S-H21:5265$^{m}$&S-H21:1402$^{m}$\newline GAIA3D:1490(146/137)&GAIA2:0.83\newline TIC82:0.97$^{m}$&S-H21:0.8$^{m}$\newline TIC82:0.95$^{m}$&GSC242:Star\newline TIC82:DWARF\\
			$^{*}$143$^{e}$&\;\;21h01m57.2s\;(315$^{\circ}$.488) +34d00m32.0s\;(34$^{\circ}$.009)&ASASSN:V0.22\newline WISEV:V0.25\newline GAIA1:N\newline WISE:N\newline IBVS81:V0.25\newline ZTFVS:V0.245(5)$^{m}$&GCVS:0.487\newline VSX:0.487\newline ATLAS:0.487\newline ASASSN:0.487\newline WISEV:0.486(1)\newline ZTFVS:0.487&BAI:6478(220)\newline GAIA2:5917\newline S-H21:6229&ASASSN:1535\newline S-H21:1481\newline GAIA3D:1487(33/29)&GAIA2:2.06\newline TIC82:1.98&S-H21:1.14\newline TIC82:1.23&ATLAS:SINE\newline VSX:EW\newline WISEV:EW\newline ZTFVS:EW\newline IBVS81:EW\newline GCVS:EW\newline GSC242:Star\newline TIC82:DWARF\\
			$^{*}$177&\;\;21h02m21.7s\;(315$^{\circ}$.591) +33d59m43.2s\;(33$^{\circ}$.995)&GAIA1:N\newline WISE:N&&BAI:5857(337)\newline GAIA2:5068\newline S-H21:5898&S-H21:908\newline GAIA3D:908(71/52)&GAIA2:1.67\newline TIC82:1.52&S-H21:0.9\newline TIC82:0.96&GSC242:Star\newline TIC82:DWARF\\
			183$^{f}$&\;\;21h02m01.4s\;(315$^{\circ}$.506) +33d59m29.8s\;(33$^{\circ}$.992)&ASASSN:V0.45\newline GAIA1:N\newline VSX:V2.9&VSX:0.118\newline RKCat:0.118&GAIA2:5338&ASASSN:2564\newline GAIA3D:2053(1084/402)&GAIA2:0.95\newline TIC82:0.92&TIC82:1.02&RKCat:NL/AM\newline VSX:AM\newline GSC242:Star\newline TIC82:DWARF\\
			227&\;\;21h02m09.6s\;(315$^{\circ}$.540) +33d58m09.0s\;(33$^{\circ}$.969)&GAIA1:N\newline WISE:N&&BAI:5967(265)\newline GAIA2:5342\newline S-H21:6179&S-H21:2718\newline GAIA3D:2857(306/313)&GAIA2:1.49\newline TIC82:1.39&S-H21:1.17\newline TIC82:1.02&GSC242:Star\newline TIC82:DWARF\\
			280&\;\;21h02m06.9s\;(315$^{\circ}$.529) +33d56m44.0s\;(33$^{\circ}$.946)&GAIA1:N\newline WISE:N&&BAI:5719(273)\newline GAIA3:5127\newline S-H21:5790&GAIA3:1226\newline S-H21:1304\newline GAIA3D:1329(48/47)&TIC82:1.39\newline GAIAp3:1.29(6)$^{m}$&GAIAp3:0.9\newline S-H21:1.0\newline TIC82:0.92&GSC242:Non-Star\newline TIC82:DWARF\\
			285&\;\;21h02m07.7s\;(315$^{\circ}$.532) +33d56m35.5s\;(33$^{\circ}$.943)&GAIA1:N\newline WISE:N&&BAI:4900(100)\newline GAIA2:4874\newline RCS:4909(95)\newline S-H21:4889&S-H21:3156\newline GAIA3D:3266(198/198)&GAIA2:4.9\newline TIC82:5.34&S-H21:1.24&GSC242:Star\newline TIC82:GIANT\\
			287&\;\;21h01m53.0s\;(315$^{\circ}$.471) +33d56m30.2s\;(33$^{\circ}$.942)&GAIA1:N\newline WISE:N&&BAI:6431(333)\newline GAIA2:5759\newline S-H21:6424&S-H21:2597\newline GAIA3D:2581(211/177)&GAIA2:1.79\newline TIC82:1.67&S-H21:1.3\newline TIC82:1.21&GSC242:Star\newline TIC82:DWARF\\
			$^{*}$315&\;\;21h02m11.4s\;(315$^{\circ}$.548) +33d55m29.3s\;(33$^{\circ}$.925)&GAIA1:N\newline WISE:N&&BAI:5296(294)\newline PIC:5114\newline RCS:4969(84)\newline GAIAe3:5500\newline S-H21:5005&PIC:1074\newline S-H21:1017\newline GAIA3D:1015(14)$^{m}$&GAIA2:4.22\newline PIC:4.26\newline TIC82:4.23&PIC:1.52\newline S-H21:1.13&GSC242:Non-Star\newline TIC82:GIANT\\
			$^{*}$356&\;\;21h01m52.9s\;(315$^{\circ}$.470) +33d54m24.8s\;(33$^{\circ}$.907)&GAIA1:N\newline WISE:N&&GAIA2:5324\newline S-H21:6233&S-H21:3692\newline GAIA3D:4683(1573/887)&TIC82:1.38&S-H21:1.32\newline TIC82:1.05&GSC242:Star\newline TIC82:DWARF\\
			$^{*}$361&\;\;21h01m38.1s\;(315$^{\circ}$.409) +33d54m19.8s\;(33$^{\circ}$.905)&GAIA1:N\newline WISE:N&&BAI:5906(293)\newline GAIA2:5077\newline S-H21:5724&S-H21:2992\newline GAIA3D:2650(340/221)&TIC82:1.24&S-H21:0.93\newline TIC82:1.0&GSC242:Star\newline TIC82:DWARF\\
			\hline\hline
		\end{tabularx}
	\end{center}
	\footnotesize{
		$^{*}$ Peculiar type light curve.\\
		$^{a}$ Catalog fullname listed in Table~\ref{tab8}.\\
		$^{b}$ Variability flag: N-Not available, C-Constant, V-Variable. The number behind V flag is the amplitude of variation in the unit of magnitude.\\
		$^{c}$ GAIA3D: Mean of the geometric and photogeometric distances.\\
		$^{d}$ SIMBAD name: V2746\,Cyg\\
		$^{e}$ SIMBAD name: V2743\,Cyg\\
		$^{f}$ Program star: RX\,J2102.0+3359\\
	}
\end{table}

\begin{table}[htpb]
	\renewcommand\arraystretch{1.3} 
	\begin{center}
		\caption{14 stars in time series of Paloma\,J0524+4244$^{a}$.}
		\label{tab5}\fontsize{6pt}{6.2pt}\selectfont
		\begin{tabularx}{18.4cm}{>{\setlength{\hsize}{0.2cm}}X>{\setlength{\hsize}{2.25cm}}X>{\setlength{\hsize}{1.6cm}}X>{\setlength{\hsize}{1.45cm}}X>{\setlength{\hsize}{1.9cm}}X>{\setlength{\hsize}{2.45cm}}X>{\setlength{\hsize}{1.6cm}}X>{\setlength{\hsize}{1.38cm}}X>{\setlength{\hsize}{2.35cm}}X}
			\hline\hline
			\multirow{2}{*}{No.}\centering&RAJ2000\hspace*{3pt}(degree)\centering&\multirow{2}{*}{Variability$^{b}$}\centering&\multirow{2}{*}{Period\hspace*{3pt}(day)}\centering&\multirow{2}{*}{T$_{\rm eff}$\hspace*{3pt}(K)}\centering&\multirow{2}{*}{Distance$^{c}$\hspace*{3pt}(pc)}\centering&\multirow{2}{*}{Radius\hspace*{3pt}(R$_{\odot}$)}\centering&\multirow{2}{*}{Mass\hspace*{3pt}(M$_{\odot}$)}\centering&\multirow{2}{*}{Classification}\\
			&DEJ2000\hspace*{3pt}(degree)\centering&&&&&&&\\\hline
			113&\;\;5h24m20.9s\;(81$^{\circ}$.087) +42d51m59.7s\;(42$^{\circ}$.867)&GAIA1:N\newline WISE:N&&BAI:6723(296)\newline GAIA3:7471\newline S-H21:6573&GAIA3:1852\newline S-H21:2049\newline GAIA3D:2199(174/156)&TIC82:1.76\newline GAIAp3:1.7(2)$^{m}$&GAIAp3:1.66\newline S-H21:1.42\newline TIC82:1.27&UKIDSS6:Non-Star\newline GSC242:Star\newline TIC82:DWARF\\
			130$^{d}$&\;\;5h24m45.0s\;(81$^{\circ}$.187) +42d51m54.9s\;(42$^{\circ}$.865)&ASASSN:V0.17\newline GAIA1:N\newline WISE:N\newline VSX:V0.15\newline ZTFVS:V0.18(3)$^{m}$&VSX:0.048\newline ATLAS:0.095\newline ASASSN:0.048\newline ZTFVS:0.048&BAI:7397(570)\newline GAIA3:12161\newline S-H21:7331&ASASSN:1253\newline GAIA3:1515\newline S-H21:1148\newline GAIA3D:1164(31/30)&TIC82:1.49\newline GAIAp3:1.3(1)$^{m}$&S-H21:1.39\newline TIC82:1.72&ATLAS:MPULSE\newline GAIAv3:DSCT\newline\hspace*{26pt}GDOR\newline\hspace*{26pt}SXPHE\newline VSX:DSCT\newline ZTFVS:DSCT\newline UKIDSS6:Non-Star\newline GSC242:Non-Star\newline TIC82:DWARF\\
			159&\;\;5h25m06.0s\;(81$^{\circ}$.275) +42d51m05.0s\;(42$^{\circ}$.851)&GAIA1:N\newline WISE:N&&BAI:6777(288)\newline GAIA3:8381\newline S-H21:6477&GAIA3:1498\newline S-H21:2229\newline GAIA3D:4426(1215/1037)&TIC82:2.34\newline GAIAp3:1.6(4)$^{m}$&GAIAp3:1.9\newline S-H21:1.51\newline TIC82:1.38&GAIAv3:DSCT\newline\hspace*{26pt}GDOR\newline\hspace*{26pt}SXPHE\newline UKIDSS6:Non-Star\newline GSC242:Star\newline TIC82:DWARF\\
			182&\;\;5h24m33.5s\;(81$^{\circ}$.139) +42d48m50.5s\;(42$^{\circ}$.814)&GAIA1:N\newline WISE:N&&BAI:7526(316)\newline GAIA3:7797\newline S-H21:6611&GAIA3:2518\newline S-H21:2962\newline GAIA3D:3397(405/313)&TIC82:2.57\newline GAIAp3:1.95(1)$^{m}$&GAIAp3:1.76\newline S-H21:1.57\newline TIC82:1.5&UKIDSS6:Non-Star\newline GSC242:Star\newline TIC82:DWARF\\
			$^{*}$222&\;\;5h24m50.9s\;(81$^{\circ}$.212) +42d47m20.3s\;(42$^{\circ}$.789)&GAIA1:N\newline WISE:N\newline TASS:C\newline ZTFPVS:V0.08(7)$^{m}$&ZTFPVS:0.5(5)$^{m}$&BAI:4953(152)\newline RCS:4768(89)\newline GAIA3:6788$^{m}$\newline S-H21:5907$^{m}$&GAIA3:2296$^{m}$\newline S-H21:3302$^{m}$\newline GAIA3D:3889(1018/700)&TIC82:5.91$^{m}$\newline GAIAp3:4.5(8)$^{m}$&GAIAp3:2.8\newline S-H21:1.33$^{m}$\newline TIC82:1.4&UKIDSS6:Non-Star\newline GSC242:Star\newline TIC82:GIANT\\
			$^{*}$233&\;\;5h23m35.8s\;(80$^{\circ}$.899) +42d44m40.3s\;(42$^{\circ}$.745)&GAIA1:N\newline WISE:N&&BAI:7809(364)\newline GAIA3:7998\newline S-H21:7341&GAIA3:3915\newline S-H21:2469\newline GAIA3D:2496(222/177)&TIC82:1.56\newline GAIAp3:2.0(4)$^{m}$&GAIAp3:1.68\newline S-H21:1.4\newline TIC82:1.49&UKIDSS6:Non-Star\newline GSC242:Star\newline TIC82:DWARF\\
			239&\;\;5h23m32.4s\;(80$^{\circ}$.885) +42d44m24.2s\;(42$^{\circ}$.740)&GAIA1:N\newline WISE:N&&BAI:7904(389)\newline S-H:6786$^{m}$\newline GAIA3:9908&S-H:2211$^{m}$\newline GAIA3:5490\newline GAIA3D:3038(248/229)&TIC:3.35\newline GAIAp3:3.1(6)$^{m}$&S-H:1.38$^{m}$\newline TIC:2.12\newline GAIAp3:2.52&TIC:DWARF\newline UKIDSS6:Non-Star\newline GSC242:Star\\
			294&\;\;5h25m45.7s\;(81$^{\circ}$.441) +42d45m29.1s\;(42$^{\circ}$.758)&GAIA1:N\newline WISE:N&&GAIA3:7469\newline S-H21:9051&GAIA3:2680\newline S-H21:6017\newline GAIA3D:11500(2500/2500)&GAIAp3:2.64\newline TIC82:4.71&S-H21:3.25\newline TIC82:1.94&UKIDSS6:Non-Star\newline GSC242:Star\newline TIC82:DWARF\\
			$^{*}$306$^{e}$&\;\;5h25m33.6s\;(81$^{\circ}$.390) +42d44m37.5s\;(42$^{\circ}$.744)&GAIA1:N\newline WISE:N&ATLAS:1.07&LAMOST5:6855(310)\newline GAIA3:7156\newline S-H21:7274&GAIA3:3234\newline S-H21:2050\newline GAIA3D:2380(310/210)&TIC82:4.53\newline GAIAp3:5.43(9)$^{m}$&GAIAp3:2.54\newline S-H21:1.97\newline TIC82:1.38&ATLAS:MSINE\newline UKIDSS6:Non-Star\newline GSC242:Star\newline TIC82:DWARF\\
			381&\;\;5h23m52.8s\;(80$^{\circ}$.970) +42d39m02.7s\;(42$^{\circ}$.651)&GAIA1:N\newline WISE:N\newline TASS:C\newline GAIAsv3:V0.04&GAIAsv3:0.117&BAI:7400(378)\newline GAIA3:6529\newline S-H21:6880&GAIA3:2991\newline S-H21:1925\newline GAIA3D:2060(76/62)&TIC82:3.06\newline GAIAp3:3.7(8)$^{m}$&GAIAp3:2.03\newline S-H21:1.66\newline TIC82:1.28&GAIAv3:DSCT\newline\hspace*{26pt}GDOR\newline\hspace*{26pt}SXPHE\newline UKIDSS6:Non-Star\newline GSC242:Star\newline TIC82:DWARF\\
			$^{*}$442&\;\;5h25m35.5s\;(81$^{\circ}$.398) +42d38m19.6s\;(42$^{\circ}$.639)&GAIA1:N\newline WISE:N&&BAI:7349(258)\newline GAIA2:6506\newline S-H21:6488$^{m}$&GAIA3:1335\newline S-H21:2493$^{m}$\newline GAIA3D:2832(573/517)&GAIAp3:0.51\newline TIC82:1.9$^{m}$&S-H21:1.21$^{m}$\newline TIC82:1.33$^{m}$&GAIAv3:DSCT\newline\hspace*{26pt}GDOR\newline\hspace*{26pt}SXPHE\newline UKIDSS6:Non-Star\newline GSC242:Star\newline TIC82:DWARF\\
			470&\;\;5h25m12.0s\;(81$^{\circ}$.300) +42d36m45.6s\;(42$^{\circ}$.613)&GAIA1:N\newline WISE:N&&BAI:8153(140)\newline GAIA3:8259\newline S-H21:7609&GAIA3:1709\newline S-H21:2104\newline GAIA3D:2456(615/511)&TIC82:2.02\newline GAIAp3:1.8(1)$^{m}$&GAIAp3:2.11\newline S-H21:1.69\newline TIC82:1.66&GAIAv3:DSCT\newline\hspace*{26pt}GDOR\newline\hspace*{26pt}SXPHE\newline UKIDSS6:Non-Star\newline S-F:Radio+Opt\newline GSC242:Star\newline TIC82:DWARF\\
			$^{*}$587&\;\;5h24m54.0s\;(81$^{\circ}$.225) +42d30m10.5s\;(42$^{\circ}$.503)&GAIA1:N\newline WISE:N&&BAI:7079(333)\newline GAIA3:6829\newline S-H21:6531&GAIA3:1651\newline S-H21:1762\newline GAIA3D:1801(71/81)&TIC82:1.44\newline GAIAp3:1.42(6)$^{m}$&GAIAp3:1.41\newline S-H21:1.22\newline TIC82:1.22&UKIDSS6:Non-Star\newline GSC242:Star\newline TIC82:DWARF\\
			641$^{f}$&\;\;5h24m30.5s\;(81$^{\circ}$.127) +42d44m48.2s\;(42$^{\circ}$.747)&GAIA1:N\newline WISE:N\newline R-HR:N\newline GAIAsv3:V0.91\newline VSX:V2.4\newline XMM11:V&VSX:0.109\newline ATLAS:0.996\newline 2MASSCV:0.109\newline RKCat:0.109\newline GAIAsv3:0.016&LAMOST5:3850(178)&2MASSCV:1887\newline GAIA3D:585(27/24)&&&RKCat:NL/AM\newline ATLAS:dubious\newline VLAM:AM\newline 2MASSCV:NL\newline S-A:CV\newline VSX:AM/DQ\newline DOWNCV:AM/DQ\newline UKIDSS6:Non-Star\newline MORX:Other\newline GSC242:Non-Star\newline LAMOSTCV2:CV\newline TIC82:DWARF\\
			\hline\hline
		\end{tabularx}
	\end{center}
	\footnotesize{
		$^{*}$ Peculiar type light curve.\\
		$^{a}$ Catalog fullname listed in Table~\ref{tab8}.\\
		$^{b}$ Variability flag: N-Not available, C-Constant, V-Variable. The number behind V flag is the amplitude of variation in the unit of magnitude.\\
		$^{c}$ GAIA3D: Mean of the geometric and photogeometric distances.\\
		$^{d}$ SIMBAD name: ATO\,J081.1876+42.8659\\
		$^{e}$ SIMBAD name: ATO\,J081.3906+42.7440\\
		$^{f}$ Program star: Paloma\,J0524+4244\\
	}
\end{table}

\begin{table}[htpb]
	\renewcommand\arraystretch{1.3} 
	\begin{center}
		\caption{Multi-band brightness of 18 stars in time series of RX\,J2102.0+3359$^{a}$.}
		\label{tab6}\fontsize{7pt}{7.2pt}\selectfont
		\begin{tabularx}{18.4cm}{>{\setlength{\hsize}{0.15cm}}X>{\setlength{\hsize}{1cm}}X>{\setlength{\hsize}{1.6cm}}X>{\setlength{\hsize}{1.25cm}}X>{\setlength{\hsize}{1.7cm}}X>{\setlength{\hsize}{1.75cm}}X>{\setlength{\hsize}{1.75cm}}X>{\setlength{\hsize}{1.6cm}}X>{\setlength{\hsize}{1.65cm}}X>{\setlength{\hsize}{1.9cm}}X}
			\hline\hline
			No.\centering&FUVmag$^{b}$\centering&NUVmag$^{c}$\centering&Umag$^{d}$\centering&Bmag$^{e}$\centering&Vmag$^{f}$\centering&Rmag$^{g}$\centering&Imag$^{h}$\centering&NIRmag$^{i}$\centering&\hspace{0.5cm}MIRmag$^{j}$\\
			\hline
			10&&GUV:20.3(2)\newline UVOT:20.66UM2&UVOT:17.179&APASS:16.498\newline FON:16.4(3)\newline NOMAD:15.59\newline PSTAR:15.745(4)\newline GAIAe3:15.655(3)\newline GSC242:16.1(3)$^{m}$\newline USNOB1:15.9(1)$^{m}$&APASS:15.55\newline NOMAD:14.71\newline GAIAe3:15.311(3)\newline GSC242:15.55&NOMAD:14.69\newline PSTAR:15.329(6)\newline GAIAe3:14.798(4)\newline GSC242:14.585\newline USNOB1:14.5(2)$^{m}$&USNOB1:14.07\newline GSC242:14.385\newline PSTAR:15.189(5)$^{m}$&2MASS:14.16(3)J\newline \hspace*{21pt}13.92(4)H\newline \hspace*{21pt}13.88(6)K\newline PSTAR:15.110(7)\newline GSC:14.39N&WISE:13.75(3)IW1\newline \hspace*{16pt}13.78(4)IW2\newline \hspace*{16pt}12.512IW3\newline \hspace*{16pt}9.25IW4\\
			26&GUV:19.9(2)&GUV:17.78(5)\newline UVOT:17.808UM2&UVOT:16.129&APASS:15.10(3)$^{m}$\newline FON:13.47(6)$^{m}$\newline NOMAD:14.46\newline PSTAR:14.771(3)\newline GAIAe3:14.776(3)\newline GSC242:15.00(4)$^{m}$\newline USNOB1:15.2(3)$^{m}$&APASS:14.667$^{m}$\newline GSC:14.58\newline NOMAD:14.1\newline GAIAe3:14.632(3)&NOMAD:14.58\newline PSTAR:14.731(5)\newline GAIAe3:14.376(5)\newline GSC242:14.648\newline USNOB1:14.3(2)$^{m}$&USNOB1:14.37\newline GSC242:14.309\newline PSTAR:14.789(4)$^{m}$&2MASS:13.94(3)J\newline \hspace*{21pt}13.74(4)H\newline \hspace*{21pt}13.64(4)K\newline PSTAR:14.752(4)\newline GSC:14.31N&WISE:13.58(3)IW1\newline \hspace*{16pt}13.59(3)IW2\newline \hspace*{16pt}11.979IW3\newline \hspace*{16pt}8.831IW4\\
			30&&UVOT:22.342UM2&UVOT:19.047&NOMAD:17.35\newline PSTAR:17.551(5)\newline GAIAe3:17.385(5)\newline GSC242:18.13(2)$^{m}$\newline USNOB1:17.9(2)$^{m}$&GSC:17.14\newline NOMAD:16.7\newline GAIAe3:17.318(3)&NOMAD:16.57\newline PSTAR:17.042(5)\newline GAIAe3:16.387(5)\newline GSC242:16.708\newline USNOB1:16.4(2)$^{m}$&USNOB1:16.27\newline GSC242:16.476\newline PSTAR:16.787(7)$^{m}$&2MASS:15.75(7)J\newline \hspace*{21pt}15.39(9)H\newline \hspace*{21pt}15.0(1)K\newline PSTAR:16.73(1)\newline GSC:16.48N&WISE:15.23(4)IW1\newline \hspace*{16pt}15.6(1)IW2\newline \hspace*{16pt}12.307IW3\newline \hspace*{16pt}9.305IW4\\
			61&&UVOT:22.87UM2&UVOT:18.989&NOMAD:17.21\newline PSTAR:17.632(7)\newline GAIAe3:17.465(7)\newline GSC242:17.8(1)$^{m}$\newline USNOB1:17.8(3)$^{m}$&GSC:17.02\newline NOMAD:16.61\newline GAIAe3:17.078(3)&NOMAD:16.73\newline PSTAR:17.145(3)\newline GAIAe3:16.528(5)\newline GSC242:16.422\newline USNOB1:16.5(2)$^{m}$&USNOB1:16.2\newline GSC242:16.342\newline PSTAR:16.93(1)$^{m}$&2MASS:15.76(9)J\newline \hspace*{21pt}15.6(1)H\newline \hspace*{21pt}15.8(2)K\newline PSTAR:16.88(1)\newline GSC:16.34N&\\
			96&&GUV:20.8(3)\newline UVOT:21.479UM2&UVOT:17.022&APASS:15.76(5)\newline FON:16.2(2)\newline NOMAD:15.5\newline PSTAR:15.436(2)\newline GAIAe3:15.310(3)\newline GSC242:15.82(4)$^{m}$\newline USNOB1:16.1(3)$^{m}$&APASS:15.13\newline NOMAD:14.44\newline GAIAe3:14.929(3)\newline GSC242:15.13&NOMAD:15.14\newline PSTAR:14.967(2)\newline GAIAe3:14.378(4)\newline GSC242:14.9\newline USNOB1:14.8(4)$^{m}$&USNOB1:14.5\newline GSC242:14.342\newline PSTAR:14.772(4)$^{m}$&2MASS:13.80(3)J\newline \hspace*{21pt}13.45(3)H\newline \hspace*{21pt}13.33(3)K\newline PSTAR:14.686(5)\newline GSC:14.34N&WISE:13.34(3)IW1\newline \hspace*{16pt}13.42(3)IW2\newline \hspace*{16pt}12.776IW3\newline \hspace*{16pt}9.181IW4\\
			102&&UVOT:22.056UM2&UVOT:18.962&NOMAD:17.08\newline PSTAR:17.18(3)\newline ZTFVS:17.173\newline GAIAe3:17.01(1)\newline GSC242:17.6(2)$^{m}$\newline USNOB1:17.5(1)$^{m}$&GSC:16.77\newline NOMAD:16.32\newline IBVS81:16.3$^{m}$\newline GAIAe3:16.524(5)&NOMAD:16.13\newline PSTAR:16.59(3)\newline ZTFVS:16.44\newline GAIAe3:15.86(1)\newline GSC242:15.988\newline USNOB1:15.9(2)$^{m}$&USNOB1:15.91\newline GSC242:15.666\newline PSTAR:16.22(2)$^{m}$&2MASS:14.99(3)J\newline \hspace*{21pt}14.55(5)H\newline \hspace*{21pt}14.54(9)K\newline PSTAR:16.20(3)\newline GSC:15.67N&WISE:14.46(3)IW1\newline \hspace*{16pt}14.83(6)IW2\newline \hspace*{16pt}12.865IW3\newline \hspace*{16pt}9.333IW4\\
			136&&&UVOT:19.862&NOMAD:18.59\newline PSTAR:17.891(6)\newline GAIAe3:17.701(7)\newline GSC242:18.5(2)$^{m}$\newline USNOB1:18.3(3)$^{m}$&GSC:17.43\newline GAIAe3:17.229(3)&NOMAD:17.09\newline PSTAR:17.309(4)\newline GAIAe3:16.615(5)\newline GSC242:17.128\newline USNOB1:16.8(3)$^{m}$&USNOB1:16.72\newline GSC242:16.511\newline PSTAR:17.004(8)$^{m}$&2MASS:15.76(6)J\newline \hspace*{21pt}15.5(1)H\newline \hspace*{21pt}15.5(2)K\newline PSTAR:16.93(2)\newline GSC:16.51N&WISE:15.59(5)IW1\newline \hspace*{16pt}16.9(4)IW2\newline \hspace*{16pt}13.029IW3\newline \hspace*{16pt}9.388IW4\\
			142&&&UVOT:18.595$^{m}$&NOMAD:16.525$^{m}$\newline PSTAR:17.67(1)$^{m}$\newline GAIAe3:17.33(1)$^{m}$\newline GSC242:17.2(2)$^{m}$\newline USNOB1:17.2(1)$^{m}$&GSC:16.195$^{m}$\newline NOMAD:15.62$^{m}$\newline GAIAe3:16.823(3)$^{m}$&NOMAD:15.86$^{m}$\newline PSTAR:18.10(3)$^{m}$\newline GAIAe3:16.099(7)$^{m}$\newline GSC242:15.707$^{m}$\newline USNOB1:15.6(3)$^{m}$&USNOB1:15.18$^{m}$\newline GSC242:15.374$^{m}$\newline PSTAR:17.58(3)$^{m}$&2MASS:14.83(4)J$^{m}$\newline \hspace*{21pt}14.45(6)H$^{m}$\newline \hspace*{21pt}14.22(8)K$^{m}$\newline PSTAR:17.14(7)$^{m}$\newline GSC:15.375N$^{m}$&WISE:13.45(3)IW1\newline \hspace*{16pt}13.51(3)IW2\newline \hspace*{16pt}12.645IW3\newline \hspace*{16pt}8.875IW4\\
			143&&GUV:18.70(7)\newline UVOT:18.68UM2&UVOT:15.713&APASS:15.016\newline FON:14.5(4)$^{m}$\newline NOMAD:14.31\newline PSTAR:14.34(1)\newline ZTFVS:14.369\newline GAIAe3:14.31(1)\newline GSC242:14.7(2)$^{m}$\newline USNOB1:14.9(4)$^{m}$&APASS:14.024\newline ASASSN:14.07\newline NOMAD:13.81\newline IBVS81:13.64$^{m}$\newline GAIAe3:14.007(4)\newline GSC242:14.024&NOMAD:14.05\newline PSTAR:14.03(2)\newline ZTFVS:13.972\newline GAIAe3:13.52(1)\newline GSC242:13.674\newline USNOB1:13.8(3)$^{m}$&USNOB1:13.99\newline GSC242:13.597\newline PSTAR:13.87(6)$^{m}$&2MASS:12.87(2)J\newline \hspace*{21pt}12.60(2)H\newline \hspace*{21pt}12.55(2)K\newline PSTAR:13.934(8)\newline GSC:13.6N&WISE:12.57(2)IW1\newline \hspace*{16pt}12.60(3)IW2\newline \hspace*{16pt}12.118IW3\newline \hspace*{16pt}9.312IW4\\
			177&&GUV:20.8(3)\newline UVOT:21.399UM2&UVOT:16.657&FON:15.5(2)\newline NOMAD:14.98\newline PSTAR:14.861(2)\newline GAIAe3:14.715(3)\newline GSC242:15.3345(5)$^{m}$\newline USNOB1:15.4(2)$^{m}$&GSC:14.36\newline NOMAD:14.02\newline GAIAe3:14.301(3)&NOMAD:14.26\newline PSTAR:14.312(2)\newline GAIAe3:13.692(4)\newline GSC242:13.997\newline USNOB1:14.0(3)$^{m}$&USNOB1:13.71\newline GSC242:13.741\newline PSTAR:14.073(4)$^{m}$&2MASS:12.98(2)J\newline \hspace*{21pt}12.59(2)H\newline \hspace*{21pt}12.50(2)K\newline PSTAR:13.947(4)\newline GSC:13.74N&WISE:12.48(2)IW1\newline \hspace*{16pt}12.54(3)IW2\newline \hspace*{16pt}12.242IW3\newline \hspace*{16pt}9.295IW4\\
			183$^{x}$&GUV:19.0(1)&GUV:18.62(7)\newline UVOT:18.393UM2&UVOT:17.581&NOMAD:17.045$^{m}$\newline PSTAR:19.29(4)$^{m}$\newline GAIAe3:18.99(8)$^{m}$\newline GSC242:17.3(7)$^{m}$\newline USNOB1:17.7(2)$^{m}$&ASASSN:15.66\newline NOMAD:14.59\newline RKCat:16.6\newline GAIAe3:18.565(9)$^{m}$\newline GSC242:15.352&NOMAD:16.32$^{m}$\newline PSTAR:18.86(4)$^{m}$\newline VSX:18.05$^{m}$\newline GAIAe3:17.88(6)$^{m}$\newline GSC242:17.49$^{m}$\newline USNOB1:15.4(9)$^{m}$&USNOB1:16.415$^{m}$\newline GSC242:16.03$^{m}$\newline PSTAR:18.36(4)$^{m}$&2MASS:13.85(2)J\newline \hspace*{21pt}13.48(2)H\newline \hspace*{21pt}13.40(3)K\newline PSTAR:16.99(3)$^{m}$\newline GSC:16.035N$^{m}$&\\
			227&&GUV:21.8(4)\newline UVOT:22.061UM2&UVOT:18.386&NOMAD:16.74\newline PSTAR:16.800(2)\newline GAIAe3:16.656(4)\newline GSC242:17.265(7)$^{m}$\newline USNOB1:17.2(2)$^{m}$&GSC:16.31\newline NOMAD:15.84\newline GAIAe3:16.251(3)&NOMAD:16.13\newline PSTAR:16.301(3)\newline GAIAe3:15.671(4)\newline GSC242:15.919\newline USNOB1:15.8(3)$^{m}$&USNOB1:15.87\newline GSC242:15.732\newline PSTAR:16.057(3)$^{m}$&2MASS:14.93(3)J\newline \hspace*{21pt}14.59(5)H\newline \hspace*{21pt}14.56(9)K\newline PSTAR:15.962(7)\newline GSC:15.73N&WISE:14.56(3)IW1\newline \hspace*{16pt}14.69(6)IW2\newline \hspace*{16pt}12.782IW3\newline \hspace*{16pt}9.154IW4\\
			\hline\hline
		\end{tabularx}
	\end{center}
\end{table}

\begin{table}[htpb]
	\addtocounter{table}{-1}
	\renewcommand\arraystretch{1.3} 
	\begin{center}
		\caption{Continued.}
		\fontsize{7pt}{8pt}\selectfont
		\begin{tabularx}{18.4cm}{>{\setlength{\hsize}{0.15cm}}X>{\setlength{\hsize}{1cm}}X>{\setlength{\hsize}{1.6cm}}X>{\setlength{\hsize}{1.25cm}}X>{\setlength{\hsize}{1.7cm}}X>{\setlength{\hsize}{1.75cm}}X>{\setlength{\hsize}{1.75cm}}X>{\setlength{\hsize}{1.6cm}}X>{\setlength{\hsize}{1.65cm}}X>{\setlength{\hsize}{1.9cm}}X}
			\hline\hline
			No.\centering&FUVmag$^{b}$\centering&NUVmag$^{c}$\centering&Umag$^{d}$\centering&Bmag$^{e}$\centering&Vmag$^{f}$\centering&Rmag$^{g}$\centering&Imag$^{h}$\centering&NIRmag$^{i}$\centering&\hspace{0.5cm}MIRmag$^{j}$\\
			\hline
			280&&GUV:21.6(4)\newline UVOT:22.65UM2&UVOT:18.001&NOMAD:16.07\newline PSTAR:16.069(2)\newline GAIA3:15.912\newline GSC242:16.6(1)$^{m}$\newline USNOB1:16.6(3)$^{m}$&GSC:15.51\newline NOMAD:14.89\newline GAIA3:15.451&NOMAD:15.47\newline PSTAR:15.488(2)\newline GAIA3:14.824\newline GSC242:15.324\newline USNOB1:15.2(3)$^{m}$&USNOB1:14.56\newline GSC242:14.953\newline PSTAR:18.35(5)$^{m}$&2MASS:14.07(2)J\newline \hspace*{21pt}13.67(3)H\newline \hspace*{21pt}13.56(4)K\newline PSTAR:17.2(1)$^{m}$\newline GSC:14.95N&WISE:13.64(7)IW1\newline \hspace*{16pt}13.75(7)IW2\newline \hspace*{16pt}12.226IW3\newline \hspace*{16pt}8.626IW4\\
			285&&&UVOT:17.908&FON:16.2(2)\newline NOMAD:15.49\newline PSTAR:15.488(2)\newline GAIAe3:15.252(3)\newline GSC242:16.0(1)$^{m}$\newline USNOB1:16.2(2)$^{m}$&GSC:14.87\newline NOMAD:14.43\newline GAIAe3:14.654(3)&NOMAD:14.64\newline PSTAR:14.705(2)\newline GAIAe3:13.916(4)\newline GSC242:14.368\newline USNOB1:14.4(3)$^{m}$&USNOB1:14.25\newline GSC242:13.901\newline PSTAR:14.265(5)$^{m}$&2MASS:12.92(2)J\newline \hspace*{21pt}12.37(2)H\newline \hspace*{21pt}12.25(2)K\newline PSTAR:14.045(4)\newline GSC:13.9N&WISE:12.20(2)IW1\newline \hspace*{16pt}12.28(2)IW2\newline \hspace*{16pt}12.0(3)IW3\newline \hspace*{16pt}9.21IW4\\
			287&&GUV:20.1(2)\newline UVOT:20.572UM2&UVOT:17.566&FON:16.86\newline NOMAD:18.09\newline PSTAR:16.226(2)\newline GAIAe3:16.114(3)\newline GSC242:16.51(7)$^{m}$\newline USNOB1:16.4(3)$^{m}$&GSC:15.6\newline NOMAD:16.56\newline GAIAe3:15.770(3)&NOMAD:15.62\newline PSTAR:15.839(2)\newline GAIAe3:15.249(4)\newline GSC242:15.628\newline USNOB1:15.4(2)$^{m}$&USNOB1:15.21\newline GSC242:15.155\newline PSTAR:15.660(5)$^{m}$&2MASS:14.68(3)J\newline \hspace*{21pt}14.31(4)H\newline \hspace*{21pt}14.30(7)K\newline PSTAR:15.599(6)\newline GSC:15.15N&WISE:14.36(3)IW1\newline \hspace*{16pt}14.47(6)IW2\newline \hspace*{16pt}12.719IW3\newline \hspace*{16pt}9.147IW4\\
			315&&GUV:20.1(2)\newline UVOT:20.894UM2&UVOT:15.327&FON:13.7(2)\newline NOMAD:14.0\newline PSTAR:13.743(1)\newline GAIAe3:13.103(3)\newline GSC242:13.51(6)$^{m}$\newline USNOB1:13.53(8)$^{m}$&GAIAe3:12.593(3)\newline GSC:12.4&NOMAD:12.37\newline PSTAR:12.770(1)\newline GAIAe3:11.924(4)\newline GSC242:12.287\newline USNOB1:12.2(1)$^{m}$&USNOB1:11.49\newline PSTAR:12.600(1)\newline GSC242:11.882&2MASS:11.06(2)J\newline \hspace*{21pt}10.60(2)H\newline \hspace*{21pt}10.47(1)K\newline PSTAR:12.106(1)\newline GSC:11.88N&WISE:10.41(2)IW1\newline \hspace*{16pt}10.48(2)IW2\newline \hspace*{16pt}10.30(7)IW3\newline \hspace*{16pt}8.631IW4\\
			356&&&UVOT:19.171&NOMAD:17.34\newline PSTAR:17.60(1)\newline GAIAe3:17.471(6)\newline GSC242:18.1(2)$^{m}$\newline USNOB1:17.9(1)$^{m}$&GSC:17.16\newline NOMAD:16.57\newline GAIAe3:17.046(3)&NOMAD:16.95\newline PSTAR:17.110(3)\newline GAIAe3:16.458(5)\newline GSC242:16.904\newline USNOB1:16.7(3)$^{m}$&USNOB1:16.48\newline GSC242:16.286\newline PSTAR:16.847(9)$^{m}$&2MASS:15.68(5)J\newline \hspace*{21pt}15.44(9)H\newline \hspace*{21pt}15.1(1)K\newline PSTAR:16.74(2)\newline GSC:16.29N&WISE:15.40(4)IW1\newline \hspace*{16pt}15.9(1)IW2\newline \hspace*{16pt}12.496IW3\newline \hspace*{16pt}9.012IW4\\
			361&&GUV:21.9(4)\newline UVOT:22.606UM2&UVOT:18.49&NOMAD:16.86\newline PSTAR:17.097(5)\newline GAIAe3:16.938(4)\newline GSC242:17.6(2)$^{m}$\newline USNOB1:17.6(3)$^{m}$&GSC:16.59\newline NOMAD:15.83\newline GAIAe3:16.510(3)&NOMAD:16.65\newline PSTAR:16.579(4)\newline GAIAe3:15.910(5)\newline GSC242:16.529\newline USNOB1:16.2(4)$^{m}$&USNOB1:15.78\newline GSC242:15.807\newline PSTAR:16.287(6)$^{m}$&2MASS:15.16(5)J\newline \hspace*{21pt}14.68(6)H\newline \hspace*{21pt}14.60(8)K\newline PSTAR:16.202(7)\newline GSC:15.81N&WISE:14.62(3)IW1\newline \hspace*{16pt}14.83(8)IW2\newline \hspace*{16pt}12.656IW3\newline \hspace*{16pt}9.01IW4\\
			\hline\hline
		\end{tabularx}
	\end{center}
	\footnotesize{
		$^{a}$ Catalog fullname listed in Table~\ref{tab8}.\\
		$^{b}$ GUV: AB magnitude in the wavelength of 100-200\,nm.\\
		$^{c}$ GUV: AB magnitude in the wavelength of 200-300\,nm; UVOT: UM2-band AB magnitudes in the wavelengths of 224.6\,nm.\\
		$^{d}$ UVOT: U-band AB magnitude in the wavelength of 346.5\,nm.\\
		$^{e}$ PSTAR: Mean AB magnitude in the wavelength of 486.6\,nm; GAIA(e)3: Integrated BP mean magnitude in the wavelength range 330-680\,nm; GSC242: Mean magnitude in Bj, O, and B photographic band.\\
		$^{f}$ GAIA(e)3: G-band mean magnitude of GAIA.\\
		$^{g}$ PSTAR: Mean AB magnitude in the wavelength of 621.5\,nm; GAIA(e)3: Integrated RP mean magnitude in the wavelength range 640-1050\,nm; GSC242: Mean magnitude in F and E photographic band.\\
		$^{h}$ GSC242: Magnitude in N photographic band in the wavelength of 700-1000\,nm; PSTAR: Mean AB magnitude of i- and z-band magnitudes in the wavelength of 754.5\,nm and 867.9\,nm, respectively.\\ 
		$^{i}$ GSC: N-band magnitude in the wavelength range 0.8-1.5\,$\mu$m; PSTAR: y-band magnitude in the wavelength of 0.96\,$\mu$m; 2MASS: J-, H-, and K-band magnitudes in the wavelengths of 1.25\,$\mu$m, 1.65\,$\mu$m, and 2.17\,$\mu$m, respectively.\\
		$^{j}$ WISE: IW1-, IW2-, IW3-, and IW4-band magnitudes in the wavelengths of 3.35\,$\mu$m, 4.6\,$\mu$m, 11.6\,$\mu$m, and 22.1\,$\mu$m, respectively.\\
		$^{m}$ Mean of multi-observations.\\
		$^{x}$ The X-ray observations include ROSAT: the count rate of 0.13(2) in the total energy range 0.1-2.4\,keV; 2SXPS: the mean count rate of 0.080(4) in the energy range 0.3-10\,keV; 2RXS: the background corrected source counts of ROSAT 119(12).\\
	}
\end{table}

\begin{table}[htpb]
	\renewcommand\arraystretch{1.3} 
	\begin{center}
		\caption{Multi-band brightness of 14 stars in time series of Paloma\,J0524+4244$^{a}$.}
		\label{tab7}\fontsize{7pt}{7pt}\selectfont
		\begin{tabularx}{18.4cm}{>{\setlength{\hsize}{0.05cm}}X>{\setlength{\hsize}{1.25cm}}X>{\setlength{\hsize}{1.7cm}}X>{\setlength{\hsize}{1.3cm}}X>{\setlength{\hsize}{1.6cm}}X>{\setlength{\hsize}{1.6cm}}X>{\setlength{\hsize}{1.75cm}}X>{\setlength{\hsize}{1.7cm}}X>{\setlength{\hsize}{1.5cm}}X>{\setlength{\hsize}{1.9cm}}X}
			\hline\hline
			No.\centering&FUVmag$^{b}$\centering&NUVmag$^{c}$\centering&Umag$^{d}$\centering&Bmag$^{e}$\centering&Vmag$^{f}$\centering&Rmag$^{g}$\centering&Imag$^{h}$\centering&NIRmag$^{i}$\centering&\hspace{0.5cm}MIRmag$^{j}$\\
			\hline
			113&&&IGAPS:16.34&APASS:16.02(4)\newline FON:16.42(3)$^{m}$\newline NOMAD:15.04\newline PSTAR:15.731(4)\newline IGAPS:15.79\newline GAIA3:15.609\newline GSC242:15.97(4)$^{m}$\newline USNOB1:16.2(1)$^{m}$&APASS:15.48(8)\newline NOMAD:14.61\newline GAIA3:15.195\newline GSC242:15.477&IGAPS:14.85H$_{\rm\alpha}$\newline \hspace*{21pt}15.05rI\newline \hspace*{21pt}15.08rU\newline NOMAD:15.49\newline PSTAR:15.261(1)\newline GAIA3:14.603\newline GSC242:14.916\newline USNOB1:15.23(3)$^{m}$&USNOB1:14.7\newline IGAPS:14.62\newline GSC242:14.613\newline PSTAR:14.979(3)$^{m}$&2MASS:13.90(3)J\newline \hspace*{21pt}13.59(4)H\newline \hspace*{21pt}13.50(5)K\newline PSTAR:14.827(7)\newline GSC:14.61N\newline UKIDSS6:13.486(4)K&WISE:13.41(3)IW1\newline \hspace*{16pt}13.48(4)IW2\newline \hspace*{16pt}12.216IW3\newline \hspace*{16pt}8.941IW4\\
			130&&&IGAPS:14.16&APASS:14.22(7)\newline FON:14.19(9)\newline NOMAD:14.005$^{m}$\newline PSTAR:13.97(1)\newline IGAPS:18.15(9)$^{m}$\newline ZTFVS:13.933\newline GAIA3:13.923\newline GSC242:13.97(4)$^{m}$\newline USNOB1:16(2)$^{m}$&APASS:13.90(7)$^{m}$\newline ASASSN:13.58\newline NOMAD:13.34\newline GAIA3:13.699\newline GSC:13.95&IGAPS:17.3(1)H$_{\rm\alpha}^{m}$\newline \hspace*{21pt}17.73(9)rI$^{m}$\newline \hspace*{21pt}13.55rU\newline NOMAD:13.88\newline PSTAR:13.83(1)\newline ZTFVS:13.714\newline VSX:13.71\newline GAIA3:13.312\newline GSC242:13.46\newline USNOB1:14.0(1)$^{m}$&USNOB1:13.67\newline IGAPS:17.1(1)$^{m}$\newline GSC242:13.372\newline PSTAR:15.41(4)$^{m}$&2MASS:12.81(2)J\newline \hspace*{21pt}12.65(3)H\newline \hspace*{21pt}12.57(2)K\newline PSTAR:16.5(1)$^{m}$\newline GSC:13.37N\newline UKIDSS6:12.583(2)K&WISE:12.56(2)IW1\newline \hspace*{16pt}12.61(3)IW2\newline \hspace*{16pt}12.302IW3\newline \hspace*{16pt}9.074IW4\\
			159&UVOT:20.794$^{m}$&UVOT:21.892UM2\newline \hspace*{16pt}19.524UW1$^{m}$&UVOT:17.814\newline IGAPS:16.45(1)&APASS:16.3(1)\newline FON:16.1(1)\newline NOMAD:15.28\newline PSTAR:15.854(3)\newline UVOT:16.103\newline IGAPS:15.92\newline GAIA3:17.559$^{m}$\newline GSC242:16.1(2)$^{m}$\newline USNOB1:16.1(3)$^{m}$&APASS:15.6(1)\newline NOMAD:14.79\newline UVOT:15.543\newline GAIA3:17.226$^{m}$\newline GSC242:15.58&IGAPS:16.79(2)H$_{\rm\alpha}^{m}$\newline \hspace*{21pt}16.793(3)rI$^{m}$\newline \hspace*{21pt}15.19rU\newline NOMAD:14.81\newline PSTAR:15.333(3)\newline GAIA3:16.493$^{m}$\newline GSC242:14.945\newline USNOB1:15.0(2)$^{m}$&USNOB1:14.52\newline IGAPS:16.37(1)$^{m}$\newline GSC242:14.388\newline PSTAR:15.933(6)$^{m}$&2MASS:13.88(3)J\newline \hspace*{21pt}13.57(3)H\newline \hspace*{21pt}13.49(4)K\newline PSTAR:14.789(6)\newline GSC:14.39N\newline UKIDSS6:15.38(3)K$^{m}$&WISE:13.37(3)IW1\newline \hspace*{16pt}13.55(4)IW2\newline \hspace*{16pt}12.688IW3\newline \hspace*{16pt}8.822IW4\\
			182&&&IGAPS:16.09&APASS:15.99(8)\newline FON:16.1(2)\newline NOMAD:15.14\newline PSTAR:15.643(4)\newline IGAPS:15.7\newline GAIA3:15.533\newline GSC242:15.87(8)$^{m}$\newline USNOB1:15.9(2)$^{m}$&APASS:15.38(3)\newline NOMAD:14.8\newline GAIA3:15.169\newline GSC242:15.379&IGAPS:14.89H$_{\rm\alpha}$\newline \hspace*{21pt}15.07rI\newline \hspace*{21pt}15.07rU\newline NOMAD:15.34\newline PSTAR:15.238(1)\newline GAIA3:14.621\newline GSC242:15.042\newline USNOB1:15.2(1)$^{m}$&USNOB1:17.725$^{m}$\newline IGAPS:14.65\newline GSC242:14.575\newline PSTAR:15.003(3)$^{m}$&2MASS:13.98(3)J\newline \hspace*{21pt}13.65(4)H\newline \hspace*{21pt}13.48(4)K\newline PSTAR:14.855(5)\newline GSC:14.58N\newline UKIDSS6:13.576(4)K&WISE:13.54(3)IW1\newline \hspace*{16pt}13.60(3)IW2\newline \hspace*{16pt}12.466IW3\newline \hspace*{16pt}9.074IW4\\
			222&&&IGAPS:16.74(1)&APASS:15.69(3)\newline FON:15.8(2)\newline NOMAD:14.75\newline PSTAR:14.982(3)\newline IGAPS:15.02\newline ZTFPVS:14.965\newline GAIA3:14.659\newline GSC242:15.5(2)$^{m}$\newline USNOB1:15.5(5)$^{m}$&APASS:14.35(5)\newline NOMAD:13.93\newline TASS:14.0(2)\newline GAIA3:13.903\newline GSC242:14.35&IGAPS:13.47H$_{\rm\alpha}$\newline \hspace*{21pt}13.84rI\newline \hspace*{21pt}17.50(6)rU$^{m}$\newline NOMAD:13.87\newline PSTAR:14.010(2)\newline ZTFPVS:13.858\newline NSVS:13.8(2)\newline GAIA3:13.055\newline GSC242:13.491\newline USNOB1:13.82(5)$^{m}$&TASS:12.9(2)\newline USNOB1:15.75\newline PSTAR:13.2515(9)\newline IGAPS:13.13\newline GSC242:12.937&2MASS:11.82(2)J\newline \hspace*{21pt}11.19(3)H\newline \hspace*{21pt}11.04(2)K\newline PSTAR:13.052(2)\newline GSC:12.94N\newline UKIDSS6:11.021(1)K&WISE:10.91(2)IW1\newline \hspace*{16pt}11.01(2)IW2\newline \hspace*{16pt}10.7(1)IW3\newline \hspace*{16pt}8.365IW4\\
			233&&&IGAPS:16.48&APASS:16.2(2)\newline FON:16.2(1)$^{m}$\newline NOMAD:15.22\newline PSTAR:15.861(3)\newline IGAPS:15.96\newline GAIA3:15.794\newline GSC242:16.15(6)$^{m}$\newline USNOB1:16.50(7)$^{m}$&APASS:15.63(9)\newline NOMAD:14.93\newline GAIA3:15.475\newline GSC242:15.629&IGAPS:15.25H$_{\rm\alpha}$\newline \hspace*{21pt}15.37rI\newline \hspace*{21pt}15.37rU\newline NOMAD:15.29\newline PSTAR:15.542(1)\newline GAIA3:14.983\newline GSC242:15.274\newline USNOB1:15.5(2)$^{m}$&USNOB1:16.97\newline IGAPS:14.99\newline GSC242:14.913\newline PSTAR:15.39(1)$^{m}$&2MASS:14.40(3)J\newline \hspace*{21pt}14.20(4)H\newline \hspace*{21pt}14.07(6)K\newline PSTAR:15.254(5)\newline GSC:14.91N\newline UKIDSS6:14.088(6)K&WISE:13.99(3)IW1\newline \hspace*{16pt}13.94(4)IW2\newline \hspace*{16pt}12.075IW3\newline \hspace*{16pt}8.999IW4\\
			239&&&IGAPS:15.63&APASS:15.40(7)\newline FON:15.5(1)\newline NOMAD:14.45\newline PSTAR:15.101(1)\newline IGAPS:15.22\newline GAIA3:15.088\newline GSC242:15.3(1)$^{m}$\newline USNOB1:15.78(8)$^{m}$&APASS:14.89(5)\newline NOMAD:14.17\newline GAIA3:16.027$^{m}$\newline GSC242:14.894&IGAPS:16.09(1)H$_{\rm\alpha}^{m}$\newline \hspace*{21pt}16.92(1)rI$^{m}$\newline \hspace*{21pt}14.76rU\newline NOMAD:14.63\newline PSTAR:14.8564(9)\newline GAIA3:14.39\newline GSC242:14.537\newline USNOB1:14.73(9)$^{m}$&USNOB1:16.4\newline IGAPS:15.695(5)$^{m}$\newline GSC242:14.106\newline PSTAR:14.654(4)$^{m}$&2MASS:13.68(3)J\newline \hspace*{21pt}13.47(5)H\newline \hspace*{21pt}13.32(4)K\newline PSTAR:14.483(9)\newline GSC:14.11N\newline UKIDSS6:13.693(4)K&WISE:13.07(3)IW1\newline \hspace*{16pt}13.09(3)IW2\newline \hspace*{16pt}11.946IW3\newline \hspace*{16pt}8.722IW4\\
			294&&&IGAPS:15.47&APASS:15.31(7)\newline FON:15.5(2)\newline NOMAD:14.35\newline PSTAR:14.949(6)\newline IGAPS:15.395$^{m}$\newline GAIA3:14.865\newline GSC242:15.0(2)$^{m}$\newline USNOB1:15.5(2)$^{m}$&APASS:14.67(8)\newline NOMAD:14.09\newline GAIA3:14.548\newline GSC242:14.67&IGAPS:14.25H$_{\rm\alpha}$\newline \hspace*{21pt}14.43rI\newline \hspace*{21pt}14.83rU$^{m}$\newline NOMAD:14.17\newline PSTAR:14.612(2)\newline GAIA3:14.038\newline GSC242:14.054\newline USNOB1:14.3(2)$^{m}$&USNOB1:13.4\newline IGAPS:14.07\newline GSC242:14.044\newline PSTAR:14.425(5)$^{m}$&2MASS:13.38(3)J\newline \hspace*{21pt}13.17(4)H\newline \hspace*{21pt}13.10(3)K\newline PSTAR:14.299(4)\newline GSC:14.04N\newline UKIDSS6:13.104(3)K&WISE:12.97(3)IW1\newline \hspace*{16pt}12.99(3)IW2\newline \hspace*{16pt}11.8(3)IW3\newline \hspace*{16pt}8.633IW4\\
			306&&&IGAPS:14.53&APASS:14.37(7)\newline FON:14.37(4)\newline NOMAD:13.7\newline PSTAR:13.986(3)\newline IGAPS:14.06\newline LAMOST7:13.89\newline GAIA3:13.899\newline GSC242:14.1(2)$^{m}$\newline USNOB1:14.2(5)$^{m}$&APASS:13.73(6)\newline NOMAD:13.34\newline GAIA3:13.548\newline GSC242:13.734&IGAPS:13.25H$_{\rm\alpha}$\newline \hspace*{21pt}13.46rI\newline \hspace*{21pt}13.43rU\newline NOMAD:13.24\newline PSTAR:13.611(2)\newline LAMOST7:13.46\newline GAIA3:12.997\newline GSC242:13.225\newline USNOB1:13.5(2)$^{m}$&USNOB1:15.45\newline PSTAR:13.320(6)\newline IGAPS:13.04\newline GSC242:12.816\newline LAMOST7:13.25&2MASS:12.31(2)J\newline \hspace*{21pt}12.03(3)H\newline \hspace*{21pt}11.88(2)K\newline PSTAR:13.21(1)\newline GSC:12.82N\newline UKIDSS6:11.861(1)K&WISE:11.76(2)IW1\newline \hspace*{16pt}11.76(2)IW2\newline \hspace*{16pt}11.5(2)IW3\newline \hspace*{16pt}8.543IW4\\
			\hline\hline
		\end{tabularx}
	\end{center}
\end{table}

\begin{table}[htpb]\tiny
	\addtocounter{table}{-1}
	\renewcommand\arraystretch{1.3} 
	\begin{center}
		\caption{Continued.}
		\fontsize{7pt}{7.5pt}\selectfont
		\begin{tabularx}{18.4cm}{>{\setlength{\hsize}{0.05cm}}X>{\setlength{\hsize}{1.2cm}}X>{\setlength{\hsize}{1cm}}X>{\setlength{\hsize}{1.2cm}}X>{\setlength{\hsize}{1.7cm}}X>{\setlength{\hsize}{1.7cm}}X>{\setlength{\hsize}{1.9cm}}X>{\setlength{\hsize}{1.8cm}}X>{\setlength{\hsize}{1.7cm}}X>{\setlength{\hsize}{1.9cm}}X}
			\hline\hline
			No.\centering&FUVmag$^{b}$\centering&NUVmag$^{c}$\centering&Umag$^{d}$\centering&Bmag$^{e}$\centering&Vmag$^{f}$\centering&Rmag$^{g}$\centering&Imag$^{h}$\centering&NIRmag$^{i}$\centering&\hspace{0.5cm}MIRmag$^{j}$\\
			\hline
			381&&&IGAPS:15.16&FON:14.4(1)$^{m}$\newline NOMAD:13.96\newline PSTAR:14.521(4)\newline IGAPS:14.55\newline GAIA3:14.404\newline GSC242:14.7(2)$^{m}$\newline USNOB1:15.0(4)$^{m}$&NOMAD:13.65\newline TASS:12.8(1)\newline KELT:13.438Sp\newline GAIA3:14.021\newline GSC:14.26&IGAPS:13.76H$_{\rm\alpha}$\newline \hspace*{21pt}13.93rI\newline \hspace*{21pt}13.88rU\newline NOMAD:13.9\newline PSTAR:14.0696(1)\newline NSVS:12.77(3)\newline GAIA3:13.46\newline GSC242:13.711\newline USNOB1:14.1(2)$^{m}$&TASS:11.95(8)\newline USNOB1:14.55\newline IGAPS:13.5\newline GSC242:13.422\newline PSTAR:13.828(3)$^{m}$&2MASS:12.76(2)J\newline \hspace*{21pt}12.46(3)H\newline \hspace*{21pt}12.39(3)K\newline PSTAR:13.659(3)\newline GSC:13.42N\newline UKIDSS6:12.342(2)K&WISE:12.25(2)IW1\newline \hspace*{16pt}12.29(2)IW2\newline \hspace*{16pt}11.6(3)IW3\newline \hspace*{16pt}8.828IW4\\
			442&&&IGAPS:14.38&APASS:14.22(6)\newline FON:14.20(6)\newline NOMAD:13.64\newline PSTAR:13.919(4)\newline IGAPS:13.96\newline GAIA3:13.843\newline GSC242:14.0(1)$^{m}$\newline USNOB1:14.5(2)$^{m}$&APASS:13.69(6)\newline NOMAD:13.34\newline GAIA3:13.562\newline GSC242:13.688&IGAPS:13.3H$_{\rm\alpha}$\newline \hspace*{21pt}13.45rI\newline \hspace*{21pt}13.43rU\newline NOMAD:13.65\newline PSTAR:13.6236(7)\newline GAIA3:13.123\newline GSC242:13.352\newline USNOB1:13.655(5)$^{m}$&USNOB1:13.93\newline IGAPS:13.15\newline GSC242:13.07\newline PSTAR:13.525(3)$^{m}$&2MASS:12.62(2)J\newline \hspace*{21pt}12.39(3)H\newline \hspace*{21pt}12.35(2)K\newline PSTAR:13.429(2)\newline GSC:13.07N\newline UKIDSS6:12.334(1)K&WISE:12.28(2)IW1\newline \hspace*{16pt}12.28(2)IW2\newline \hspace*{16pt}12.0(3)IW3\newline \hspace*{16pt}9.123IW4\\
			470$^{r}$&&&IGAPS:14.72&APASS:14.60(6)\newline FON:14.66(7)\newline NOMAD:13.88\newline PSTAR:14.313(5)\newline IGAPS:14.38\newline GAIA3:14.279\newline GSC242:14.4(1)$^{m}$\newline USNOB1:14.8(3)$^{m}$&APASS:14.16(5)\newline NOMAD:13.66\newline GAIA3:14.052\newline GSC242:14.164&IGAPS:13.87H$_{\rm\alpha}$\newline \hspace*{21pt}13.97rI\newline \hspace*{21pt}13.96rU\newline NOMAD:14.26\newline PSTAR:14.144(2)\newline GAIA3:15.711$^{m}$\newline GSC242:13.789\newline USNOB1:14.20(6)$^{m}$&USNOB1:14.71\newline IGAPS:13.7\newline GSC242:13.678\newline PSTAR:14.082(5)$^{m}$&2MASS:13.19(3)J\newline \hspace*{21pt}13.04(4)H\newline \hspace*{21pt}12.97(3)K\newline PSTAR:13.988(8)\newline GSC:13.68N\newline UKIDSS6:14.91(3)K$^{m}$&WISE:12.90(2)IW1\newline \hspace*{16pt}12.93(3)IW2\newline \hspace*{16pt}12.557IW3\newline \hspace*{16pt}8.499IW4\\
			587&&&IGAPS:15.92&APASS:15.93(9)\newline FON:16.1(2)\newline NOMAD:15.0\newline PSTAR:15.575(4)\newline IGAPS:15.64\newline GAIA3:15.472\newline GSC242:15.7(1)$^{m}$\newline USNOB1:16.0(1)$^{m}$&APASS:15.3(1)\newline NOMAD:15.01\newline GAIA3:15.124\newline GSC242:15.304&IGAPS:14.84H$_{\rm\alpha}$\newline \hspace*{21pt}15.01rI\newline \hspace*{21pt}15.01rU\newline NOMAD:15.37\newline PSTAR:15.193(2)\newline GAIA3:14.593\newline GSC242:14.801\newline USNOB1:15.2(1)$^{m}$&USNOB1:14.5\newline IGAPS:14.62\newline GSC242:14.462\newline PSTAR:14.979(3)$^{m}$&2MASS:13.97(3)J\newline \hspace*{21pt}13.76(4)H\newline \hspace*{21pt}13.63(5)K\newline PSTAR:14.878(5)\newline GSC:14.46N\newline UKIDSS6:13.638(4)K&WISE:13.58(3)IW1\newline \hspace*{16pt}13.62(4)IW2\newline \hspace*{16pt}12.048IW3\newline \hspace*{16pt}9.113IW4\\
			641$^{x}$&&&IGAPS:16.17&NOMAD:16.375$^{m}$\newline PSTAR:17.57(3)\newline IGAPS:17.21\newline MORX:16.9\newline LAMOST7:17.15\newline GAIA3:17.177\newline GSC242:17(1)$^{m}$\newline USNOB1:16.5(3)$^{m}$&GSC:17.61\newline NOMAD:16.41\newline RKCat:16.5\newline VSX:17.3$^{m}$\newline GAIA3:17.115&IGAPS:16.47(1)H$_{\rm\alpha}$\newline \hspace*{21pt}17.27(1)rI\newline \hspace*{21pt}17.06(1)rU\newline NOMAD:18.06$^{m}$\newline PSTAR:17.33(6)\newline MORX:16.9\newline LAMOST7:17.3\newline GAIA3:16.826\newline GSC242:17.016\newline USNOB1:17.2(9)$^{m}$&USNOB1:18.43\newline IGAPS:16.99(1)\newline GSC242:17.361\newline LAMOST7:17.33\newline PSTAR:19.4(2)$^{m}$&2MASS:16.3(1)J\newline \hspace*{21pt}15.7(1)H\newline \hspace*{21pt}15.809K\newline PSTAR:17.25(8)\newline GSC:17.36N\newline UKIDSS6:15.49(2)K&WISE:15.39(4)IW1\newline \hspace*{16pt}15.5(1)IW2\newline \hspace*{16pt}12.005IW3\newline \hspace*{16pt}8.855IW4\\
			\hline\hline
		\end{tabularx}
	\end{center}
\footnotesize{
	$^{a}$ Catalog fullname listed in Table~\ref{tab8}.\\
	$^{b}$ UVOT: The UVW2-band AB magnitude in the wavelength of 192.8\,nm.\\
	$^{c}$ UVOT: The UW1- and UM2-band AB magnitudes in the wavelengths of 260.0\,nm and 224.6\,nm, respectively.\\
	$^{d}$ UVOT: U-band AB magnitude in the wavelength of 346.5\,nm; IGAPS: U\_RGO-band magnitude.\\
	$^{e}$ PSTAR: Mean AB magnitude in the wavelength of 486.6\,nm; GAIA3: Integrated BP mean magnitude in the wavelength range 330-680\,nm; GSC242: Mean magnitude in Bj, O, and B photographic band; IGAPS: g-band magnitude.\\
	$^{f}$ GAIA3: G-band mean magnitude of GAIA; Sp: Photographic magnitude in optical part of the spectrum.\\
	$^{g}$ PSTAR: Mean AB magnitude in the wavelength of 621.5\,nm; GAIA3: Integrated RP mean magnitude in the wavelength range 640-1050\,nm; GSC242: Mean magnitude in F and E photographic band; IGAPS: Magnitude in rU-, rI-, and H$_{\rm\alpha}$-band.\\
	$^{h}$ IGAPS: i-band magnitude; GSC242: Magnitude in N photographic band in the wavelength of 700-1000\,nm; PSTAR: Mean AB magnitude of i- and z-band magnitudes in the wavelength of 754.5\,nm and 867.9\,nm, respectively.\\
	$^{i}$ GSC: N-band magnitude in the wavelength range 0.8-1.5\,$\mu$m; PSTAR: y-band magnitude in the wavelength of 0.96\,$\mu$m; 2MASS: J-, H-, and K-band magnitudes in the wavelengths of 1.25\,$\mu$m, 1.65\,$\mu$m, and 2.17\,$\mu$m, respectively; UKIDSS6: K-band magnitudes in the wavelengths of 2.20\,$\mu$m.\\
	$^{j}$ WISE: IW1-, IW2-, IW3-, and IW4-band magnitudes in the wavelengths of 3.35\,$\mu$m, 4.6\,$\mu$m, 11.6\,$\mu$m, and 22.1\,$\mu$m, respectively.\\
	$^{m}$ Mean of multi-observations.\\
	$^{r}$ The integrated 1.4 GHz flux density of the radio observations include NVSS: 7($\pm$2) and S-F: 6.8($\pm$2.0).\\
	$^{x}$ The X-ray observations include XMMSL2: the count rates of 3.196, 1.538, and 1.311 in the total energy range 0.2-2\,keV, 2-12\,keV, and 0.2-12\,keV, respectively; S-E: 1.3(2)$\times$10$^{-4}$\,photon/cm2/s, 4(1)$\times$10$^{-5}$\,photon/cm2/s, 4(1)$\times$10$^{-5}$\,photon/cm2/s, 2(1)$\times$10$^{-5}$\,photon/cm2/s,4$\times$10$^{-6}$\,photon/cm2/s in 14-20\,keV, 20-24\,keV, 24-35\,keV, 35-50\,keV, 75-100\,keV; respectively; ROSAT: the count rate of 0.08(1) in the total energy range 0.1-2.4\,keV; WGA: the background subtracted count rate of ROSAT, 0.094(3) accumulated in PHA bins 24 to 224 (0.24-2.0keV); 2RXS: the background corrected source counts of ROSAT, 31(6); S-A: the flux 5(+3/-2)$\times$10$^{-12}$\,erg/s/cm$^{2}$ in 4-12\,keV; XMM11: the mean flux 6.99(4)$\times$10$^{-12}$\,erg/s/cm$^{2}$ in the energy range 0.2-12keV.\\
}
\end{table}

\begin{table}[htpb]
	\renewcommand\arraystretch{1.3} 
	\begin{center}
		\caption{The queried VizieR Online Data Catalogs.}
		\label{tab8}\fontsize{9pt}{8.5pt}\selectfont
		\begin{tabularx}{18.4cm}{>{\setlength{\hsize}{1.7cm}}X>{\setlength{\hsize}{11.1cm}}X>{\setlength{\hsize}{3.5cm}}X>{\setlength{\hsize}{0.2cm}}X}
			\hline\hline
			Abbreviation$^{a}$\centering&Catalog fullname\centering&Catalog Designation$^{b}$\centering&Ref.$^{c}$\\
			\hline
			2MASS&Two-Micron All Sky Survey&II/246&(1)\\
			2MASSCV&2MASS photometry of CVs&J/other/NewA/13.133/table1&(2)\\
			2RXS&The 2nd R\"{o}entgen satellite (ROSAT) all-sky survey source catalog&J/A+A/588/A103/cat2rxs&(3)\\
			2SXPS&The Swift-XRT Point Source catalog&IX/58/2sxps&(4)\\
			APASS&The American Association of Variable Star Observers (AAVSO) Photometric All Sky Survey&II/336/apass9&(5)\\
			ASASSN&The All-Sky Automated Survey for Supernovae Catalog of variable stars&II/366/catalog&(6)\\
			ATLAS&The Asteroid Terrestrial-impact Last Alert System&J/AJ/156/241/table4&(7)\\
			BAI&Machine-learning Regression of T$_{\rm eff}$ derived from the 2nd Data Release of Global Astrometric Interferometer For Astrophysics (GAIA)&J/AJ/158/93/table2&(8)\\
			DOWNCV&The catalog of CVs&V/123A/cv&(9)\\
			FON&The Northern Sky Survey&I/342/f3&(10)\\
			GAIA1&The 1st Data Release of GAIA&I/337/gaia&(11)\\
			GAIA2&The 2nd Data Release of GAIA&I/345/gaia2&(12)\\
			GAIAe3&The early 3rd Data Release of GAIA&I/350/gaiaedr3&(13)\\
			GAIA3&The 3rd Data Release Part 1 Main source of GAIA&I/355/gaiaedr3&(14)\\
			GAIAp3&1D astrophysical parameters of GAIA3&I/355/paramp&(15)\\
			GAIAv3&Classification of variability in GAIA3&I/358/vclassre&(16)\\
			GAIAsv3&Short-timescale sources in Gaia DR3 Part 4. Variability&I/358/vst&(17)\\
			GAIA3D&The estimated distance catalog corresponding to GAIA3&I/352/gedr3dis&(18)\\
			GCVS&The General Catalog of Variable Stars&B/gcvs/gcvs\_cat&(19)\\
			GSC&The Guide Star Catalog&I/305&(20)\\
			GSC242&The Guide Star Catalog V2.4.2&I/353/gsc242&(20)\\
			GUV&The Galaxy Evolution Explorer catalog of UV sources&II/335/galex\_ais&(21)\\
			IBVS81&The 81th name-list of variable stars. II.&J/other/IBVS/6155/nl81\_2&(22)\\
			IGAPS&The Isaac Newton Telescope (INT) Galactic Plane Survey&V/165/igapsdr1&(23)\\
			KELT&The Kilodegree Extremely Little Telescope survey&J/AJ/155/39/table5&(24)\\
			LAMOST5&Large Sky Area Multi-Object Fiber Spectroscopy Telescope (LAMOST) DR5&J/ApJS/245/34/catalog&(25)\\
			LAMOST7&LAMOST DR7&V/156/dr7lrs&(26)\\
			LAMOSTCV2&A catalog of 323 CVs from LAMOST DR6.&J/ApJS/257/65/cand&(27)\\
			MORX&The Million Optical Radio/X-ray Associations catalog&V/148/morx&(28)\\
			NOMAD&The Naval Observatory Merged Astrometric Dataset&I/297/out&(29)\\
			NSVS&The Northern Sky Variability Survey&II/287/skydot&(30)\\
			NVSS&The National Radio Astronomy Observatory (NRAO) Very Large Array (VLA) Sky Survey&VIII/65/nvss&(31)\\
			PIC&The all-sky ESA PLAnetary Transits and Oscillations of stars (PLATO) mission input catalog&J/A+A/653/A98/aspic1\_1&(32)\\
			PSTAR&The Pan-STARRS survey&II/349/ps1&(33)\\
			R-HR&The ROSAT HRI Pointed Observations&IX/28A/hricat&(34)\\
			RCS&A photometric Sample of 2.6 million red clump stars&J/MNRAS/495/3087/catalog&(35)\\
			RKCat&The Cataclysmic Binaries, LMXBs, and related objects Catalogs, V7.24&B/cb/cbdata&(36)\\
			ROSAT&The ROSAT All-Sky Survey Bright Source Catalog&IX/10A/1rxs&(37)\\
			S-A&SRG/ART-XC all-sky X-ray survey catalog&J/A+A/661/A38/catalog&(38)\\
			S-F&The 2nd release of the radio continuum spectra (SPECFIND) catalog&VIII/85A/waste&(39)\\
			S-H&The catalog of astrophysical parameters of stars in GAIA DR2 determined by using the python code StarHorse&I/349/starhorse&(40)\\
			S-H21&The catalog of astrophysical parameters of stars in GAIA EDR3 determined by using the python code StarHorse&I/354/starhorse2021&(41)\\
			S-E&Onboard catalog of known X-ray sources for SVOM-ECLAIRs&J/A+A/645/A18/table&(42)\\
			TASS&The Amateur Sky Survey Mark IV Photometric Survey of the Northern Sky&II/271A/patch2&(43)\\
			TIC&The Transiting Exoplanet Survey Satellite (TESS) Input Catalog&IV/38/tic&(44)\\
			\hline\hline
		\end{tabularx}
	\end{center}
\end{table}

\begin{table}[htpb]
	\addtocounter{table}{-1}
	\renewcommand\arraystretch{1.3} 
	\begin{center}
		\caption{Continued.}
		\fontsize{9pt}{9pt}\selectfont
		\begin{tabularx}{18.4cm}{>{\setlength{\hsize}{1.7cm}}X>{\setlength{\hsize}{11.1cm}}X>{\setlength{\hsize}{3.5cm}}X>{\setlength{\hsize}{0.2cm}}X}
			\hline\hline
			Abbreviation$^{a}$\centering&Catalog fullname\centering&Catalog Designation$^{b}$\centering&Ref.$^{c}$\\
			\hline
			TIC82&The TESS Input Catalog V8.2&IV/39/tic82&(45)\\
			UKIDSS6&The United Kingdom Infrared Telescope (UKIRT) Infrared Deep Sky Survey Galactic Plane Survey (GPS) Release 6&II/316/gps6&(46)\\
			UVOT&The Swift's Ultraviolet/Optical Serendipitous Source Catalog&II/339/uvotssc1&(47)\\
			USNOB1&The U.S. Naval Observatory catalog&I/284/out&(48)\\
			VLAM&The Jansky Very Large Array (VLA) survey of magnetic cataclysmic variable stars. I.&J/AJ/154/252/table2&(49)\\
			VSX&The AAVSO International Variable Star Index&B/vsx/vsx&(50)\\
			WGA&A point source catalog from all ROSAT Position Sensitive Proportional Counter (PSPC) pointed observations&IX/31/wgacat&(51)\\
			WISE&The Wide-field Infrared Survey Explorer&II/328/allwise&(52)\\
			WISEV&The Wide-field Infrared Survey Explorer Catalog of Periodic Variable Stars&J/ApJS/237/28/table2&(53)\\
			XMM11&The XMM-Newton serendipitous survey. IX. (4XMM-DR11)& IX/65/xmm4d11s&(54)\\
			XMMSL2&XMM-Newton slew survey Source Catalog, V2.0&IX/53/xmmsl2c&(55)\\
			ZTFPVS&The Zwicky Transient Facility (ZTF) catalog of periodic variable stars&J/ApJS/249/18/table2&(56)\\
			ZTFVS&The ZTF catalog of suspected variable stars&J/ApJS/249/18/table3&(56)\\
			\hline\hline
		\end{tabularx}
	\end{center}
	\normalsize{
		$^{a}$ Abbreviation of catalog name.\\
		$^{b}$ https://vizier.cds.unistra.fr/viz-bin/VizieR-2\\
		$^{c}$ (1) \cite{cut03}; (2) \cite{ak08}; (3) \cite{bol16}; (4) \cite{eva20}; (5) \cite{hen15}; (6) \cite{jay18}; (7) \cite{hei18}; (8) \cite{bai19}; (9) \cite{dow01}; (10) \cite{and16}; (11) \cite{gai16}; (12) \cite{gai18}; (13) \cite{gai20}; (14) \cite{gai22a}; (15) \cite{gai22b}; (16) \cite{gai22c}; (17) \cite{gai22d}; (18) \cite{bai21}; (19) \cite{sam17}; (20) \cite{las08}; (21) \cite{bia17}; (22) \cite{kaz15}; (23) \cite{mon20}; (24) \cite{oel18}; (25) \cite{xia19}; (26) \cite{luo22}; (27) \cite{sun21}; (28) \cite{fle16}; (29) \cite{zac04}; (30) \cite{woz04}; (31) \cite{con98}; (32) \cite{mon21}; (33) \cite{cham16}; (34) \cite{ros00}; (35) \cite{luc20}; (36) \cite{rit03}; (37) \cite{vog99}; (38) \cite{pav22}; (39) \cite{vol10}; (40) \cite{and19}; (41) \cite{and22}; (42) \cite{dag21}; (43) \cite{dro06}; (44) \cite{sta19}; (45) \cite{pae21}; (46) \cite{luc08}; (47) \cite{yer14}; (48) \cite{mon03}; (49) \cite{bar17}; (50) \cite{wat06}; (51) \cite{whi00}; (52) \cite{cut13}; (53) \cite{che18}; (54) \cite{web20}; (55) \cite{xmm17}; (56) \cite{che20}.\\
	}
\end{table}

\begin{table}[htpb]
	\begin{center}
		\caption{Stars with the periodical light curves.}\label{tab9}
		\begin{tabular}{cccccc}
			\hline\hline
			Time series$^{a}$&Star No.&Star name$^{b}$&Period$_{lcp}^{c}$&Period$_{sin}^{d}$&Amplitude$^{e}$\\
			&&&day&day&mag\\
			\hline
			1&10&J210217+340417&0.082&0.085&0.06\\
			1&183&RX\,J2102.0+3359&0.095&0.103&1.08\\
			2&113&J052421+425160&0.195&0.169&0.024\\
			2&130&J052445+425155&0.047&0.047&0.17\\
			2&159&J052506+425105&0.047&0.046&0.02\\
			2&182&J052434+424851&0.041&0.042&0.03\\
			2&294&J052546+424529&0.124&0.118(2)&0.025\\
			2&381&J052353+423903&0.151&0.137&0.036\\
			2&470&J052512+423646&0.039&0.039&0.017\\
			\hline\hline
		\end{tabular}
	\end{center}
	$^{a}$ The same as Table~\ref{tab3}.\\
	$^{b}$ The celestial coordinates (i.e., RA and Dec) were designed to be the star name for that without the SIMBAD name.\\
	$^{c}$ Period derived by the LSP method.\\
	$^{d}$ The sinusoidal fitting period.\\
	$^{e}$ Full amplitude, i.e., the magnitude difference between the maximal and minimal light.\\
\end{table}

\begin{table}[htpb]
	\begin{center}
		\caption{Stars with the transient light curves.}\label{tab10}
		\begin{tabular}{cccccc}
			\hline\hline
			Time series$^{a}$&Star No.&Star name$^{b}$&Type$^{c}$&Amplitude&Duration\\
			&&&&mag&minute\\
			\hline
			1&26&J210141+340327&dip&-0.04&21\\
			1&30&J210219+340327&hump&0.54&49\\
			1&61&J210138+340231&hump&0.79&26\\
			1&96&J210140+340145&dip&-0.056&23\\
			1&136&J210144+340044&hump&1.37&27\\
			1&227&J210210+335809&dip&-0.16&102\\
			1&280&J210207+335644&hump&0.13&69\\
			1&285&J210208+335636&hump&0.07&47\\
			1&287&J210153+335630&dip&-0.12&86\\
			2&239&J052332+424424&hump&0.022&56\\
			2&641&Paloma\,J0524+4244&dip&-1.43&75\\
			\hline\hline
		\end{tabular}
	\end{center}
	$^{a}$ The same as Table~\ref{tab3}.\\
	$^{b}$ Name designation is the same as Table~\ref{tab9}.\\
	$^{c}$ The hump and dip is corresponding to the positive and negative amplitude, respectively.\\
\end{table}

\clearpage

\appendix                  

\section{Five Steps of the SPDE Program}
\label{app1}

\begin{table}[htpb]
	\begin{center}
		\caption{Quality grades of time series.}\label{taba1}
		\begin{tabular}{ccc}
			\hline\hline
			Grade$^{a}$&F$_{f}$&Operation$^{b}$\\
			\hline
			A&0&N\\
			B&$<$20\%&N\\
			C&20\%-80\%&PM\\
			D&$>$80\%&R/S\\
			\hline\hline
		\end{tabular}
	\end{center}
	$^{a}$ A: Excellent; B: Good; C: Normal; D: Bad.\\
	$^{b}$ N: None operation; PM: Plot and Mark the ``doubtful" images; R/S: Reset the DAOPHOT parameters, or skip to the next time series.\\
\end{table}

\begin{itemize}[itemindent=15pt]
	\item[\textbf{Step1}]{\textbf{Classification:}\\
		Since most of photometry data packets contain many types of calibration images (i.e., bias, dark\footnote{Some types of CCD are able to lower the operating temperature to significantly suppress the counts caused by thermally excited electrons. In this case, the darks are not necessary.}, and flats in different band), the SPDE program provides two options for an automatic classification of the astronomical and calibration images. If there is a keyword  (e.g., IMAGETYPE) listed in the FITS header specifying the image types, then the 1st option is the check of the filenames and keywords. But if not, the 2nd option is a self-classification method combining the unsupervised (K-means) clustering method and Pearson correlation coefficient \citep{egg03}. The outputs of this step are many classified datasets: several raw time series and some calibration images.}	
	\item[\textbf{Step2}]{\textbf{Pre-processing:}\\
		In the past, it is necessary to manually match the corresponding calibration images for each raw time series before the formal data reductions, based on the night log normally recorded during the observations. At present, the SPDE program is able to automatically skip and report the erroneous time series lack of the corresponding calibration images by checking the results in Step1. For the correct time series with the required calibration images, a sequence of calibrated time series can be obtained. This means that the night log now is just a reference used for the manual verification in case of some questionable time series. To some extent, this step can largely reduce the troubles caused by the possible mistakes of manual records in the night log.}	
	\item[\textbf{Step3}]{\textbf{Quality Justification:}\\
		In this step, the data quality of time series is examined, and the reference image with a catalog of the indexed stars is prepared. On the basis of the output matrix of the DAOPHOT \footnote{A famous routine with the documentation on the webpage: \url{http://www.star.bris.ac.uk/~mbt/daophot/}, proposed by \citet{ste87} for crowded-field stellar photometry, is capable of simultaneously identifying and marking the positions and magnitudes of all the stellar objects satisfying the finding algorithms on each of astronomical image.}, the SPDE program automatically justify the quality of time series. Since the night sky cannot always guarantee constant lasting the whole time series length, the changing seeing and transparency may cause the quantitative discrepancy of the star detected by the DAOPHOT on each of images. Based on a criteria that the less detected stars mean the worse image quality, the image with the most of stars (N$_{\rm max}$) detected by the DAOPHOT is the best image of a time series (i.e., the reference image). A lower limit of the ratio of number of the detected stars on an image to N$_{\rm max}$, F$_{d}$, is arbitrarily preset to be 20\%. This means that all images with F$_{d}<$\,20\% are marked with the ``poor-quality'' images and directly removed from the time series avoiding the following matching failures. Assuming that the images with F$_{d}>$\,80\% are the ``good-quality'' images, the ratio of number of the``median-quality" images with F$_{d}$ in a range of 20-80\% to that of all images in a time series (N$_{\rm f}$), F$_{f}$, is used to justify the quality of time series. Table~\ref{taba1} lists the details of this justification.}	
	\item[\textbf{Step4}]{\textbf{Automatic Matching:}\\
		Based on the coordinates matrix of N$_{\rm ci}$ stars on No.i image of the grade-A/B/C time series obtained in Step3, the SPDE program automatically selects the best group of stars for the matching processes (the details of algorithm are presented in Appendix~\ref{app2}). Note that N$_{\rm ci}$ is not the same for the different image in time series, but depending on the quality of No.\,i image. N$_{\rm ci}$=N$_{\rm max}$ is for the reference image, while N$_{\rm ci}<$\,3 is for the worst image since 3 stars are the lowest request of the differential photometry. The matching success is able to determine N$_{\rm cm}$ ($\leq$N$_{\rm max}$) indexed stars ready for the following annulus aperture photometry. To know what each of star is, the common world coordinates (i.e., Right Ascension (RA) and Declination(Dec)) are desired. Using the World Coordinate Systems (WCS) keywords and supplementary data listed in the FITS header, the pixels of the reference image are mapped onto the celestial sphere. However, for the lack of the WCS keywords, the reference image would be uploaded to the web service of Astrometry.net\footnote{\url{http://nova.astrometry.net}}, or directly searched using the downloadable Astrometry.net software package installed in the local computer \citep{lan10}, obtaining the required WCS keywords and supplementary data\footnote{In order to completely get rid of the WCS keywords dependency, an image comparison method between the reference image and finding chart downloaded from the SIMBAD (\url{http://simbad.cds.unistra.fr/simbad/}) is configuring in the future version.} Based on a position list of all stars on the sky (i.e., a sequence of RA and Dec), a cross-identification with the SIMBAD is able to find and mark all N$_{\rm cm}$ stellar objects on the reference image.}
	\item[\textbf{Step5}]{\textbf{Annulus Aperture Photometry:}\\
		Using the catalog of N$_{\rm cm}$ stars output from Step4, the SPDE program is capable of batching annulus aperture photometry for the time series. At the beginning of formal calculations, the FWHM of the radial intensity profile of program star close to the center of each image is estimated using three methods: interpolation, Gaussian-, and Moffat-fit. The former method directly calculating the FWHM of star and background noise level in average is much simpler, while the latter two classical methods are the same as those performed by the IRAF. Then, an applicable aperture shape can be decided by a ratio of FWHM along two directions (X- and Y-axis of CCD frame), R$_{\rm fwhm}$, and a sequence of aperture sizes (N$_{\rm a}$) is preset to locate in a default range of 1$\sim$5\,FWHM. This means that the annulus aperture photometry for one of CCD image produces a N$_{\rm a}\times$N$_{\rm cm}\times$(N$_{\rm cm}$-1)/2 triangular matrix of differential magnitudes. For a time series containing N$_{\rm f}$ images, an indexed triangular matrix with a size of N$_{\rm f}\times$N$_{\rm a}\times$N$_{\rm cm}\times$(N$_{\rm cm}$-1)/2 shown in the schematic diagram (Figure~\ref{figa1}) can be obtained. Each of star has N$_{\rm a}\times$(N$_{\rm cm}$-1) differential light curves, each of which matches with N$_{\rm cm}$-2 reference light curves derived from N$_{\rm cm}$-1 stars. All the optimal N$_{\rm cm}$ differential light curves corresponding to an appropriate aperture with the matched shape and size can be automatically confirmed by searching out their corresponding optimal reference light curves with the minimal scatters from a total of N$_{\rm a}\times$(N$_{\rm cm}$-1)$\times$(N$_{\rm cm}$-2)/2 indexed differential light curves\footnote{To further improve the photometry precision, an auto-adaptive aperture size and shape for each star on each image is configuring in the future version.}. Despite some of stars may share the same reference light curve, the required computations cannot be alleviated. The standard deviation of the corresponding optimal reference light curve, RLC$_{\rm std}$, denotes the uncertainty of annulus aperture photometry. Compared with the previous 4 steps, this step requires the most of the memory and takes the longest runtime. To avoid the case of the memory exhaustion, all inputs and outputs are automatically grouped into several small sections. And each of section is easily interruptable and recoverable at any time. An uneven four-dimension indexed matrix, composed of the star identifier, optimal differential light curve, corresponding reference light curve, and parameters of annulus aperture photometry, is the final data production of the SPDE program.}
\end{itemize}

\begin{figure}
	\centering
	\includegraphics[width=15cm,angle=0]{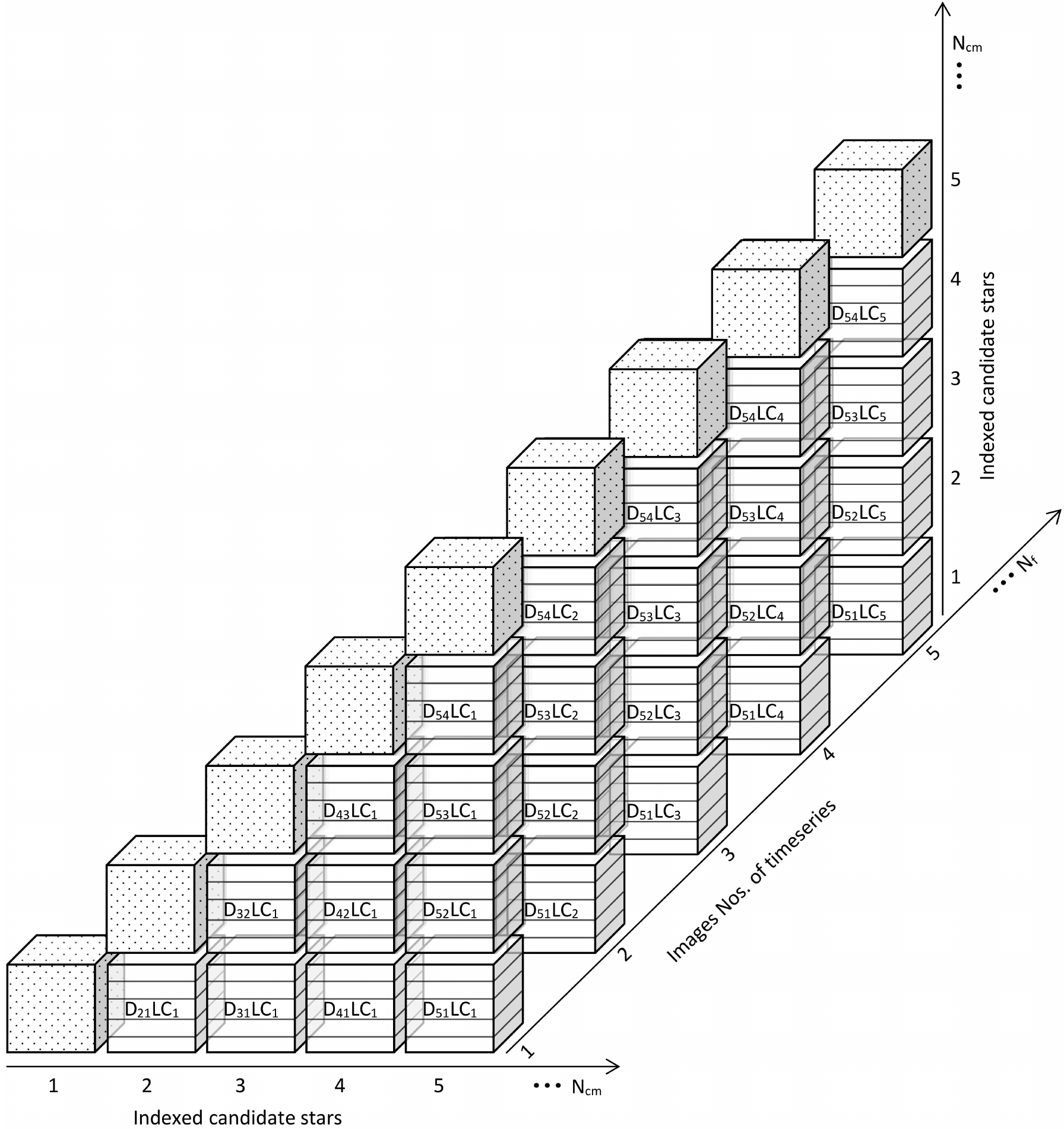}
	\caption{The schematic diagram of a N$_{\rm f}\times$N$_{\rm cm}\times$N$_{\rm cm}$ data cube of N$_{\rm cm}\times$(N$_{\rm cm}$-1)/2 differential light curves produced in Setp5 Annulus Aperture Photometry. Each piece of this data cube is layered by a preset range of N$_{\rm a}$ aperture sizes (e.g., N$_{\rm a}$=5 displays 5 layers in each cube here). The identifier (ID) marked on each piece of data cube denotes the differential magnitude between two indexed candidate stars (two index Nos. are the two subscript digits below ``D") detected in an image of time series (image No. is the subscript digit below ``LC").}
	\label{figa1}
\end{figure}

\section{Two Key Functionalities of the LCA Program}
\label{app2}

\begin{itemize}[itemindent=15pt]
	\item[\textbf{Mod1}]{\textbf{Separations:}\\
		Considering that a variety of variability in photometry light curve are hard to be exactly identified by a general data reduction program, the variations in light curves are roughly separates into periodical-, transient-, and peculiar-type in morphology. To distinguish the light curve profile, 1-D smoothing spline fit (i.e., the scipy API UnivariateSpline) is first used to derive a general variation trend of the sigma-clipped light curve (i.e., a ''clean" light curve by removing outliers). Then, a ratio of R$_{\rm lc}$=Amp/Std, where Amp and Std are the amplitude of fitting light curve and the standard deviation of fitting residuals, respectively, is set to be a variation indicator. For a monotonic light curve, due to Amp=0 derived using the scipy.signal package, R$_{\rm lc}$ becoming an invalid parameter is neglected and Amp is replaced with the magnitude range of fitting light curve with Std$<$S$_{\rm t}$. This monotonic light curve is directly labeled to be peculiar type. Hence, the three parameters, R$_{\rm lc}$, Amp, and Std are combined to carry out a 3-step light curve morphological separation with the following experiential criteria with 6 adjustable thresholds listed in Table~\ref{taba2} depending on the quality of light curve.
		\begin{itemize}
			\item[1.]{The light curves with R$_{\rm lc}>$\;f$_{\rm t}$ are preliminarily filtered from all the optimal N$_{\rm cm}$ light curves.}
			\item[2.]{The light curves with R$_{\rm lc}>$\;f$_{\rm g}$, or R$_{\rm lc}\leq$\;f$_{\rm g}$ but Amp$>$A$_{\rm g}$ are tentatively identified to be the periodical type, while the monotonic (i.e., Amp=0) and irregular light curves with Std$<$\;S$_{\rm t}$ is classified to the peculiar type. All the leftover light curves with the humps/dips satisfying height larger than h$_{\rm t}$ and duration longer than l$_{\rm t}$ are preliminarily labeled to the short-timescale transient type.}
			\item[3.]{Considering that the morphological and brief criteria definitely cause many misclassifications and redundancy checks, the output results are the light curves with the suspected variations including many false-alarm variable light curves. Thus, a manual verification is imperative. Although the limited manual operations lower the degree of automation to some extent, the robustness of this separation routine is significantly strengthened.}
		\end{itemize}
		N$_{\rm lc}$ indexed light curves are finally singled out from the data production of the SPDE program.}
	\item[\textbf{Mod2}]{\textbf{Cross-identifications:}\\
		Based on the celestial coordinates of N$_{\rm lc}$ stars, the query tool of ViezieR \footnote{\url{https://vizier.cds.unistra.fr/viz-bin/VizieR}}, which is the most complete on-line library of published astronomical catalogs \citep{och00}, is able to provide an associated data catalog for each suspected ``variable" star. Considering the possible uncertainties of the mapped coordinates, all the celestial objects listed in the queried VizieR on-line data catalogs locating in a 10\,arcsec region around each of star are recorded. To figure out their basic characteristics for reference, several physical parameters (e.g., modulation period/amplitude, distance, stellar mass/radius/temperature, and so on), multi-band brightness spanning from X-ray to radio, and the queried VizieR on-line catalogs are collected and compiled into three Tables. Due to the sporadic observations in X-ray and radio for most of stars, the output table specifying the multi-band brightness of stars only list the observations from far-ultraviolent (FUV) to mid-infrared (MIR), while the occasional X-ray and radio observations for some special stars are exclusively given as a footnote at the bottom of table.}
\end{itemize}

\begin{table}[htpb]
	\begin{center}
		\caption{Six adjustable thresholds for light curve separation.}\label{taba2}
		\begin{tabular}{ccc}
			\hline\hline
			Parameters&Default$^{a}$&Statements\\
			\hline
			f$_{\rm t}$&1.5&R$_{\rm lc}$ threshold of fitting light curve\\
			f$_{\rm g}$&2.0&lower limit of good R$_{\rm lc}$\\
			A$_{\rm g}$&0.2&lower limit of good Amp\\
			S$_{\rm t}$&0.02&Std threshold of fitting light curve\\
			h$_{\rm t}$&6&hump/dip amplitude threshold in unit of RLC$_{\rm std}$\\
			l$_{\rm t}$&0.05&hump/dip duration threshold in unit of time series length\\
			\hline\hline
		\end{tabular}
	\end{center}
	$^{a}$ On the basis of the average level of data quality observed using a typical meter-class ground-based telescope.\\
\end{table}

\section{Automatic matching algorithm used in the SPDE program}
\label{app3}

Compared with the known matching functions provided by C-Munipack, IRAF, and the other programs commonly requiring manual selections of the landmark stars used for the matching processes, our algorithm automatically picking out the landmark stars from a list of stars indexed and sorted via magnitude is much simpler. Furthermore, since the manual selections of the landmark stars usually face a problem how to select the best group of stars reducing the possibility of matching failure as much as possible, a significant advantage of our matching algorithm is the technical realization that almost all stars (typically from dozens to hundreds) appearing on the reference image can be concerned with the matching processes obviously improving its stability and success rate.

As we know that three non-collinear points (i.e., a triangle) are sufficient to determine a plane, a triangle composed of three bright (landmark) stars approaching to the center of the reference image is singled out to be a standard triangle (e.g., the triangles shown in the left-hand panels of Figure~\ref{fig2}). There are two reasons for the selection of a triangle in our automatic matching algorithm. For one thing, more complex geometric shape composed of more stars only increases the calculation time, rather than the matching success possibility. For another thing, the three bright stars guarantee the largest possibility of appearing on all CCD images of time series, avoiding the possible matching failure caused by the image excursions translating any landmark star outside the CCD frame, or the image blurring resulting from the poor seeing.

For the time series photometry using the small/medium aperture ground-based telescope, it may be still a reasonable assumption that the CCD images covering a small patch of night sky (FOV$<$50$^{'}$ in most cases) taken by the same telescope and the attached CCD instrument have no rotation and scaling, but a bit of translation (i.e., the image excursion) during a few hours at night. This is able to significantly reduce the matching complexity, and correspondingly improve the computing speed. Since the main purpose of the SPDE program is the detection of potential variations in brightness, the magnitude overlap (i.e., a mandatory pre-definition for the constant landmark stars) is not regarded as a judgment criterion. Therefore, the core of our matching algorithm is simplified to be a geometric searching of congruent triangle among a batch of indexed bright stars (the number of matching stars is changeable depending on the output catalog of the DAOPHOT sorted by magnitude) appearing in each of images. 

A massive of our testing trials indicates that a strict judgment criteria of two congruent triangles can significantly lower the possibility of two or more congruent triangles appearing on the same CCD image (i.e., degeneration). Hence, an uniform tolerant deviation of 5\% between two triangles is preset to be a criterion of congruent triangle. This is appropriate to further simplify our matching calculations \footnote{Our matching algorithm is able to automatically adjust the criterion (i.e., tolerant deviation) judging a pair of congruent triangles in a range of 1-10\% to guarantee an optimal triangle appearing on an image.}. Although two judgment criteria: the distance and slopes of the three sides of triangle are enough to determine a pair of congruent triangles, a degeneration case caused by flipping along any side of triangle can cause the matching failure. To eliminate this degeneration, the sign of a designed parameter R$_{\rm i}$\footnote{R$_{\rm i}$=y$_{\rm i}$-[y$_{\rm k}$(x$_{\rm i}$-x$_{\rm j}$)-y$_{\rm j}$(x$_{\rm i}$-x$_{\rm k}$)]/(x$_{\rm k}$-x$_{\rm j}$), where the subscripts of i, j and k denote the three vertices of a triangle, S$_{\rm i}$(x$_{\rm i}$, y$_{\rm i}$), S$_{\rm j}$(x$_{\rm j}$, y$_{\rm j}$) and S$_{\rm k}$(x$_{\rm k}$, y$_{\rm k}$), respectively.}, defined as the difference of y-axis value between a vertex S$_{\rm i}$(x$_{\rm i}$, y$_{\rm i}$) and the point S$_{\rm i}^{'}$(x$_{\rm i}$, y$_{\rm i}^{'}$) on the side L$_{\rm i}$ against this vertex at the same x-axis value, can be an extra judgment criterion during the matching process, since R$_{\rm i}$ exclusively determines the position of the vertex S$_{\rm i}$ against the side L$_{\rm i}$. By adding R$_{\rm i}$ as the third judgment criterion, a templated triangle on the reference image is uniquely defined to search out the optimal pairs of congruent triangles on a sequence of CCD images.

Provided that all three judgment criteria within a default tolerant deviation of 5\% are satisfied for an image, the average deviations of three landmark stars coordinates in pixels between a CCD image and the reference image is regarded as the centroid shifting of this CCD image translation uniformly applying to all the indexed stars appearing on this CCD image. Considering that the star coordinates and the corresponding uncertainties on the CCD image derived by the DAOPHOT are projected from the celestial sphere to the CCD plane, a default tolerant deviation of 2 pixels for the same indexed star on the matched and reference image is preset to carry out an imperative verification, whether the centroid shifting of the image translation is appropriate for all the indexed stars. If the number of stars on an image passing through this verification is more than 80\%\;N$_{\rm max}$, then this image is judged to be matching success, otherwise, be temporary matching failure labeling with the already matched stars plotting for manual verification. In most of cases, the matching failure is caused by missing many indexed stars on bad-quality images resulting from the possible occultation by the patched cloud. To reduce the invalid calculations of annulus aperture photometry in Step5 of the SPDE program, our matching algorithm does not output all the indexed stars appearing on the reference image, but assemble the stars successfully matched in a sequence of images spanning over 2/3 time series length to be a group of re-indexed candidate stars for the following measurements.

\label{lastpage}

\end{document}